\documentclass[12pt]{article}

\usepackage{epsfig}
\usepackage{color}
\usepackage{array}
\usepackage{amsmath}
\usepackage{amssymb}

\newcommand{\bea}{\begin{eqnarray}}
\newcommand{\eea}{\end{eqnarray}}

 \makeatletter
\def\alt{\mathrel{\mathpalette\gl@align<}}
\def\agt{\mathrel{\mathpalette\gl@align>}}
\def\gl@align#1#2{\lower.6ex\vbox{\baselineskip\z@skip\lineskip\z@
\ialign{$\m@th#1\hfil##\hfil$\crcr#2\crcr\sim\crcr}}} \makeatother

 \makeatletter
\def\alt{\mathrel{\mathpalette\gl@align<}}
\def\agt{\mathrel{\mathpalette\gl@align>}}
\def\gl@align#1#2{\lower.6ex\vbox{\baselineskip\z@skip\lineskip\z@
\ialign{$\m@th#1\hfil##\hfil$\crcr#2\crcr\sim\crcr}}} \makeatother

\catcode`@=11
\long\def\@caption#1[#2]#3{\par\addcontentsline{\csname
  ext@#1\endcsname}{#1}{\protect\numberline{\csname
  the#1\endcsname}{\ignorespaces #2}}\begingroup
    \small
    \@parboxrestore
    \@makecaption{\csname fnum@#1\endcsname}{\ignorespaces #3}\par
  \endgroup}
\catcode`@=12
\newcommand{\newc}{\newcommand}
\newc\order{{\cal O}}
\newc\CO{\order}
\begin{document}
\begin{flushright}
OSU-HEP-09-08
\end{flushright}
\vspace*{3.0cm}

\begin{center}
\baselineskip 20pt {\Large\bf TASI Lectures on Flavor Physics}
\vspace{0.5cm}

{\large \bf K.S. Babu\footnote{Email: babu@okstate.edu}}

\vspace{0.2cm}

{\it Department of Physics, Oklahoma State University, Stillwater, OK 74078, USA}

\begin{abstract}
This review is based on lectures on flavor physics given at TASI 2008.  First I summarize
our present knowledge on the fundamental parameters of the flavor sector.  Then I discuss various
scenarios going beyond the standard model which attempt to explain aspects of the ``flavor puzzle".
Relating quark masses and mixing angles via flavor symmetry is explored. Explaining the mass hierarchy
via the Froggatt--Nielsen mechanism is reviewed and illustrated.  Grand unification ideas are pursued to
seek a pattern in the observed masses and mixings of quarks and leptons. Generating light fermion masses
as radiative corrections is explained and illustrated.  The popular solutions to the strong CP problem
are summarized.  Finally, specific processes in $B$ meson system where significant new flavor contributions
can arise are discussed.
\end{abstract}

\end{center}

\newpage

\addtocounter{page}{-1}

\tableofcontents
\newpage

\baselineskip 18pt

\section{Overview\label{sec1}}


This set of lectures will focus on the flavor sector of the
Standard Model (SM). As you know, most of the free parameters of
the SM reside in this sector.  In case you have not thought about it
lately, let me remind you of the counting of parameters of the SM.  Not
including neutrino masses, there are 19 parameters in the SM. Five
of these are flavor universal -- the three gauge couplings $(g_1,~g_2,~g_3)$, one
Higgs quartic coupling $\lambda$, and one Higgs mass-squared
$\mu^2$, while the remaining fourteen are parameters associated with
the flavor sector. Six quark masses, three charged lepton masses, four quark mixing
angles (including one weak CP violating phase) make up thirteen,
while the strong CP violating parameter $\overline{\theta}$, which
is intimately related to the quark masses, is the fourteenth
flavor parameter.  If we include small neutrino masses and
mixing angles into the SM, as needed to explain neutrino oscillation data
from a variety of experiments, an additional nine
parameters will have to be introduced (three neutrino masses,
three neutrino mixing angles and three CP violating phases, in the
case of Majorana neutrinos).  You see that twenty three of the twenty eight parameters
describe flavor physics in the SM.

While there is abundant information on the numerical values these
parameters take, a fundamental understanding of the origin of these parameters is
currently lacking.  Why are there three families of quarks and leptons
in the first place? Are the flavor parameters all arbitrary, or are they inter-connected? Why do
the charged fermion masses exhibit a strong hierarchical structure spanning
some six orders of magnitude? Why are the mixing angles in the quark sector
hierarchical?  Are the mixing parameters related to
the mass ratios? Why is $\overline{\theta} < 10^{-9}$?   Why are neutrino
masses so much smaller than the charged fermion masses?  What causes (at
least two of) the neutrino mixing angles to be much larger than the
corresponding quark mixing angles? What is the origin of CP violation?
The lack of a fundamental understanding of such
issues is often referred to as the ``flavor puzzle".

Various solutions to this puzzle have been proposed, inevitably leading to
physics beyond the Standard Model, for within the SM these parameters can only be
accommodated, and not explained. Forthcoming experiments, especially at the
LHC, have the potential to confirm or refute some, but not all, of these proposed
non--standard scenarios.  If the new flavor dynamics occurs near the TeV scale, it is
potentially accessible to the LHC, but if it occurs at a much
higher scale, then it will not be directly accessible.  It should be mentioned at
the outset that there is no compelling reason for the flavor dynamics to occur
near the TeV scale, most puzzles can be explained even when the dynamics takes place near
the Planck scale.  This is because the small parameters of the flavor sector are
quite stable under radiative corrections, owing to chiral symmetries.  If the smallness
of a certain parameter has an explanation from Planck scale physics, it is an equally
good explanation at the low energy scale.
Testing such high scale theories would be more difficult in general.  In some
cases, for example, with low energy supersymmetry, information from the high scale flavor dynamics
will be carried by particles which survive to the TeV scale (the SUSY particles), in
which case flavor physics may be tested at colliders.  Processes such as lepton flavor violating
$\mu \rightarrow e \gamma$ decay and  $b \rightarrow s \gamma$ transition appear to be promising
setups to test such scenarios.

The Higgs boson is waiting to be discovered at the LHC.
Its production and decay rates can be significantly modified relative to the
SM expectations  in some of the flavor--extensions of the SM. I will describe explicit models
in this category.  Very little is known about the top quark properties currently.
LHC will serve as a top quark factory where modifications in the top sector arising from flavor--extensions can
be studied.  These include flavor changing decays of the top and its possible anomalous couplings
to the gauge bosons.  We have learned a lot about the $B$ meson system from the $B$ factories lately, but
there are still many open issues and some puzzles which will be probed at the LHC.  These include
precise determination of the CP violating parameters, rare processes allowed in the SM
but not yet observed, and new physics processes in $B$ decays that require modification
of the SM structure.

In Sec.~\ref{sec2}, we will take a tour of the flavor parameters of the
SM and review how these are measured and interpreted.  Various
ideas attempting to understand aspects of the flavor puzzle will
then be introduced and their experimental consequences
outlined.  In Sec.~\ref{sec3} we will seek inter--relations between quark masses
and mixing angle.  Sec.~\ref{sec4} will be devoted to an understanding of the fermion
mass and mixing hierarchies based on the Frogatt--Nielsen mechanism.
In Sec.~\ref{sec5} we will develop grand unification as a possible clue to the
flavor puzzle.  Sec.~\ref{sec6} discusses radiative fermion mass generation, Sec.~\ref{sec7}
summarizes the suggested solutions to the strong CP problem, and in Sec.~\ref{sec8}
we introduce specific beyond the SM scenarios for the flavor sector and study
their experimental manifestations at the LHC.

\section{Flavor structure of the Standard Model \label{sec2}}

Because of the chiral structure of weak interactions, bare fermion masses are
not allowed in the Standard Model. Fermion masses arise via Yukawa interactions given by the Lagrangian
\begin{equation}
\label{Yukawa}
{\cal L}_{\rm Yukawa} = Q^T Y_u u^c H - Q^T Y_d d^c \tilde{H} - L^T Y_\ell e^c \tilde{H} + h.c.
\end{equation}
Here I have used the standard notation for quark ($Q, u^c, d^c$) and lepton ($L, e^c)$ fields.
$(Q,~L)$ are $SU(2)_L$ doublets, as is the Higgs field $H$ and its conjugate $\tilde{H} = i \tau_2 H^*$, while
the ($u^c,~d^c,~e^c$) fields are $SU(2)_L$ singlets.  All fermion fields are left--handed, a
charge conjugation matrix $C$ is understood to be sandwiched between all of the fermion bi-linears in Eq. (\ref{Yukawa}). Contraction of the color indices is not displayed, but should be obvious.  $Y_{u,d,\ell}$ are
the Yukawa couplings matrices spanning generation space which are complex and non--Hermitian.  $SU(2)_L$ contraction
between the fermion doublet and Higgs doublet involves the matrix $i \tau_2$.  Explicitly, we have
(for a family labeled by index $i$)
\begin{eqnarray}
\label{def1}
Q_i = \left(\begin{matrix}u_i \\ d_i\end{matrix}\right)\,;~ L_i = \left(\begin{matrix}\nu_i \\ e_i\end{matrix}\right)\,;~
H = \left(\begin{matrix}H^+ \\ H^0\end{matrix}\right)\,;~ \tilde{H} = \left(\begin{matrix}{H^{0*}} \\ -H^-\end{matrix}\right)\,,
\end{eqnarray}
so that Eq. (\ref{Yukawa}) expands to
\begin{equation}
\label{Yukawa1}
{\cal L}_{\rm Yukawa} =
(Y_u)_{ij} [u_i u^c_j H^0 - d_i u^c_j H^+] + (Y_d)_{ij} [u_i d^c_j H^- + d_i d^c_j {H^{0*}}]
+ (Y_\ell)_{ij} [ \nu_i e^c_j H^- + e_i e^c_j {H^{0*}}] + h.c.
\end{equation}
The neutral component of $H$ acquires a vacuum expectation value (VEV) $\left\langle H^0 \right\rangle = v$, spontaneously breaking the electroweak symmetry ($v \simeq 174$ GeV).  The Higgs field can then be parametrized
in the unitary gauge as $H^0= (\frac{h} {\sqrt{2}} + v)$ where $h$ is a real physical field (the Higgs boson).  In this gauge $H^\pm$, which are eaten up by the $W^\pm$ gauge bosons, and the phase of $H^0$, which is eaten up by the $Z^0$ gauge boson, do not appear.

The VEV of $H^0$ generates the following fermion mass matrices:
\begin{equation}
\label{Mude}
M_u = Y_u v\,,~~~~~M_d = Y_d v\,,~~~~~M_\ell = Y_\ell v\,.
\end{equation}
The Yukawa coupling matrices contained in $(Y_u)_{ij}/\sqrt{2} (u
u^c h)$, etc in each of the up, down and charged lepton sector
becomes proportional to the corresponding mass matrix.  Once the
mass matrices are brought to diagonal forms, the Yukawa coupling
matrices will be simultaneously diagonal.  There is thus no
tree--level flavor changing current mediated by the neutral Higgs
boson in the Standard Model.  This is a feature that is
generally lost as we extend the SM  to address the flavor issue
(for example by introducing multiple Higgs doublets or extra fermions).

We make unitary rotations on the quark fields in family space. Unitarity of these rotations will ensure that
the quark kinetic terms remain canonical.  Specifically, we define mass eigenstates ($u^0, u^{c 0}, d^0, d^{c 0}$) via
\begin{eqnarray}
\label{def2}
u &=& V_u ~u^0,~~u^c = V_{u^c}~ u^{c 0}\,, \nonumber\\
d &=& V_d ~d^0,~~d^c = V_{d^c}~ d^{c 0}\,,
\end{eqnarray}
and we choose the unitary matrices such that
\begin{eqnarray}
\label{trans}
V_u^T (Y_u v) V_{u^c} = \left(\begin{matrix}m_u & ~ & ~ \\ ~ & m_c & ~ \\ ~ & ~ & m_t\end{matrix}\right)\,,~~
V_d^T (Y_d v) V_{d^c} = \left(\begin{matrix}m_d & ~ & ~ \\ ~ & m_s & ~ \\ ~ & ~ & m_b\end{matrix}\right)\,.
\end{eqnarray}
We have assumed here that the number of families is three, but the procedure applies to
any number of families. Bi-unitary transformations such as the ones in Eq. (\ref{trans}) can diagonalize non--Hermitian matrices. The same transformations should be applied to all interactions of the quarks.  As already noted,
these transformations will bring the Yukawa interactions of quarks with the Higgs boson into diagonal forms.  The couplings of the $Z^0$ boson and the photon to quarks will have the original diagonal form even after this rotation.
For example, $(\overline{u} \gamma_\mu I u) Z^\mu$ where $I$ is the identity matrix acting on
family space will transform to $(\overline{u^0} \gamma_\mu (V_u^\dagger I V_u) u^0) Z^\mu$, which
is identical to $(\overline{u^0} \gamma_\mu  I u^0) Z^\mu$.  Similarly, $(\overline{u^c} \gamma_\mu I u^c) Z^\mu$
will transform to $(\overline{u^{c0}} \gamma_\mu I u^{c0}) Z^\mu$.  We see that there is no tree level
flavor changing neutral current (FCNC) mediated by the $Z^0$ boson and the photon in the SM.

Most significantly, the transformations of Eq. (\ref{trans}) will bring the charged current quark interaction,
which originally is of the form ${\cal L}_{cc} = g/\sqrt{2}(\overline{u} \gamma_\mu d) W^{+ \mu} + h.c.$, into the form
\begin{equation}
\label{cc}
{\cal L}_{cc} = \frac{g} {\sqrt{2}} [\overline{u^0} \gamma_\mu V d^0]~ W^{\mu +} + h.c.
\end{equation}
where
\begin{equation}
\label{CKM}
V = V_u^\dagger V_d
\end{equation}
is the quark mixing matrix, or the Cabibbo--Kobayashi--Maskawa (CKM) matrix \cite{C,KM}.  In the SM, all the flavor
violation is contained in $V$.  Being product of unitary matrices, $V$ is itself unitary.  This feature
has thus far withstood experimental scrutiny, with further scrutiny expected from LHC experiments.

Note that the right--handed rotation matrices $V_{u^c}$ and
$V_{d^c}$ have completely disappeared, a result of the purely
left--handed nature of charged weak current.

We can repeat this process in the leptonic sector.  We define, in analogy with Eq. (\ref{def2}),
\begin{equation}
\label{def3}
\nu = V_\nu ~\nu^0\,,~~~ e = V_e ~e^0\,,~~~e^c =  V_{e^c}~ e^{c 0}\,.
\end{equation}
We choose $Y_e$ and $Y_{e^c}$ such that
\begin{equation}
\label{translep}
Y_e^T (Y_\ell v) Y_{e^c} = \left(\begin{matrix}m_e & ~ & ~ \\ ~ & m_\mu & ~ \\ ~ & ~ & m_\tau\end{matrix}\right)\,.
\end{equation}

Note that there is no right--handed neutrino in the SM.  If the Yukawa Lagrangian is
as given in Eq. (\ref{Yukawa}), there is no neutrino mass.  In that case one can choose $V_\nu = V_e$,
so that the charged current weak interactions will remain flavor diagonal.  However, it is now well
established that neutrinos have small masses.  Additional terms must be added to Eq. (\ref{Yukawa}) in order
to accommodate them.  The simplest possibility is to add a non--renormalizable term
\begin{equation}
\label{eqseesaw}
{\cal L}_{\nu-{\rm mass}} = \frac{(L^T Y_\nu L) HH} {2M_*} + h.c.
\end{equation}
where the $SU(2)_L$ contraction between the $H$ fields is in the triplet channel and $Y_\nu$ is a
complex symmetric matrix in generation space.  Here $M_*$
is a mass scale much above the weak interaction scale.  Eq. (\ref{eqseesaw}) can arise by integrating out
some heavy fields with mass of order $M_*$.  The most celebrated realization of this is the
seesaw mechanism, where $M_*$ corresponds to the mass of the right--handed neutrino \cite{refseesaw}.  The neutrino
masses are suppressed, compared to the charged fermion masses, because of the inverse dependence
on the heavy scale $M_*$.  Right--handed neutrinos, if they exist, are complete singlets of the
SM gauge symmetry, and can possess bare SM invariant mass terms, unlike any other fermion of the
SM.  This is an elegant explanation of why the neutrinos are much lighter than other fermions,
relying only on symmetry principles and dimensional analysis.  Eq. (\ref{eqseesaw}) leads to a light neutrino
mass matrix given by
\begin{equation}
\label{def4}
M_\nu = Y_\nu \frac{v^2} {M_*}\,.
\end{equation}
Now we choose $V_\nu$ so that
\begin{equation}
\label{transnu}
V_\nu^T Y_\nu \frac{v^2}  {M_*} V_\nu =  \left(\begin{matrix}m_1 & ~ & ~ \\ ~ & m_2 & ~ \\ ~ & ~ & m_3\end{matrix}\right)\,,
\end{equation}
with $m_{1,2,3}$ being the tiny masses of the three light neutrinos.  The leptonic charge current interaction now
becomes
\begin{equation}
\label{cclep}
{\cal L}_{cc}^{\ell} = \frac{g} {\sqrt{2}} [\overline{e^0} \gamma_\mu U \nu^0] ~ W^{- \mu} + h.c.
\end{equation}
where
\begin{equation}
\label{MNS}
U = V_e^\dagger V_\nu
\end{equation}
is the leptonic mixing matrix, or the Pontecorvo--Maki--Nakagawa--Sakata (PMNS) matrix \cite{PMNS}.  As $V$, $U$ is
also unitary.  Neutrino oscillations observed in experiments are attributed to the off--diagonal
entries of the matrix $U$.  We assumed here that the neutrino mass generation mechanism violated
total lepton number by two units.  While this is very attractive, it should be mentioned that neutrinos
could acquire masses very much like the quarks.  That would require the right--handed $\nu^c$ states
to be part of the low energy theory.  $M_\nu$ will then be similar to $M_\ell$ of Eq. (\ref{translep}).  Neutrino
oscillation phenomenology will be identical to the case of $L$--violating neutrino masses.  In this case, however,
the neutrino Yukawa couplings will have to be extremely tiny to accommodate the observed masses.  Furthermore,
some global symmetries, such as total lepton number, will have to be assumed in order to forbid gauge
invariant mass terms for the right--handed neutrinos.

The fermionic states $(e_i^0)$  are simply the physical electron, the muon, and the tau lepton states. Similarly,
the quark fields with a superscript $^0$ are the mass eigenstates.  It is conventional to drop these
superscripts, which we shall do from now on.

\subsection{Lepton masses}

Conceptually charged lepton masses are the easiest to explain.  Leptons are propagating states, and their
masses are simply the poles in the propagators.  Experimental information on charged lepton masses is
rather accurate \cite{PDG}:
\begin{eqnarray}
\label{leptonmasses}
m_e &=& 0.510998902 \pm 0.000000021 ~{\rm MeV}\,, \nonumber \\
m_\mu &=& 105.658357 \pm 0.000005 ~ {\rm MeV}\,, \nonumber \\
m_\tau &=& 1777.03^{+0.30}_{-0.26} ~ {\rm MeV}\,.
\end{eqnarray}

The direct kinematic limits on the three neutrino masses are \cite{PDG}:
\begin{equation}
m_{\nu_e} \leq 3~ {\rm eV}\,,~ m_{\nu_\mu} \leq 0.19 ~ {\rm MeV}\,,~ m_{\nu_\tau} \leq 18.2 ~ {\rm MeV}\,.
\end{equation}
Neutrino oscillation experiments have provided much more accurate determinations of the squared mass
differences $\Delta m_{ij}^2 = m_i^2 - m_j^2$.  Solar and atmospheric neutrino oscillation experiments,
when combined with accelerator and reactor neutrino experiments,
suggest the following allowed values (with $2 \sigma$ error quoted) \cite{valle}:
\begin{eqnarray}
\Delta m^2_{21} &=& (7.25 - 8.11) \times 10^{-5} ~ {\rm eV}^2\,, \nonumber \\
\Delta m^2_{31} &=& \pm(2.18 - 2.64) \times 10^{-3} ~ {\rm eV}^2\,.
\end{eqnarray}
While this still leaves some room for the absolute masses, when combined with the direct limit
on $m_{\nu_e} \leq 3 ~{\rm eV}$, the options become limited.  Current data allow for two possible
ordering of the mass hierarchies: (i) normal hierarchy where $m_1 \leq m_2 \ll m_3$, and
(ii) inverted hierarchy where $m_1 \simeq m_2 \gg m_3$.  More specifically, $\nu_e$ is mostly
in the lightest eigenstate in the case of normal hierarchy, while it is mostly in the
heavier eigenstate in the case of inverted hierarchy.  The sign of $\Delta m_{31}^2$ is not known
at the moment, which gives these two ordering options.  On the other hand, the sign of $\Delta m_{21}^2$
is fixed from the condition that MSW resonance occurs inside the Sun.

\subsection{Leptonic mixing matrix}


The PMNS matrix $U$, being unitary, has $N^2$ independent components for $N$ families of leptons.
Out of these, $N(N-1)/2$ are Euler angles, while the remaining $N(N+1)/2$ are phases.  Many of
these phases can be absorbed into the fermionic fields and removed.  If one writes $U = Q \hat{U} P$,
where $P$ and $Q$ are diagonal phase matrices, then by redefining the phases of $e$ fields as
$e \rightarrow Q e$, the $N$ phases in $Q$ can be removed.  $P$ has only $N-1$ non--removable phases (an
overall phase is irrelevant).  For $N= 3$,  $P = diag.(e^{i \alpha},~ e^{i \beta},~ 1)$.
$\alpha, \beta$ are called the Majorana phases.  (If the neutrino masses are of the Dirac type,
these phases can also be removed by redefining the $\nu^c$ fields.)  $\hat{U}$ will then have
$N(N+1)/2 - (2N-1) = \frac{1}  {2} (N-1) (N-2)$ phases.  For $N=3$, there is a single ``Dirac" phase
in $U$.  This single phase will be relevant for neutrino oscillation phenomenology.  The two
Majorana phases $(\alpha,~\beta$) do not affect neutrino oscillations, but will be relevant for
neutrino-less double beta decay.

In general, the PMNS matrix for three families of leptons can be written as
\begin{eqnarray}
\label{U}
U = \left(\begin{matrix}U_{e1} & U_{e2} & U_{e3} \\ U_{\mu 1} & U_{ \mu 2} & U_{\mu 3} \\ U_{\tau 1} &
U_{\tau 2} & U_{\tau 3}\end{matrix} \right)\,.
\end{eqnarray}
To enforce the unitarity relations it is convenient to adopt specific parametrizations.
The Euler angles, as you know, can be parametrized in many different ways.  Furthermore, the
Dirac phase can be chosen to appear in different ways (by field redefinitions).  The ``standard
parametrization" that is now widely used \cite{PDG} has $U_{PMNS} = U.P$ where
\begin{eqnarray}
\label{UMNS}
U = \left(\begin{matrix}c_{12}c_{13} &  s_{12}c_{13} &   s_{13}e^{-i\delta}  \\
-s_{12}c_{23}-c_{12}s_{23}s_{13}e^{i\delta}&  c_{12}c_{23}-s_{12}s_{23}s_{13}e^{i\delta} & s_{23}c_{13} \\
s_{12}s_{23}-c_{12}c_{23}s_{13}e^{i\delta} &  -c_{12}s_{23}-s_{12}c_{23}s_{13}e^{i\delta}& c_{23}c_{13}\end{matrix}\right)\,.
\end{eqnarray}
Here $s_{ij} = \sin \theta_{ij}$, $c_{ij} = \cos \theta_{ij}$.

Our current understanding of these mixing angles arising from neutrino oscillations can be summarized
as follows ($2 ~ \sigma$ error bars quoted) \cite{valle}:
\begin{eqnarray}
\label{numixing}
\sin^2\theta_{12} &=& 0.27 - 0.35\,, \nonumber \\
\sin^2\theta_{23} &=& 0.39 - 0.63\,, \nonumber \\
\sin^2\theta_{13} &\leq& 0.040\,.
\end{eqnarray}
Here $\theta_{12}$ limit arises from solar neutrino data (when combined with KamLand reactor neutrino data), $\theta_{23}$ from atmospheric neutrinos (when combined with MINOS accelerator neutrino data),
and $\theta_{13}$ from reactor neutrino data.

It is intriguing that the current understanding of leptonic mixing can be parametrized by the unitary
matrix
\begin{eqnarray}
\label{tribimaximal}
U_{TB} = \left( \begin{matrix}\sqrt{\frac{2}  {3}} & \frac{1}  {\sqrt{3}} &
0 \\ -\frac{1}  {\sqrt{6}} & \frac{1}  {\sqrt{3}} & -\frac{1}
{\sqrt{2}} \\ -\frac{1} {\sqrt{6}} & \frac{1}  {\sqrt{3}} & \frac{1}
{\sqrt{2}} \end{matrix} \right)P\,.
\end{eqnarray}
This mixing is known as tri-bimaximal mixing \cite{tribi}. This nomenclature is based on the numerology
$\sin^2\theta_{12} = 1/3,~\sin^2\theta_{23} = 1/2,~\sin^2\theta_{13} = 0$ that follows from
Eq. (\ref{tribimaximal}).  As we will see, such
a geometric structure is far from being similar to the quark
mixing matrix. Note that currently $\theta_{13}$ is allowed to be
zero, in which case the Dirac phase $\delta$ becomes irrelevant.
We also have no information on the Majorana phases ($\alpha,~\beta$ in $P$), which can only
be tested in neutrino-less double beta decay experiments.

There have been considerable activity in the literature in trying the reproduce
the tri-bimaximal mixing matrix of Eq. (\ref{tribimaximal}) based on symmetries.
The most popular idea has been to adopt the non-Abelian flavor symmetry $A_4$, which
is the symmetry group of a regular tetrahedron.  It is also the group of even permutations
of four letters. This finite group has twelve elements, which fall into one three--dimensional
(${\bf 3}$) and three one--dimensional $({\bf 1+1'+1''})$ irreducible representations.
$A_4$ is the simplest symmetry group with a triplet representation.  Assigning the lepton doublets
to the ${\bf 3}$, and the three charged lepton singlets to the the $({\bf 1+1'+1''})$, it is possible,
assuming a specific vacuum structure, to reproduce the ``geometric" form of the leptonic mixing
matrix \cite{A4}.

\subsection{Quark masses}

Unlike the leptons, quarks are not propagating particles.  So their masses have to
be inferred indirectly from properties of hadrons.  There are various techniques to do this.  Let me
illustrate this for the light quark masses ($u,~d,~s$) by the method of chiral perturbation theory \cite{gasser}.

Consider the QCD Lagrangian at low energy scales.  Electroweak
symmetry has already been broken, and heavy quarks $(t, b, c$)
have decoupled.  The Lagrangian for the light quarks $(u,~d,~s)$ and the gluon fields takes the form
\begin{equation}
\label{QCD}
{\cal L} = \sum_{k = 1}^{N_F=3} \overline{q}_k (i
\displaystyle{\not}D   - m_k)q_k - \frac{1}  {4} G_{\mu \nu} G^{\mu
nu}\,,
\end{equation}
where $G_{\mu \nu}$ is the gluon field strength and $\displaystyle{\not}D$ is the covariant
derivative.  $m_k$ is the mass of the $k$-th quark and $q_k$ denotes the quark field.  This Lagrangian
has a chiral symmetry in the limit where the quark masses vanish.  The three left--handed quarks can be
rotated into one another, and the three right--handed quarks can be rotated independently.  The symmetry is $SU(3)_L
\times SU(3)_R \times U(1)_V$, with the axial $U(1)_A$ (of the classical symmetry $U(3)_L \times U(3)_R$) explicitly broken by anomalies.  The
$U(1)_V$ is baryon number, which remains unbroken even after QCD dynamics.  QCD dynamics
breaks the $SU(3)_L \times SU(3)_R$ symmetry down to the diagonal subgroup $SU(3)_V$.  In the limit of vanishing
quark masses, there must be 8 Goldstone bosons corresponding to this symmetry breaking.
These Goldstone bosons are identified as the pseudoscalar mesons, which are however, not
exactly massless. The (small) quark masses actually break the chiral symmetry explicitly and thus generate
small masses for the mesons.

Chiral perturbation theory is a systematic expansion in $p/\Lambda_\chi$, where $p$ is
the particle momentum and $\Lambda_\chi \sim 1$ GeV is the chiral symmetry breaking scale.
Since the masses of the light quarks ($u,~d,~s$) are smaller than $\Lambda_\chi$, we can
treat them as small perturbations and apply chiral expansion.  The explicit breaking of chiral symmetry
occurs via the mass term
\begin{eqnarray}
\label{chiral1}
M = \left(\begin{matrix}m_u & ~ & ~ \\ ~ & m_d & ~ \\ ~ & ~ & m_s\end{matrix}\right)\,.
\end{eqnarray}
$M$ can be thought of as a spurion field which breaks the chiral symmetry spontaneously.  Under
$SU(3)_L \times SU(3)_R$ symmetry $q_L \rightarrow U_L ~q_L,~q_R \rightarrow U_R~ q_R$, while $M \rightarrow
U_L~ M ~U_R^\dagger$.  That is, $M$ transforms as a $(3,3^*)$ of this group.  Under the unbroken
diagonal $SU(3)_V$ subgroup, both $q_L$ and $q_R$ transform as triplets, while $M$ splits into
a ${\bf 1+8}$. Thus $M$ can be written as $M = M_1 + M_8$, where $M_1$ is a singlet of $SU(3)_V$, while $M_8$
is an octet:
\begin{eqnarray}
\label{chiral2}
M_1 &=& \frac{(m_u + m_d + m_s)}  {3}\left(\begin{matrix}1 & ~ & ~ \\ ~ & 1 & ~ \\ ~ & ~ & 1\end{matrix}\right)\,,\nonumber \\
M_8 &=& \frac{(m_u-m_d)}  {2} \left(\begin{matrix}1 & ~ & ~ \\ ~ & -1 & ~ \\ ~ & ~ & 0\end{matrix}\right) +
\frac{(m_u-m_d-2m_s)}  {6}\left(\begin{matrix}1 & ~ & ~ \\ ~ & 1 & ~ \\ ~ & ~ & -2\end{matrix}\right)\,.
\end{eqnarray}
The octet (under $SU(3)_V$) of mesons can be written down as a (normalized) matrix
\begin{eqnarray}
\label{meson}
\Phi = \left(\begin{matrix}\frac{\pi^0}  {\sqrt{2}} + \frac{\eta^0}
{\sqrt{6}} & \pi^+ & K^+ \\ \pi^- & -\frac{\pi^0}  {\sqrt{2}} +
\frac{\eta^0}  {\sqrt{6}} & K^0 \\ K^- & \overline{K^0} & -\sqrt{\frac{2}
 {3}} \eta^0 \end{matrix} \right)\,.
\end{eqnarray}
The lowest order invariants involving $\Phi$ bilinear and  $M$ are
\begin{equation}
\label{chiral3}
A~ {\rm Tr} (\Phi^2) M_1 + B ~{\rm Tr}(\Phi^2 M_8)\,.
\end{equation}
Here $A$ and $B$ are arbitrary coefficients.  Eq. (\ref{chiral3}) can be
readily expanded, which will give relations for the masses of
mesons. Now, in the limit of $m_u = 0, m_d = 0, m_s \neq 0$, the
$SU(2)_L \times SU(2)_R$ chiral symmetry remains unbroken, and so
the pion fields should be massless.  Working out the mass terms,
and demanding that the pion mass vanishes in this limit, one finds a
relation $A = 2B$.  Using this relation we can write down the
pseudoscalar meson masses. In doing so, let us also recall that
electromagnetic interactions will split the masses of the neutral
and charged members.  To lowest order, this splitting will be
universal.  Then we have
\begin{eqnarray}
\label{mumd}
m_{\pi^0}^2 &=& B (m_u + m_d) \nonumber \\
m_{\pi^\pm}^2 &=& B (m_u + m_d) + \Delta_{\rm em} \nonumber \\
m_{K^0}^2 &=& m_{\overline{K^0}}^2 = B(m_d + m_s) \nonumber \\
m_{K^\pm}^2  &=& B(m_u + m_s) + \Delta_{\rm em} \nonumber \\
m_{\eta}^2 &=& \frac{1}  {3} B(m_u + m_d + 4 m_s)\,.
\end{eqnarray}
Here small $\pi^0-\eta^0$ mixing has been neglected, which vanishes in the limit $m_u-m_d$ vanishes.

Eliminating $B$ and $\Delta_{\rm em}$ from Eq. (\ref{mumd}) we obtain two relations for
quark mass ratios:
\begin{eqnarray}
\frac{m_u}  {m_d} &=& \frac{2 m_{\pi^0}^2 - m_{\pi^+}^2+ m_{K^+}^2 - m_{K^0}^2}
{m_{K^0}^2 - m_{K^+}^2 + m_{\pi^+}^2} = 0.56 \nonumber \\
\frac{m_s}  {m_d} &=& \frac{m_{K^0}^2 + m_{K^+}^2 - m_{\pi^+}^2}
{m_{K^0}^2 - m_{K^+}^2 + m_{\pi^+}^2} = 20.1
\end{eqnarray}
This is the lowest order chiral perturbation theory result for the mass ratios.  Second order
chiral perturbation theory makes important corrections to these ratios as discussed in more
detail in Ref. \cite{manohar}.  Note that the absolute masses cannot be determined in this way.  Alternative techniques, such
as QCD sum rules and lattice calculations which provide the most precise numbers have to
be applied for this.

For heavy quarks ($c$ and $b$), one can invoke another type of symmetry, the
heavy quark effective theory (HQET) \cite{neubert}.  When
the mass of the quark is heavier than the typical momentum of the partons $\Lambda \sim
m_p/3 = 330$ MeV, one can make another type of expansion.  In analogy with atomic physics,
where different isotopes exhibit similar chemical behavior, the behavior of charm hadrons
and bottom hadrons will be similar.  In fact, there will be an $SU(2)$ symmetry relating
the two, to lowest order in HQET expansion.  One consequence is that the mass splitting
between the vector and scalar mesons in the $b$ and $c$ sector should be related.  This
leads to a relations $M_{B^*} - M_B = \Lambda^2/m_b$ and $M_{D^*} - M_D = \Lambda^2/m_c$,
leading to the prediction
\begin{equation}
\label{heavy}
\frac{M_{B^*} - M_B}  {M_{D^*} - M_D} = \frac{m_c}  {m_b}\,,
\end{equation}
which is in good agreement with experiments.

The most reliable determination of light quark masses come from lattice QCD.  The QCD
Lagrangian of Eq. (\ref{QCD}) has only very few parameters, the strong coupling
constant, and the three light quark masses.  All the hadron masses and decay constants
should in principle be calculable in terms of these parameters.  Since QCD coupling
is strong at low energies, perturbation theory is not reliable.  Lattice QCD is formulated
on discrete space time lattice points, rather than in the continuum.  When the lattice spacing
takes small value, lattice QCD should reproduce continuum QCD.  No approximation is made as
regards the value of the strong coupling constant $\alpha_s$.  It is thus a non-perturbative
technique which, upon matching certain measured quantities, can be used to calculate the light quark
mass parameters.  In the last five years there has been tremendous advances in lattice QCD,
owing to improved lattice action, as well as increased computing power.  Early results on light
quark masses assumed ``quenching", i.e., ignored fermions propagating inside loops, but now full three
flavor un-quenched calculation with dynamical fermions are available.  There have been several independent
evaluations of the light quark masses, which generally are in good agreement with one another.
Conventionally these masses are presented as running masses at $q = 2$ GeV in the $\overline{\rm MS}$ scheme.

The MILC collaboration \cite{refMILC}, which adopted a partially quenched approximation, finds for the light quark masses
\begin{eqnarray}
\label{MILC}
m_u(2~{\rm GeV}) &=& 1.7 \pm 0.3 ~{\rm MeV}\,, \nonumber \\
m_d(2~{\rm GeV}) &=& 3.9 \pm 0.46 ~{\rm MeV}\,, \nonumber \\
m_s(2~{\rm GeV}) &=& 76 \pm 7.6 ~{\rm MeV}\,.
\end{eqnarray}
Here I have combined the various uncertainties (statistical, systematic, simulation, and electromagnetic)
in quadrature.
The ratios of light quark masses are thought to be more reliable, as many of the uncertainties cancel
in the ratios.  It is customary to define an average mass of up and down quarks $\hat{m} = (m_u+m_d)/2$.
The results of MILC collaboration corresponds to the following mass ratios:
\begin{eqnarray}
\frac{m_u}  {m_d} &=& 0.43 \pm 0.08\,, \nonumber \\
\frac{m_s}  {\hat{m}} &=& 27.4 \pm 4.2\,.
\end{eqnarray}

The JLQCD collaboration \cite{refJLQCD}, which includes three flavors of dynamical quarks finds
\begin{eqnarray}
\label{JLQCD}
\hat{m}(2~{\rm GeV}) &=& 3.55^{+0.65}_{-0.28}~ {\rm MeV}\,, \nonumber \\
m_s(2~{\rm GeV}) &=& 90.1^{+17.2}_{-6.1}~ {\rm MeV}\,, \nonumber \\
\frac{m_u}  {m_d} &=& 0.577 \pm 0.025\,.
\end{eqnarray}

The RBC \& UKQCD collaboration \cite{refRBC}, which includes $2+1$ dynamical domain wall quarks finds
\begin{eqnarray}
\label{RBC}
\hat{m}(2~{\rm GeV}) &=& 3.72 \pm 0.41~ {\rm MeV}\,, \nonumber \\
m_s(2~{\rm GeV}) &=& 107.3 \pm 11.7~{\rm MeV}\,, \nonumber \\
\hat{m}: m_s &=& 1: 28.8 \pm 1.65 \,.
\end{eqnarray}

And finally, the HPQCD collaboration finds \cite{refHPQCD}
\begin{eqnarray}
m_u(2~{\rm GeV}) &=& 1.9 \pm 0.24 ~{\rm MeV}\,, \nonumber\\
m_d(2~{\rm GeV}) &=& 4.4 \pm 0.34 ~{\rm MeV}\,, \nonumber\\
m_s(2~{\rm GeV}) &=& 87 \pm 5.7 ~{\rm MeV}\, \nonumber \\
\hat{m}(2~{\rm GeV}) &=& 3.2 \pm 0.89 ~{\rm MeV}\,, \nonumber \\
\frac{m_u} {m_d} &=& 0.43 \pm 0.08\,.
\end{eqnarray}

One sees that the lattice calculations are settling down, and have become quite reliable.
It should be mentioned that the same lattice QCD calculations also provide several of the
hadronic form factors which enter into the determination of the CKM mixing angles.

The masses of the $c$ and $b$ quarks can be determined in a variety of ways.  Charmonium
and Upsilon spectroscopy, in conjunction with lattice calculations seem to be the most
reliable.  We summarize the masses of these quarks thus obtained, along with the
ranges for the light quark masses \cite{PDG}.
\begin{eqnarray}
\label{masssum}
{m}_u(2~{\rm GeV}) &=& 1.5~{\rm to}~ 3.3 ~{\rm MeV}\,, \nonumber \\
{m}_d(2~{\rm GeV}) &=& 3.5~{\rm to}~ 6.0 ~{\rm MeV}\,, \nonumber \\
{m}_s(2~{\rm GeV}) &=& 105^{+25}_{-35} ~{\rm MeV}\,, \nonumber \\
\frac{m_u}{m_d} &=& 0.35~{\rm to}~ 0.60 \,, \nonumber \\
\frac{m_s}{m_d} &=& 17~{\rm to}~ 22 \,, \nonumber \\
\frac{m_s}{(m_u+m_d)/2} &=& 25~{\rm to}~ 30 \,, \nonumber \\
{m}_c(m_c) &=& 1.27^{+0.07}_{-0.11} ~{\rm GeV}\,, \nonumber \\
{m}_b(m_b) &=& 4.20^{+0.17}_{_-0.07}  ~{\rm GeV}\,.
\end{eqnarray}
Sometimes the light quark masses are quoted at $q=1$ GeV, rather at $q=2$ GeV.  There are significant
differences in these two sets of values due to the rapid running of the strong coupling in this regime.
Typically one finds for example, $m_u(1~{\rm GeV}) \simeq 1.35 ~m_u({\rm 2~GeV})$.

The top quark mass is more directly determined leading to the value \cite{PDG}
\begin{equation}
\label{top}
m_t = 171.3 \pm 1.1 \pm 1.2 ~{\rm GeV}\,.
\end{equation}
Any ambitious theory of flavor should aim to address these observed values of quark masses.

\subsection{Running quark and lepton masses}

In attempting to explain the observed masses of fermions, it will be convenient to compare their masses
at a common momentum scale $\mu$.  Usually this scale is taken to be much heavier than the QCD scale of
about 1 GeV, or even the weak scale of 246 GeV, since new flavor dynamics cannot happen at lower scales.
The measured quark and lepton masses then have to be extrapolated to a common momentum scale $\mu$.  Below
the weak scale, this extrapolation would require the renormalization group evolution of the mass parameters
caused by QCD and QED loops.  The beta functions and the gamma functions necessary to do this have been
computed to three--loop (and in some cases four--loop) accuracy \cite{ramondrge}.  In the $\overline{MS}$ scheme,
which is widely used, the contributions to the beta functions and gamma functions from a specific flavor
of fermion will decouple for momenta $\mu$ less than the mass of the particle.  Before discussing
this evolution, it is necessary to remark on the differences between ``pole mass" and ``running mass" of
a fermion.  For heavy quarks ($c,\,b,\,t$) the pole mass $M_q$ is the physical mass, which appears as the pole in the propagator.  (For light quarks ($u,\,d,\,s$) pole mass is not defined because of the non--perturbative
nature of strong interactions at their mass scales.) The running mass $m_q(M_q)$ includes corrections from QCD and QED loops.  The two are related for quarks via
\begin{equation}
\label{pole}
M_q =m_q(M_q)\left[ 1 + \frac{4}{3} \frac{\alpha_s(M_q)}{\pi} + \kappa_q^{(2)} \left(\frac{\alpha_s(M_q)}{\pi}\right)^2
+\kappa_q^{(3)} \left(\frac{\alpha_s(M_q)}{\pi}\right)^3
\right]\,,
\end{equation}
where terms of order $\alpha_s^4$ and higher have been neglected.
The two--loop and the three--loop QCD correction factors are $\{\kappa^{(2)}_c,\,\kappa^{(2)}_b,\,\kappa_t^{(2)}\}
= \{11,21,\,10.17,\,9.13\}$ and  $\{\kappa^{(3)}_c,\,\kappa^{(3)}_b,\,\kappa_t^{(3)}\}
= \{123.8,\,101.5,\,80.4\}$.  There can be significant differences between $M_q$ and $m_q(M_q)$.  For
example, using $\alpha_s(M_Z) = 0.1176$ and $M_t = 172.5$ GeV, one obtains, with QCD evolution of $\alpha_s$
from $M_Z$ to $M_t$, $\alpha_s(M_t) = 0.108$, and then from Eq. (\ref{pole}) $m_t(M_t) = 162.8$ GeV.  For $c$ and $b$ quarks the differences are even bigger.

The running masses of leptons can be defined analogously, but now the QCD corrections are replaced
by QED corrections.  Consequently the differences between the pole mass $M_\ell$ and running mass
$m_\ell(M_\ell)$ are less significant. The two masses are related via
\begin{equation}
m_\ell(\mu) = M_\ell \left[1 - \frac{\alpha}{\pi}\left\{1 + \frac{3}{2}{\rm ln} \frac{\mu} {m_\ell(\mu)}  \right\}
\right]\,.
\end{equation}

For momentum scales higher than the electroweak symmetry breaking scale, one should evolve the
Yukawa couplings of the fermions, rather then their masses.  One can define the running mass in this
momentum regime as
\begin{equation}
m_i(\mu) = Y_i(\mu) ~v \,.
\end{equation}
Here $v=174$ GeV is the VEV of the Higgs doublet evaluated at the weak scale.  Since the VEV $v$ also
is a function of momentum (owing to wave function renormalization of the Higgs filed), one could in
principle define the running mass as $m_i(\mu) = Y_i(\mu) v(\mu)$.  But this is usually not necessary, and
will not be adopted here.  The renormalization group evolution equations for the Yukawa couplings of
the SM have been worked out to two--loop accuracy \cite{ramondrge}.

While extrapolating the Yukawa coupling above the weak scale one has to specify the theory valid in that
regime.  Often it will be assumed to be the minimal supersymmetric standard model (MSSM).  In the fermion
Yukawa sector there are significant differences between the MSSM and the SM.  The main difference is that
supersymmetry requires two Higgs doublets, $H_u$ with $(Y/2) = +1/2$ and $H_d$ with $(Y/2) = -1/2$.  The
extra doublet is needed for anomaly cancelation and also for generating all fermion masses.  Recall that in the SM Yukawa interaction of Eq. (\ref{Yukawa}) we used $H$ for generating the up--type quark masses  and
its conjugate $\tilde{H}$ for
the down--type quark and charged lepton masses.  Supersymmetric Yukawa couplings must be derived from a
superpotential $W$, which is required to be holomorphic.  This means that if $H$ appears in $W$, then
$H^*$ cannot appear.  The MSSM Yukawa interactions arise from the following superpotential.
\begin{equation}
\label{YukawaMSSM}
{\cal W}_{\rm Yukawa}^{\rm MSSM} = Q^T Y_u u^c H_u - Q^T Y_d d^c H_d - L^T Y_\ell e^c H_d\,.
\end{equation}
If we denote the VEVs of $H_u$ and $H_d$ as $v_u$ and $v_d$, then the mass matrices for the three
charged fermion sectors are
\begin{equation}
\label{MudeMSSM}
M_u = Y_u v_u\,,~~~~~M_d = Y_d v_d\,,~~~~~M_\ell = Y_\ell v_d\,.
\end{equation}
The diagonalization procedure follows as in the SM.  Notably, there is no Higgs boson
mediated flavor changing couplings at tree level, in spite of having two Higgs doublets.  The constraints
of supersymmetry is the reason for its absence. (Only a single Higgs doublet couples to each one
of the three sectors.)  A new parameter appears, which is the ratio
of the two Higgs vacuum expectation values:
\begin{equation}
\tan\beta = \frac{v_u}{v_d}\,.
\end{equation}
This parameter will influence many physical processes.
$\tan\beta$ plays an important role in the RGE evolution of the Yukawa couplings.
The range of $\tan\beta$ preferred in the MSSM is $\tan\beta = (1.7 - 60)$.  When
$\tan\beta < 1.7$ the top quark Yukawa couplings blows up before the momentum scale
$\mu = \Lambda_{\rm GUT} \approx 2 \times 10^{16}$ GeV.  $\Lambda_{\rm GUT}$ is
associated with the scale of grand unification, where the three gauge couplings of
the SM appear to meet, if there is low energy supersymmetry.  For $\tan\beta > 60$ the
$b$--quark and $\tau$--lepton Yukawa couplings become non-perturbative before reaching
$\Lambda_{\rm GUT}$.

\begin{table}[t]
\begin{center}
\begin{tabular}{|c|c|c|c|c|c|c|c|}\hline \hline
$m_i \diagdown  \mu $ & $m_c(m_c)$  & 2 GeV  & $m_b(m_b)$ & $m_t(m_t)$ & 1 TeV & $\Lambda_{\rm GUT}^{\tan\beta = 10}$ & $\Lambda_{\rm GUT}^{\tan\beta=50}$ \\ \hline\hline
$m_u({\rm MeV})$ & 2.57 & {\bf 2.2} & 1.86 & 1.22 & 1.10 & 0.49 & 0.48 \\ \hline
$m_d({\rm MeV})$ & 5.85 & {\bf 5.0} & 4.22 & 2.76 & 2.50 & 0.70 & 0.51 \\ \hline
$m_s({\rm MeV})$ & 111 & {\bf 95} & 80 & 52 & 47 & 13 & 10 \\ \hline
$m_c({\rm GeV})$ & {\bf 1.25} & 1.07 & 0.901 & 0.590 & 0.532 & 0.236 & 0.237 \\ \hline
$m_b({\rm GeV})$ & 5.99 & 5.05 & {\bf 4.20} & 2.75 & 2.43 & 0.79 & 0.61 \\ \hline
$m_t({\rm GeV})$ & {\it 384.8} & {\it 318.4} & {\it 259.8} & {\bf 162.9} & 150.7 & 92.2 & 94.7 \\ \hline
$m_e({\rm MeV})$ & 0.4955 & $\sim$ & 0.4931 & 0.4853 & 0.4959 & 0.2838 & 0.206 \\ \hline
$m_\mu({\rm MeV})$ & 104.474 & $\sim$ & 103.995 & 102.467 & 104.688 & 59.903 & 43.502 \\ \hline
$m_\tau({\rm MeV})$ & 1774.90 & $\sim$ & 1767.08 & 1742.15 & 1779.74 & 1021.95 & 773.44 \\ \hline
\hline
\end{tabular}
\caption{The running masses of quarks and leptons as a function of momentum $\mu$. The last two columns
correspond to the running masses at $\Lambda_{\rm GUT} = 2 \times 10^{16}$ GeV assuming low energy MSSM spectrum
with $\tan\beta = 10$ and $50$.}
\label{Table:masses}
\end{center}
\end{table}

In Table \ref{Table:masses} we list the running masses of quarks and leptons as a function
of the momentum scale $\mu$.  We have adopted the numbers listed from Ref. \cite{xing}, but our
independent calculations show general agreement at the level of few per cent with Ref. \cite{xing}.  The input values for $(c,\,b,\,t)$ quarks are the running masses
indicated in bold.  For this Table we have used light quark masses at $\mu=2$ GeV as indicated
in bold.  For the charged lepton, we have used as input the masses given in Eq. (\ref{leptonmasses}).
The masses of all fermions are listed at momentum scale $\mu = m_t$ and $\mu = 1$ TeV assuming
the validity of the SM up to 1 TeV.  Also listed are the running masses at
$\mu = \Lambda_{\rm GUT} = 2 \times 10^{16}$ GeV assuming MSSM spectrum, for two values of
$\tan\beta$ (10 and 50).  The following input values have been used.  $\alpha_s(M_Z) = 0.1176,\,
\alpha^{-1}(M_Z) = 127.918$, and $\sin^2\theta_W(M_Z) = 0.23122$.

There are various noteworthy features in Table \ref{Table:masses}.  The light quark masses
$(m_u,\,m_d,,m_s)$ decrease by about a factor of two in going from $\mu = 2$ GeV to
$\mu = 1$ TeV.  This decrease is a result of QCD corrections.  The $d$ and $s$--quark masses decrease
by about another factor of 4 in going from $\mu=1$ TeV to $\mu = \Lambda_{\rm GUT}$, while $m_u$ decreases
by a factor of $2.3$. The net change in the values of $(m_u,\,,m_d,\,m_s$) in going from $\mu = 2$ GeV to $\mu =
\Lambda_{\rm GUT}$ for the case of $\tan\beta = 10$ is a factor $(4.9,\,7.9,\,7.3)$.
The value of $b$--quark mass decreases considerably, by a factor of $6.9$, in going
from $\mu = m_b$ to $\mu = \Lambda_{\rm GUT}$ for $\tan\beta = 10$. $m_b(\mu=\Lambda_{\rm GUT})$ is
close to the $\tau$--lepton mass $m_\tau(\mu = \Lambda_{\rm GUT})$ (to within about 20\%).  The lepton masses decrease by about a factor of 2 in going from low energies to $\Lambda_{\rm GUT}$. This decrease
occurs because of the $SU(2)_L \times U(1)_Y$ contributions to the beta functions of $Y_\ell$.
These features will be relevant when we discuss predictions for fermion masses from Grand Unified
theories in Sec. \ref{sec5}.

Sometimes the light quark masses are quoted at $\mu = 1$ GeV.  In going from $\mu = 2$ GeV down
to $\mu = 1$ GeV, the masses increase by a factor of $1.31$, if $\alpha_s(M_Z) = 0.1176$ is used.
The running factor to go from $\mu=2$ GeV down to $\mu=m_c$ is indicated in Table \ref{Table:masses},
while the additional running factor to go from $\mu=m_c$ to $\mu=1$ GeV is found to be $1.12$.  Thus,
$(m_u,\,m_d,\,m_s) = (2.2,\,5,\,95)$ MeV at $\mu=2$ GeV correspond to $(m_u,\,m_d,\,m_s) = (2.88,\,6.58,\,124)$
MeV at $\mu=1$ GeV.

In Table  \ref{Table:masses} we have also included the top quark mass at momentum scales
below $M_t$ (indicated in italics).  These values, which are un-physical, since the top
quark decouples at its mass, will be rarely used.

\subsection{Quark mixing and CP violation}

The unitary matrix $V$ of Eq. (\ref{CKM}) which appears in the charged current interactions
of Eq. (\ref{cc}) enters in a variety of processes.  A lot of information has
been gained on the matrix elements of $V$.  The general matrix can be written as
\begin{eqnarray}
V = \left(\begin{matrix}
        V_{ud} & V_{us} & V_{ub} \\
        V_{cd} & V_{cs} & V_{cb} \\
        V_{td} & V_{ts} & V_{tb} \\
        \end{matrix}\right)\,.
\end{eqnarray}
The standard parametrization of $V$ is as in Eq. (\ref{UMNS}), but
now understood to be for the quark sector.  $V$ has a single un-removable phase for three families
of quarks and leptons.  (The phases ($\alpha,\,\beta)$ which appeared in the case of Majorana neutrinos
can be removed by right--handed quark field redefinition.)  The single un-removable phase in $V$ allows for the violation of CP symmetry in the quark sector. Unlike in the leptonic sector, the quark mixing angles turn out to be small. This enables one to make a perturbative expansion of the mixing matrix a la Wolfenstein \cite{wolf}.  The small parameter is taken to be $\lambda = |V_{us}|$ in terms of which one has
\begin{eqnarray}
V = \left(\begin{matrix}
    1-\frac12\lambda^2-\frac18\lambda^4
        & \lambda
            & A\lambda^3(\rho-i\eta) \\
    -\lambda
        & 1-\frac12\lambda^2 -\frac18\lambda^4(1+4A^2)
            & A\lambda^2 \\
    A\lambda^3(1-\rho-i\eta)
        & -A\lambda^2 + \frac{1}{2}A\lambda^4 \left(1 - 2(\rho+i\eta)\right)
            & 1 - \frac{1}{2}A^2\lambda^4
   \end{matrix}\right) + {\cal O}(\lambda^5)\,.
\end{eqnarray}
Here the exact correspondence with Eq. (\ref{UMNS}) is given by
\begin{equation}
s_{12} \equiv \lambda,~~~~s_{23} \equiv A\lambda^2,~~~~ s_{13}e^{-i\delta} \equiv A\lambda^3(\rho -i\eta)\,.
\end{equation}

\begin{figure}[!htb]
\begin{center}
\includegraphics[width=0.9\linewidth]{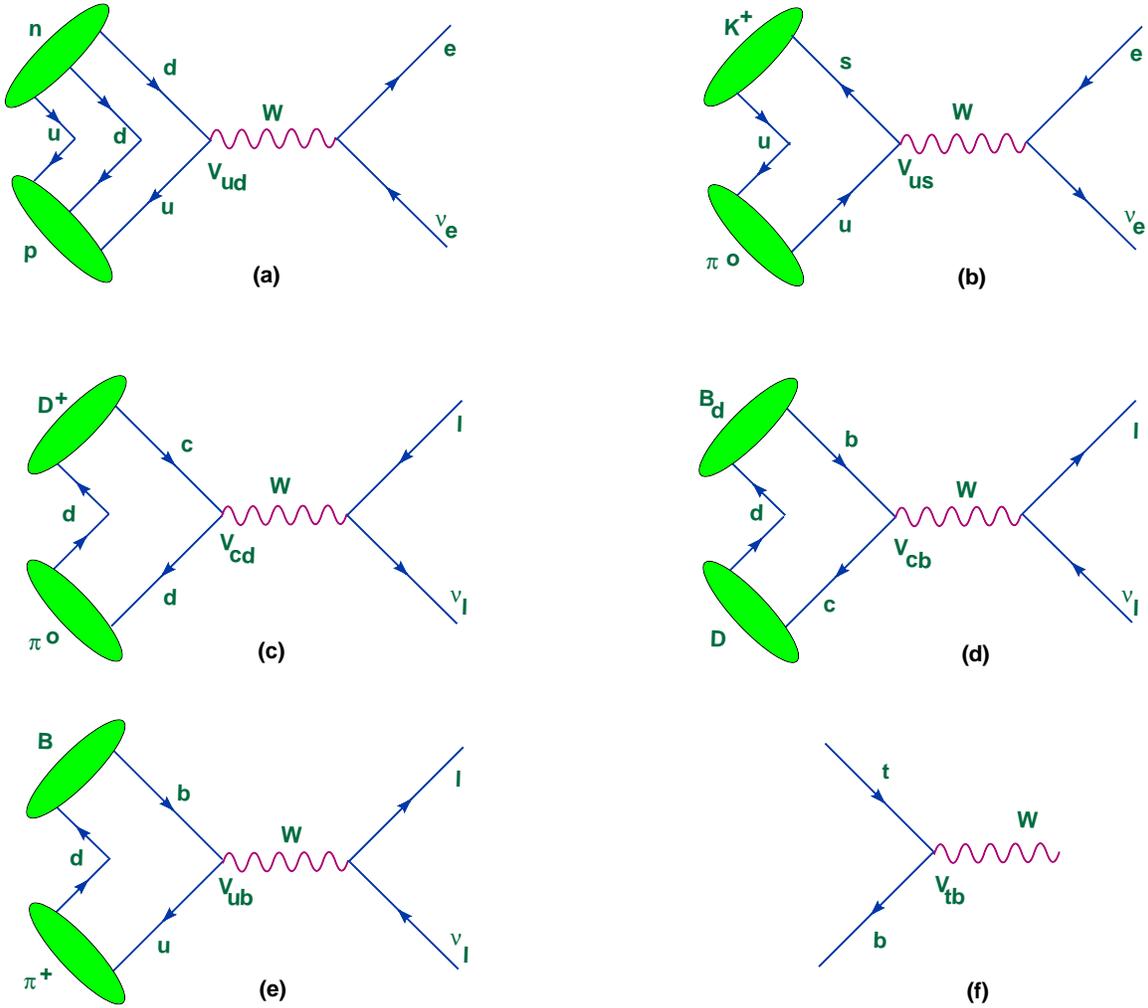}
\vspace*{0.2in} \vspace*{-0.2in}\caption{Processes determining
$|V_{ij}|$.}
\label{Fig.decay}
\end{center}
\end{figure}

Matrix elements of $V$ are determined usually via semileptonic
decays of quarks.  In Fig. \ref{Fig.decay} we have displayed the dominant
processes enabling determination of these elements. Fig. \ref{Fig.decay} (a) is
the diagram for nuclear beta decay, from which $|V_{ud}|$ has been
extracted rather accurately \cite{pdgmixing}:
\begin{equation}
|V_{ud}| = 0.97377 \pm 0.00027\,.
\end{equation}
Fig. \ref{Fig.decay} (b) shows semileptonic $K$ decay from which the Cabibbo
angle $|V_{us}|$ can be extracted.  The decays $K_L^0 \rightarrow \pi \ell \nu$
and $K^\pm \rightarrow \pi^0 \ell^\pm \nu$ ($\ell = e,\,\mu$) have been averaged to obtain for the
product $|V_{us}| f_+(0) = 0.21668 \pm 0.00045$.  Here $f_+(0)$ is the form factor
associated with this semileptonic decay evaluated at $q^2=0$.  Using $f_+(0) = 0.961 \pm 0.008$
(obtained from QCD calculations, which are in agreement with lattice QCD evaluations),
one obtains
\begin{equation}
|V_{us}| = 0.2257 \pm 0.0021\,.
\end{equation}
$|V_{cd}|$ is extracted from $ D  \rightarrow K \ell \nu$ and $ D
\rightarrow \pi \ell \nu$ decays with assistance from lattice QCD for the
computation of the relevant form factors.  $V_{cs}$ is determined from
semileptonic $D$ decays and from leptonic $D_s$ decay ($D_s^+ \rightarrow \mu^+ \nu$),
combined with lattice calculation of the decay form factor $f_{D_s}$.
Both $|V_{cd}|$ and $|V_{cs}|$ have rather large errors currently:
\begin{eqnarray}
|V_{cd}| &=& 0.230 \pm 0.011 \,,\nonumber \\
|V_{cs}| &=& 0.957 \pm 0.010\,.
\end{eqnarray}
$|V_{cb}|$ is determined from both inclusive and exclusive decays of
$B$ hadrons into charm, yielding a value
\begin{equation}
|V_{cb}| = (41.6 \pm 0.6) \times 10^{-3}\,.
\end{equation}
$|V_{ub}|$ is determined from charmless $B$ decays and gives
\begin{equation}
|V_{ub}| = (4.31 \pm 0.30) \times 10^{-3}\,.
\end{equation}

 Elements $|V_{td}|$ and $|V_{ts}|$ cannot be currently determined,
for a lack of top quark events, but can be inferred from $B$ meson
mixings where these elements appear through the box diagram. The
result is
\begin{eqnarray}
|V_{td}| = (7.4 \pm 0.8) \times 10^{-3}\,, \nonumber \\
\frac{|V_{td}|}  {|V_{ts}|} = 0.208 \pm 0.008\,.
\end{eqnarray}

Fig. \ref{Fig.decay} (f) depicts the decay of top quark into $W + b$.  It can
also decay into $W + q$ where $q$ is $d,s,b$.  By taking the ratio
of branching ratios $R = B (t \rightarrow W b)/\sum_qB(t
\rightarrow W q)$, CDF and D0 have arrived at a limit on $|V_{td}|
> 0.74$ \cite{PDG}.

\subsubsection{Heavy quark symmetry}

Heavy quark symmetry plays an important role in the determination of
$|V_{ub}|$ and $|V_{cb}|$.  While a thorough discussion of HQET
(Heavy Quark Effective Theory) is outside the scope of this write-up,
it would be useful to get a feeling of some of the ideas involved.
We refer the reader to Ref. \cite{neubert} for a thorough review,
and Ref. \cite{flei} for a pedagogical exposure.

Consider first the purely leptonic decay $B^- \rightarrow \ell^- \overline{\nu}_\ell$ for
$\ell = e,\,\mu,\,\tau$.  The transition amplitude for this decay is
\begin{equation}
T_{fi} = \frac{G_F}{\sqrt{2}} V_{ub} [\overline{u}_\ell \gamma_\mu (1-\gamma_5) u_\nu]\left\langle
0|\overline{u} \gamma^\mu (1-\gamma_5)b|B^-\right\rangle\,.
\end{equation}
Here $G_F$ is the Fermi coupling.  To compute the decay rate, the hadronic matrix
element for the transition of $B$ meson to vacuum needs to be evaluated. Note that the matrix element of vector current between pseudscalar meson and vacuum vanishes: $\left\langle 0|\overline{u} \gamma^\mu b|B^-\right\rangle\ = 0$, while the axial vector current matrix element is parametrized as $\left\langle
0|\overline{u} \gamma^\mu \gamma_5 b|B^-\right\rangle\ = i f_B q^\mu$, with
$f_B$ being the $B$ meson decay constant and $q^\mu$ the $B$ meson momentum.  With this matrix element,
the decay rate can be readily computed.  One obtains
\begin{equation}
\Gamma(B^- \rightarrow \ell^- \overline
{\nu}_\ell) = \frac{G_F^2}{8 \pi}f_B^2 |V_{ub}|^2 M_B m_\ell^2
\left(1-\frac{m_\ell^2} {M_B^2} \right)^2\,.
\end{equation}
Note the helicity suppression, which implies that the number of events in this channel will
be small.  Recently BELLE collaboration has observed the decay $B^- \rightarrow \tau^-\overline{\nu}$
with a 3.5 sigma statistical significance.  Their results can be converted to a value for the product
$|V_{ub}|f_B$  as
\begin{equation}
\label{belle}
|V_{ub}|f_B = [10.1^{+1.6}_{-1.4}({\rm stat})^{+1.3}_{-1.4}({\rm syst})]\times 10^{-4}\, {\rm GeV}\,.
\end{equation}
Using lattice evaluations of $f_B$, one can obtain the value of $|V_{ub}|$ from Eq. (\ref{belle}).
The accuracy of this determination, which is rather direct, suffers from the lack of
events for this helicity suppressed decay.

Semileptonic decays do not suffer from the helicity suppression, and are therefore more promising.
Unlike a single form factor that appears in the purely leptonic decay, now there will be two form factors.
These two can be related  via heavy quark symmetry, as we outline below.  Consider
the decay $\overline{B}^0_d \rightarrow D^+ \ell \overline{\nu}_\ell$ which proceeds via Fig. \ref{Fig.decay} (d).
The transition amplitude for this decay has the form
\begin{equation}
\label{vubTfi}
T_{fi} = \frac{G_F}{ \sqrt{2}}V_{cb} [\overline{u}_\ell \gamma_\mu (1-\gamma_5) u_\nu]\left\langle
D^+|\overline{c} \gamma^\mu (1-\gamma_5)b|\overline{B}^0_d\right\rangle\,.
\end{equation}
A similar expression is obtained for the decay $\overline{B}^0_d \rightarrow \pi^+ \ell \overline{\nu}_\ell$,
with $|V_{cb}|$ replaced by $|V_{ub}|$ in Eq. (\ref{vubTfi}).  The matrix element of axial vector current
between two pseudoscalar mesons vanishes: $\left\langle
D^+|\overline{u} \gamma^\mu \gamma_5 b|\overline{B}^0_d\right\rangle\ =0$.  The vector current matrix
element between two pseudoscalar mesons contains two form factors:
\begin{equation}
\label{form}
\left\langle
D^+(k)|\overline{u} \gamma^\mu  b|\overline{B}^0_d(p)\right\rangle =
F_1(q^2)\left[(p+k)_\mu- \frac{M_B^2 - M_D^2}{q^2}\, q_\mu\right] + F_0(q^2) \frac{M_B^2-M_D^2}{q^2} \,q_\mu\,,
\end{equation}
where $q=p-k$.

To see how HQET can relate the two form factors $F_1(q^2)$ and $F_0(q^2)$, let me briefly review the
crucial elements of HQET.  In a hadron composed of one heavy ($b$) quark and one light anti-quark $\overline{u}$ or
$\overline{d}$, the mass of $b$ is much larger than the scale of QCD dynamics, $\Lambda_{\rm QCD}$.  The
$b$ quark is then almost on-shell, moving with a velocity close to the hadron's four velocity.  We
write this as
\begin{equation}
p_Q^\mu = m_Q v^\mu + k^\mu\,,
\end{equation}
where $k \ll m_Q$ is the residual momentum, and $v^2 = 1$.  The $b$ quark interacts with the light degrees
of freedom, but such interactions can cause a change in the residual momentum by $\Delta k \sim \Lambda_{\rm QCD}
\ll m_Q$.  Thus $\Delta v \rightarrow 0$ as $\Lambda_{\rm QCD}/m_Q \rightarrow 0$.

In the heavy quark symmetry limit ($\Lambda_{\rm QCD}/m_Q \rightarrow 0$), the elastic
scattering process $\overline{B}(v) \rightarrow \overline{B}(v')$ has the amplitude
\begin{equation}
\label{heavyquark}
\frac{1}{M_B} \left\langle \overline{B}(v')|\overline{b}(v')\gamma_\mu b(v)|\overline{B}(v)
\right\rangle = \xi(v'.v)(v+v')_\mu\,.
\end{equation}
A term of the type $(v-v')_\mu$ cannot appear on the right-hand side of Eq. (\ref{heavyquark}) since
$\displaystyle{\not}v\, b_v = b_v$ and $\overline{b}_{v'} \displaystyle{\not} \, v' = \overline{b}_{v'}$.  The $1/M_B$ factor in Eq. (\ref{heavyquark}) is associated
with normalization of states, so the right-hand side of Eq. (\ref{heavyquark}) has no dependence on
the heavy quark flavor.  Current conservation implies $\xi(v'.v=1)=1$, so that the function $\xi(v.v')$, the Isgur--Wise
function \cite{isgur}, is independent of the heavy quark flavor.  Thus, in the heavy quark symmetry limit, we have
\begin{equation}
\frac{1}{\sqrt{M_DM_B}}\left\langle D(v')|\overline{c}_{v'}\gamma_\mu b_v|\overline{B}(v)\right\rangle =
\xi(v.v')(v+v')_\mu\,.
\end{equation}
This transition is now governed by a single form factor, $\xi(v'.v)$ with $\xi(1)=1$.  Comparing with
Eq. (\ref{form}), one finds
\begin{eqnarray}
F_1(q^2) &=& \frac{M_D + M_B} {2\sqrt{M_DM_B}}~\xi(w) \nonumber \\
F_0(q^2) &=& \frac{2 \sqrt{M_D M_B}} {M_D+M_B}\left(\frac{1+w}{2}\right)\xi(w)\,
\end{eqnarray}
where
\begin{equation}
w=v_D.v_B = \frac{M_D^2+M_B^2-q^2} {2M_DM_B}\,.
\end{equation}

As an application of these ideas, consider the decay $\overline{B} \rightarrow D^* \ell \nu$.  The differential
decay rate for this process can be written as
\begin{equation}
\frac{d\Gamma}{d w} = G_F^2 \,K \, F(w)^2|V_{cb}|^2\,,
\end{equation}
where $K$ is a known kinematic function and $F(w)$ is related to the Isgur--Wise function (up to
perturbative QCD corrections).  It should obey the normalization
\begin{equation}
\label{lukeeqn}
F(1) = \eta_A(\alpha_s)\left[1 + \frac{0}{m_c}
+\frac{0}{m_b} + {\cal O}\left(\frac{\Lambda_{\rm QCD}^2}{m_{b,c}^2}\right)\right]\,.
\end{equation}
Here $\eta_A(\alpha_s)$ is a perturbatively
calculable function.  Note that ${\cal O}(\Lambda_{\rm QCD}/m_{c,b})$ corrections vanish \cite{Luke}.  This
decay distribution can be measured as a function of $w$, from which $F(w)|V_{cb}|$ can be extracted.
Now, when extrapolated to zero recoil limit ($w=1)$, whence the decay rate vanishes), from Eq. (\ref{lukeeqn}),
one obtains a value of $|V_{cb}|$.

\subsubsection{CP violation}

 Charge conjugation (C) takes a particle to its antiparticle,
 Parity (spatial reflection) changes the helicity of the particle.
 Under CP, $e_L^-$ will transform to $e^+_R$.  Both C and P are
 broken symmetries in the SM, but the product CP is approximately
 conserved.  Violation of CP has been seen only in weak
 interactions.  The CKM mechanism predicts CP violation through
 a  single complex phase that appears in the CKM matrix.  Thus in the SM, various CP
 violating processes in $K$, $B$ and other systems get correlated.
 So far such correlations have been consistent with CKM
 predictions, but more precise determinations in the $B$ and $D$
 systems at the LHC may open up new physics possibilities.

\begin{figure}[htb]
\begin{center}
\includegraphics[width=0.7\linewidth]{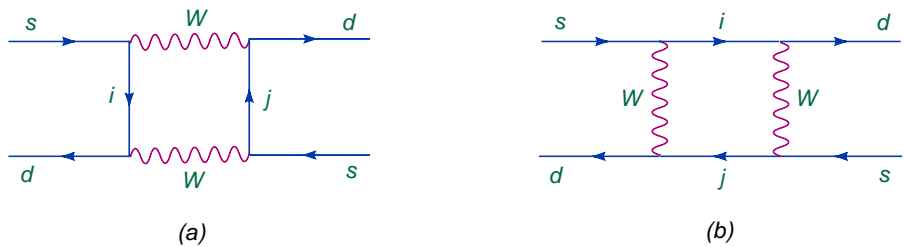}
\caption{Box diagram inducing $K^0-\overline{K^0}$ transition in
the SM.}
\label{box}
\end{center}
\end{figure}

In the $K^0-\overline{K^0}$ system, CP violation has been observed
both in mixing and in direct decays.  CP violation in mixing arises in
the SM via the $W$--boson box diagram shown in Fig. \ref{box}.  The CP asymmetry
in mixing is parametrized by $\epsilon$, which is a measure of the mixing
between the CP even and CP odd states $K_{1,2}^0 = (K^0 \pm \overline{K^0})/\sqrt{2}$.
It has been measured to be
\begin{equation}
|\epsilon| = (2.229 \pm 0.010)\times 10^{-3} \,.
\end{equation}
The measured value in in excellent agreement with expectations from the SM,
and enables us to determine the single phase of the CKM matrix.
The box diagram contribution to $\epsilon$ is given by
\begin{eqnarray}
|\epsilon| &=& \frac{G_F^2 f_k^2 m_K m_W^2}{12\sqrt{2} \pi^2 \Delta m_K} \hat{B}_K \left\{ \right. \eta_c S(x_c){\rm Im}[(V_{cs}V_{cd}^*)^2]    \nonumber \\
 &+& \eta_t S(x_t){\rm Im}[(V_{ts}V_{td}^*)^2] + 2 \eta_{ct} S(x_c,x_t){\rm Im} [V_{cs}V_{cd}^*V_{ts} V_{td}^*] \left. \right\}\,.
\end{eqnarray}
Here $S(x)$ and $S(x,y)$ are Inami--Lim functions \cite{inami} with $x_{c,t} = m^2_{c,t}/M_W^2$, and
the $\eta$ factors are QCD correction factors for the running of the effective $\Delta S = 2$
Hamiltonian from $M_W$  to the hadron mass scale.

\begin{figure}[htb]
\begin{center}
\includegraphics[width=0.3\linewidth]{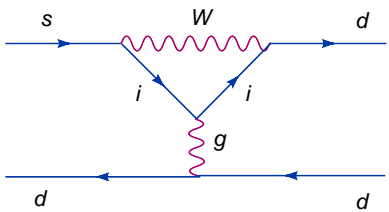}
\caption{One loop penguin diagram that generates CP violation in
direct $K \rightarrow \pi \pi$ decay.}
\label{penguin}
\end{center}
\end{figure}

The direct CP violation parameter that leads to the decay $K \rightarrow \pi \pi$ has also been measured,
leading to the value
\begin{eqnarray}
Re(\epsilon'/\epsilon) &=& (1.65 \pm 0.26)\times 10^{-3}\,.
\end{eqnarray}
These decays occur via the penguin diagrams shown in Fig. \ref{penguin}.  There are electromagnetic
penguins and gluonic penguins, which tend to cancel each other.  While the KM model predicts non-zero
value of $\epsilon'/\epsilon$, estimating this value reliably has been difficult, partly because of
this cancelation.  Most estimates are in agreement with observations.

\begin{figure}[ht]
\begin{center}
\includegraphics[width=0.6\linewidth]{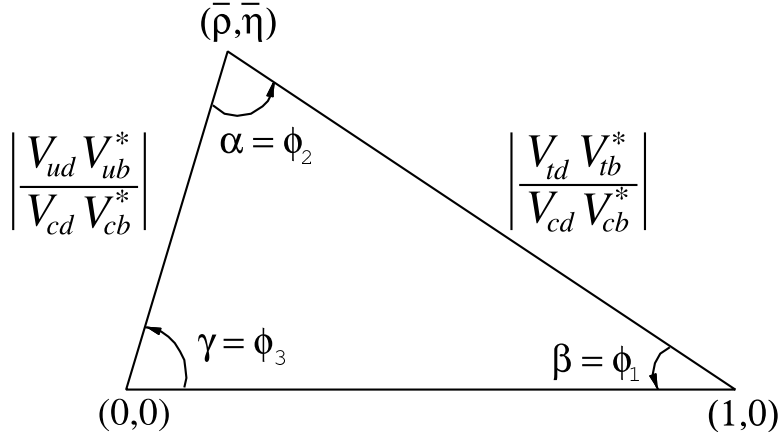}
\caption{Unitarity triangle in the CKM model.}
\label{Fig:triangle}
\end{center}
\end{figure}

A wealth of information has been gained about CP violation from the $B$ factories over the last decade.
CP violation in $B$ meson system is now well established.
Several CP violating quantities have been measured in $B_d$ meson
system \cite{flei}, all of which show consistency with the CKM mixing matrix.
Unitarity of the CKM matrix implies that $\sum_i V_{ij} V_{ik}^* = \delta_{jk}$ and
$\sum_j V_{ij} V_{kj}^* = \delta_{ik}$.  There are six vanishing combinations, which can be
expressed as triangles in the complex plane.  The areas of all of these triangles are the same.
The most commonly used triangle arises from the relation
\begin{equation}
V_{ud}V_{ub}^* + V_{cd}V_{cb}^* + V_{td}V_{tb}^* = 0\,.
\end{equation}
In the complex plane, the resulting triangle has
sides of similar length (of order $\lambda^3$).  This unitarity triangle relation
is shown in Fig. \ref{Fig:triangle}.  The
three interior angles ($\alpha, \beta, \gamma$), also referred to as $(\phi_2,\,\phi_1,\,\phi_3)$,
can be written in the CKM model as
\begin{eqnarray}
\alpha &=& {\rm arg}\left(\frac{-V_{td} V_{tb}^*}
{V_{ud}V_{ub}^*}\right) \simeq {\rm arg}\left(-\frac{1-\rho-i\eta}
{\rho + i \eta}\right)\,,\nonumber \\
 \beta &=& {\rm arg}\left(\frac{-V_{cd}
V_{cb}^*} {V_{td}V_{tb}^*}\right) \simeq {\rm arg}\left(\frac{1}
{1-\rho - i \eta}\right)\,,\nonumber \\
 \gamma &=& {\rm
arg}\left(\frac{-V_{ud} V_{ub}^*} {V_{cd}V_{cb}^*}\right) \simeq
{\rm arg}\left(\rho + i \eta\right)\,.
\end{eqnarray}
One experimental test of the CKM mechanism is the measurement of
$\alpha + \beta +\gamma = 180^0$.

The angle $\beta$ can be measured with the least theoretical uncertainty
from the decay of $B_d \rightarrow J/\psi K_S$.  It is found to be
\begin{equation}
\sin 2\beta = 0.68 \pm 0.03\,.
\end{equation}
This value is in in good agreement with the CKM prediction.

The angle $\alpha$
is measured from decay modes where $b \rightarrow u \overline{u} d$ is dominant.
Such decays includ $B \rightarrow \pi \pi$, $B \rightarrow \rho \rho$ and
$B \rightarrow \pi \rho$.  The value of $\alpha$ extracted is
\begin{equation}
\alpha = (88^{+6}_{-5})^0\,.
\end{equation}

The angle $\gamma$ does not depend on the top quark, and can in principle be
measured from tree--level decays of $B$ meson.  Strong interaction uncertainties
are rather large in decays such as $B^\pm  \rightarrow  D^0 K^\pm$.  The current
value of the angle $\gamma$ is
\begin{equation}
\gamma = (77^{+30}_{-32})^0\,.
\end{equation}

\begin{figure}[ht]
\begin{center}
\begin{tabular}{cc}
\includegraphics[width=0.4\linewidth]{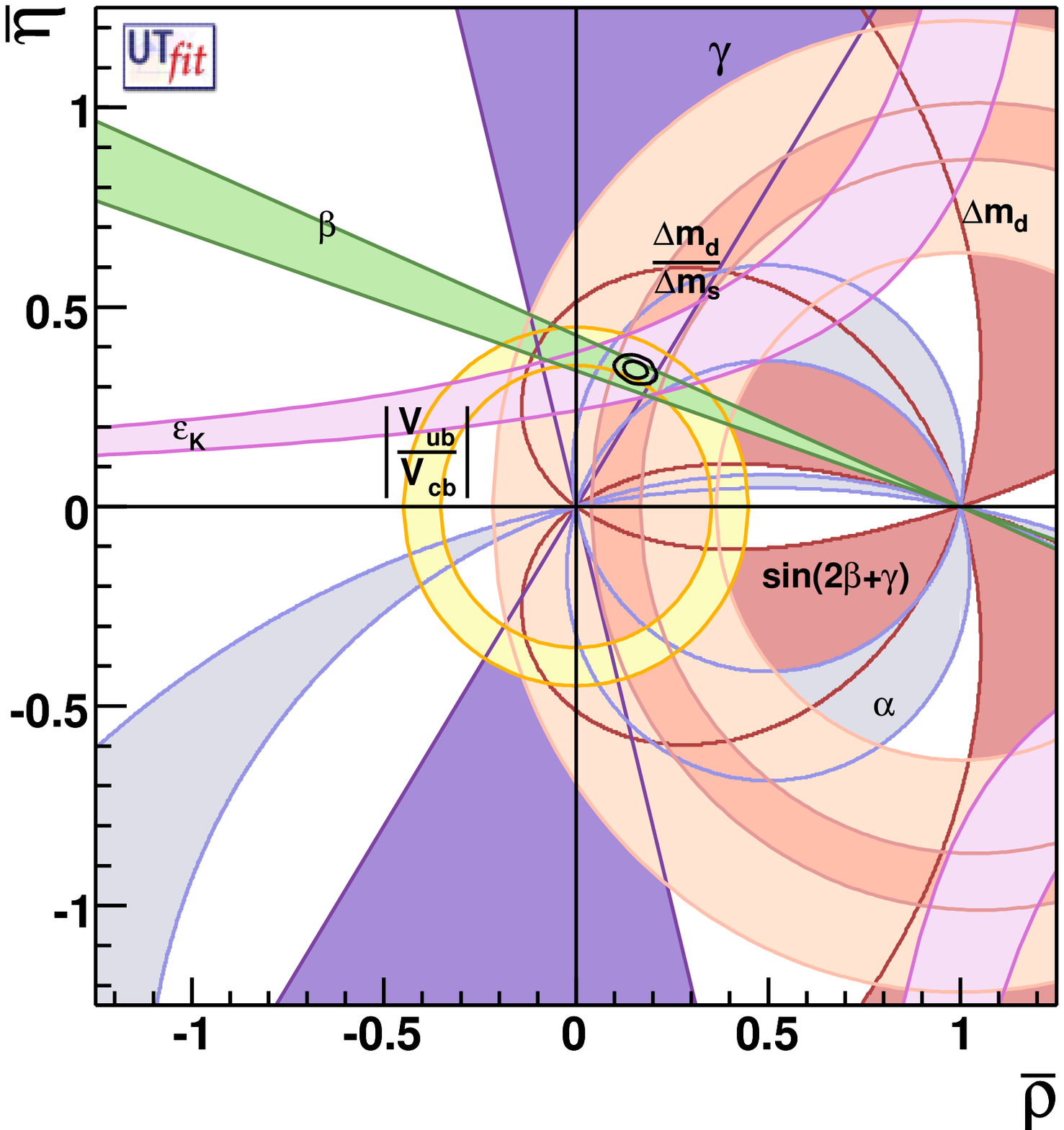} & \hspace*{0.2in}
\includegraphics[width=0.385\linewidth]{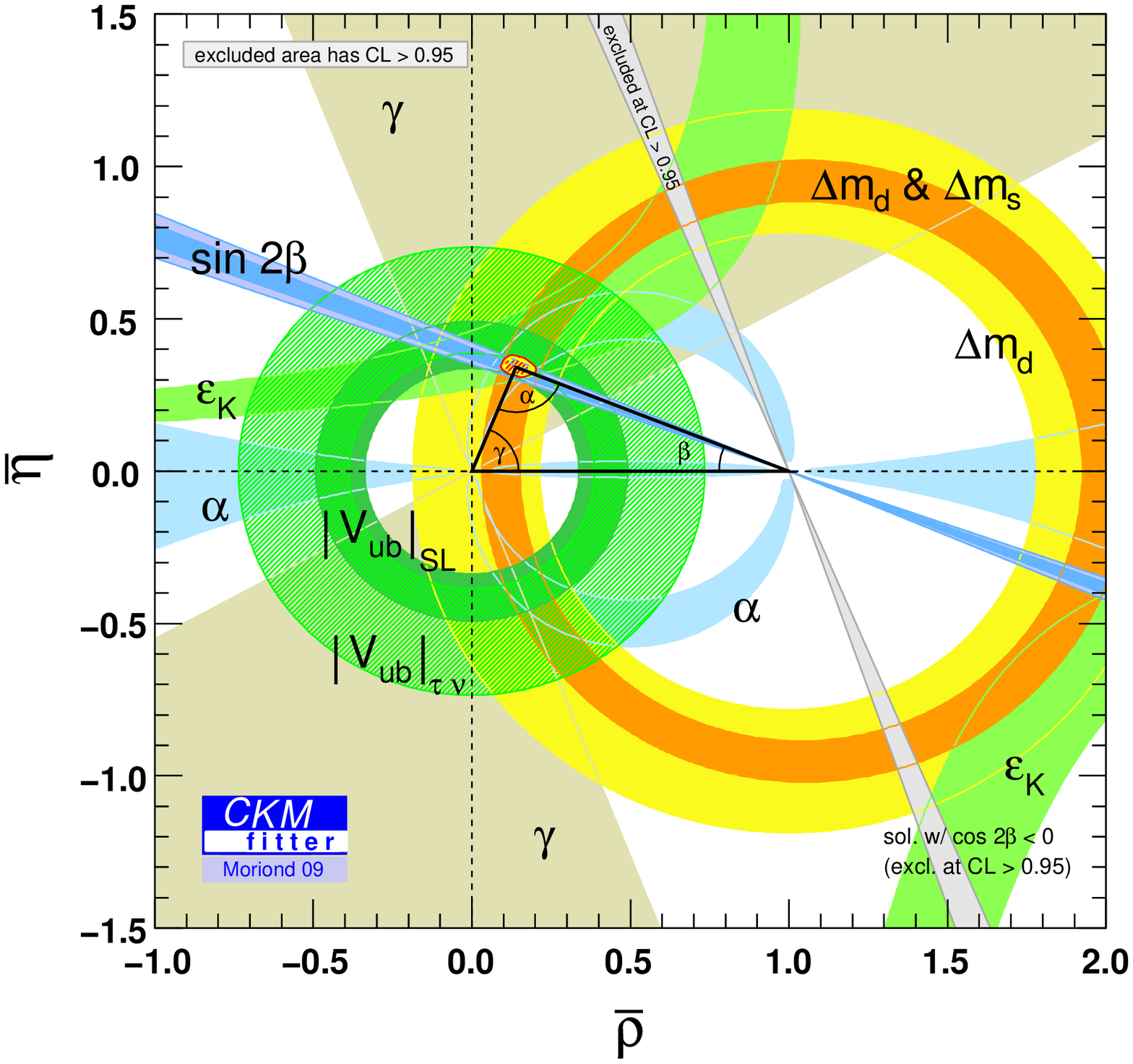} \\
\end{tabular}
\caption{Global fit to the mixing and CP violation data from the
UTFit collaboration (left panel) \cite{refUTfit} and the CKMFitter collaboration (right panel) \cite{refCKMfitter}. }
\label{Fig:UTfit}
\end{center}
\end{figure}

The current situation with the CKM mixing angles and CP violation phase
is depicted in Fig. \ref{Fig:UTfit}.  The left panel is the result of a global analysis
of flavor mixing and CP violation data by the UTFit \cite{refUTfit} collaboration, while the right panel depicts
the results from an independent CKMFitter \cite{refCKMfitter} collaboration.  The Wolfenstein
parameters $\overline{\eta}$ is plotted against  $\overline{\rho}$ in these
figures.  Here $\overline{\eta} = \eta(1-\lambda^2/2)$ and
$\overline{\rho} = \rho(1-\lambda^2/2)$. A variety of input parameters have gone into these fits.  Some of the
constraints used are explicitly indicated in these figures.  It is very non-trivial that
the various constraint curves have a common intersection.  This demonstrates
the success of the CKM mechanism of flavor mixing and CP violation. The intersection of the
various ellipses gives the best fit value for the Wolfenstein parameters $(\lambda,~A,~\rho,~\eta$),
which are as follows \cite{PDG}:
\begin{eqnarray}
\lambda = 0.2272 \pm 0.0010,~ A = 0.818_{-0.017}^{+0.007}, ~ \rho
=0.221^{+0.064}_{-0.028},~ \eta = 0.340^{+0.017}_{-0.045}\,.
\end{eqnarray}
Theories of flavor should provide an understanding of these fundamental parameters.

\section{Relating quark mixings and mass ratios \label{sec3}}

Having reviewed the fundamental flavor parameters of the Standard Model, now we
turn to attempts which explain some of the observed features.  Necessarily one
needs to invoke non--standard physics, which can be potentially tested at colliders.

We begin with a simple idea of relating quark masses and mixings by virtue of
flavor symmetries. In the quark sector we have seen that the mass ratios such as
$m_d/m_s$, $m_u/m_c$, etc are strongly hierarchical, while the
mixing angles, such as $V_{us}$ are also hierarchical, although
the hierarchy here is not as strong. Can the quark mixing angle be
computed in terms of the quark mass ratios?  Clearly such attempts
have to go beyond the SM. Here I give a simple two--family example
which assumes a flavor $U(1)$ symmetry that distinguishes the two
families.

\subsection{Prediction for Cabibbo angle in a two family model}

Consider the mass matrices for $(u,~c)$ and $(d,~s)$ quarks given
by \cite{sumrule}
\begin{eqnarray}
\label{ww}
M_u = \left(\begin{matrix}0 & A_u \\ A_u^* & B_u \end{matrix}\right)\,,~~~ M_d =
\left(\begin{matrix}0 & A_d \\ A_d^* & B_d \end{matrix}\right)\,.
\end{eqnarray}
The crucial features of these matrices are (i) the zeros in the
(1,1) entries, and (ii) their hermiticity.  Neither of these
features can be realized within the SM.  Recall that the SM symmetry
would have arbitrary non--hermitian matrices for $M_u$ and $M_d$.
The zero entries in Eq. (\ref{ww}) can be
enforced by a flavor $U(1)$ symmetry, the hermitian nature can be
obtained if the gauge sector is left--right symmetric.  Before
constructing such a model, let us examine the consequences of Eq.
(\ref{ww}). Matrices in Eq. (\ref{ww}) have factorizable phases.  That is,
$M_u = P_u \hat{M}_u P_u^*$, where $\hat{M}_u$ has the same form
as $M_u$ but with all entries real, and where $P_u = diag (e^{i
\alpha_u}, ~ 1)$ is a diagonal phase matrix. A similar
factorization applies to $M_d$ with a phase matrix $P_d = diag
(e^{i \alpha_d}, ~ 1)$.  We can absorb these phase matrices into
the quark fields, but since $\alpha_u \neq \alpha_d$, the matrix
$P_u^* P_d = diag. (e^{i \psi},~ 1)$ will appear in the charged
current matrix $(\psi = \alpha_d-\alpha_u)$.  The matrices
$\hat{M}_u$ and $\hat{M_d}$, which have all real entries, can be
diagonalized readily, yielding for the mixing angles $\theta_u$ and
$\theta_d$
\begin{eqnarray}
\tan^2\theta_u &=& \frac{m_u} {m_c}\,, \nonumber\\
\tan^2\theta_d &=& \frac{m_d}  {m_s}\,.
\end{eqnarray}
This yields a prediction for the Cabibbo angle \cite{sumrule}
\begin{equation}
\label{cabibbo}
|\sin\theta_C|\simeq \left|\sqrt{\frac{m_d}  {m_s}} - e^{i \psi}
\sqrt{\frac{m_u}  {m_c}}\right|\,.
\end{equation}
This formula works rather well, especially since even without the
second term, the Cabibbo angle is correctly reproduced.  The phase
$\psi$ is a parameter, however, its effect is rather restricted.
For example, since $\sqrt{m_d/m_s} \simeq 0.22$ and
$\sqrt{m_u/m_c} \simeq 0.07$, $|\sin\theta_C|$ must lie between
0.15 and 0.29, independent of the value of $\psi$.

Now to a possible derivation of Eq. (\ref{ww}).  Since SM interactions
do not conserve Parity, it is useful to extend the gauge sector to
the left-right symmetric group $G \equiv SU(3)_C \times SU(2)_L \times
SU(2)_R \times U(1)_{B-L}$, wherein Parity invariance can be
imposed \cite{LR}.  The (1,2) and (2,1) elements of $M_{u,d}$ being complex
conjugates of each other will then result. The left--handed and
the right--handed quarks transform as $Q_{iL}(3,2,1,1/3) +
Q_{iR}(3,1,2,1/3)$ under $G$.  Under discrete parity operation $Q_{iL}
\leftrightarrow Q_{iR}$.  This symmetry can be consistently
imposed, as $W_L \leftrightarrow W_R$ in the gauge sector under
Parity. The leptons transform as $\psi_{iL}(1,2,1,-1) + \psi_{iR}(1,1,2,-1)$ under
the gauge symmetry.  Note that $\psi_R$, which is a doublet of $SU(2)_R$,  contains
the right--handed neutrino, as the partner of $e_R$.  Thus there is a compelling
reason for the existence of $\nu_R$, unlike in the SM, where it is optional.

The Higgs field that couples to quarks should be
$\Phi(1,2,2,0)$, and under Parity $\Phi \rightarrow \Phi^\dagger$.
In matrix form $Q_{iL}, Q_{iR}, \Phi$ read as
\begin{eqnarray}
Q_{iL} = \left(\begin{matrix}u_i \\ d_i \end{matrix}\right)_L\,,~~Q_{iR} =
\left(\begin{matrix}u_i \\ d_i \end{matrix}\right)_R\,,~~ \Phi =
\left(\begin{matrix}\phi_1^0 & \phi_2^+ \\ \phi_1^- & \phi_2^0 \end{matrix}\right)\,,
\end{eqnarray}
so that the Yukawa Lagrangian for quarks
\begin{equation}
{\cal L}_{\rm Yukawa} = \overline{Q}_L\Phi Y Q_R +  \overline{Q}_L \tilde{\Phi} \tilde{Y} Q_R + h.c.
\end{equation}
 is gauge invariant.  Here $\tilde{\Phi} \equiv \tau_2 \Phi^* \tau_2$.
 Imposing Parity, we see that the Yukawa matrices $Y$ and $\tilde{Y}$
 must be hermitian, $Y = Y^\dagger$ and $\tilde{Y} = \tilde{Y}^\dagger$.  This is the
 desired result for deriving Eq. (\ref{ww}).  The VEVs $\left\langle \phi_1^0 \right\rangle$
 and $\left\langle \phi_2^0 \right\rangle$ can be complex in general, but this will not
 affect the prediction for the Cabibbo angle of Eq. (\ref{cabibbo}), since that only requires $|(M_{u,d})_{12}| =
|(M_{u,d})_{21}|$.   Additional Higgs fields, eg., $\Delta_L(1,3,1,2)+
 \Delta_R(1,1,3,2)$, would be required for breaking the left--right symmetric gauge group down
 to the SM and for simultaneously generating large $\nu_R$ Majorana masses.  However, these fields
 do not enter into the mass matrices of quarks.

 To enforce zeros in the (1,1) entries of $M_{u,d}$ of Eq. (\ref{ww}), we can employ
 the following $U(1)$ flavor symmetry: $Q_{1L}: 2,~ Q_{1R}: -2.~Q_{2L}: 1,~
 Q_{2R}:-1.~\Phi_1:
 2,~ \Phi_2: 3$.  Note that two Higgs bidoublet fields are
 needed. $\Phi_1$ generates the (2,2) entries, while $\Phi_2$
 generates the (1,2) and (2,1) entries.  There is no (1,1) entry
 generated, since there is no Higgs field with $U(1)$ charge of
 $+4$.  Note also that the $\tilde{\Phi}_{1,2}$ fields, which have
 $U(1)$ charges $(-2,\,-3)$, do not couple to the quarks.

 While we cannot determine the scale of flavor dynamics in this model,
 the $U(1)$ flavor symmetry and the left--right symmetry, which
 were crucial for the derivation of Eq. (\ref{cabibbo}), could show up as new particles
 at the LHC.  In general, one would also expect multiple Higgs
 bosons.  We should note that the full theory is more elaborate compared
 to the minimal left--right symmetry without the flavor symmetry (two, instead
 of one bi-doublet Higgs fields are needed), but the effective theory is simpler,
 with the mass matrices being predictive.

\subsection{Three family generalization}

 Eq. (\ref{ww}) can be generalized for the case of three families, a la
 Fritzsch \cite{fritzsch}.  The up and down quark mass matrices have hermitian
 nearest neighbor interaction form:
 \begin{eqnarray}
 \label{frit}
M_{u,d} = \left(\begin{matrix}0 & A & 0 \\ A^* & 0 & B \\ 0 & B^* &
C \end{matrix}\right)_{u,d}\,.
 \end{eqnarray}
 Such matrices have factorizable phases, i.e., $M_{u,d} = P_{u,d} \hat{M}_{u,d} P^*_{u,d}$,
 where $\hat{M}_{u,d}$ are the same as in Eq. (\ref{frit}), but without any phases, and $P_{u,d}$ are diagonal
 phase matrices.  Only two combinations of phases will enter into the CKM matrix, contained in the matrix
 $P = P_u^* P_d = {\rm diag}.\{e^{i \alpha},\,e^{i \beta},\, 1\}$.  The CKM matrix is then given by
\begin{equation}
\label{phase}
V= O_u^T P O_d\,,
\end{equation}
where $O_{u,d}$ are the orthogonal matrices that diagonalize $\hat{M}_{u,d}$ via
\begin{equation}
\label{Oud}
O_{u,d}^T \hat{M}_{u,d}\hat{M}_{u,d}^T O_{u,d} = {\rm diag.} \{m^2_{u,d},~m^2_{c,s},~m^2_{t,b} \}\,.
\end{equation}

In this model there are a total of eight parameters that describe quark masses, mixings
and CP violation:  six real parameters from $\hat{M}_{u,d}$ and the two phases $(\alpha,\,\beta)$.
(Note that the six mixing angles that enter into $O_u$ and $O_d$ are determined in terms of
the quark mass ratios.) These eight parameters must describe ten observables in the quark sector.  There are thus
two true predictions.  Furthermore, since  $(\alpha,\,\beta)$  are phases, they do not count as
full parameters. One finds four relations between masses and mixings \cite{babushafi}:
\begin{eqnarray}
\label{fritpred}
|V_{us}| &\simeq& \left|\sqrt{\frac{m_d}{m_s}} - e^{i \psi} \sqrt{\frac{m_u}{m_c}}\right|\,, \nonumber \\
|V_{cb}| &\simeq& \left|\sqrt{\frac{m_s}{m_b}} - e^{i \phi} \sqrt{\frac{m_c}{m_t}}\right|\,, \nonumber \\
|V_{ub}| &\simeq& \left|\frac{m_s}{m_b} \sqrt{\frac{m_d}{m_b}} + e^{i \psi}\sqrt{\frac{m_u}{m_c}}
\left(\sqrt{\frac{m_s}{m_b}} - e^{i \phi} \sqrt{\frac{m_c}{m_t}}\right) \right|\,, \nonumber \\
|V_{td}| &\simeq& \left|\frac{m_c}{m_t} \sqrt{\frac{m_u}{m_t}} + e^{i \psi}\sqrt{\frac{m_d}{m_s}}
\left(\sqrt{\frac{m_c}{m_t}} - e^{i \phi} \sqrt{\frac{m_s}{m_b}}\right) \right|\,.
\end{eqnarray}
Here the two phases $\psi$ and $\phi$ are related to the phases in the diagonal matrix $P$ as $\psi= (\alpha-\beta)$ and $\phi=\beta$.  Note that all independent elements of $V$ are determined in this model in terms of
the quark mass ratios and two phase parameters.
In the expression for $|V_{ub}|$ in Eq. (\ref{fritpred}), the first term is numerically $\simeq 6 \times 10^{-4}$,
which is about a factor of 8 less than the value of $|V_{ub}|$.  Similarly, the first term in the expression
for $|V_{td}|$ is $\sim 1 \times 10^{-5}$, which is negligible in relation to the value of $|V_{td}|$.  (For
these numerical estimates, I used the values of the running masses given in Table \ref{Table:masses} evaluated
at $\mu = 1$ TeV.) If these terms are neglected, one would have the following predictions:
\begin{equation}
\label{hallrasin}
\frac{|V_{ub}|}{|V_{cb}|}  \simeq \sqrt{\frac{m_u}{m_c}}\,, \, ~~~~~\frac{|V_{td}|}{|V_{ts}|} \simeq \sqrt{\frac{m_d}{m_s}}\,.
\end{equation}
These predictions are consistent with experimental data.

While the prediction for $|V_{us}|$ in the Fritzsch ansatz is the same as
in the two family model of Eq. (\ref{cabibbo}), which is successful, the relation for $|V_{cb}|$ will predict
the mass of the top quark to be in the range $(40-80)$ GeV, which is now excluded by data.

There have been attempts to fix the problem of Fritzsch mass
matrices by modifying its form slightly.  If  small (2,2) elements
are allowed in $M_{u,d}$, the troublesome relation for $|V_{cb}|$
will be removed. However, adding (2,2) entries
in $M_u$ and $M_d$ introduces two more complex parameters, and such a model will have no
true prediction.  The relation of Eq. (\ref{cabibbo}) will however be maintained, provided
that the (2,2) entries are not too large.  Furthermore, the relations of Eq. (\ref{hallrasin}),
$|V_{ub}|/|V_{cb}| \simeq \sqrt{m_u/m_c}$ and $|V_{td}|/|V_{ts}| \simeq \sqrt{m_d/m_s}$,
will be preserved \cite{rasin}, if the new (2,2) entries are small perturbations.

A different alternative is to make the (2,3) and (3,2) entries of Eq. (\ref{frit})
different, while maintaining the relations between (1,2) and (2,1)
entries.  This can be achieved by non--Abelian discrete
symmetries.  Again, the number of parameters will increase by two compared to the original
Fritszch ansatz.  A special case where there is still a true prediction is worth mentioning.

Consider a non--Abelian discrete subgroup $G$ of $SU(2)$ serving as a family symmetry.  $G$ is assumed
to have pseudo--real doublet representations, just as $SU(2)$.  Let the
first two families of quarks be pseudo--real doublets of $G$, while the third family quarks are singlets of $G$.
A {\it real} Higgs doublet which is a true singlet of $G$ will generate the (3,3) entries of $M_{u,d}$ as
well as (1,2) and (2,1) entries.  Note that invariance under $G$ will lead to the (1,2) entry being
the negative of the (2,1) entry, a property of the original $SU(2)$ family symmetry.  Now, if $G$
is broken by a Higgs field transforming as a doublet of $G$, then unequal (2,3) and (3,2) entries in
$M_{u,d}$ can be generated.  This is a concrete modification of the Fritszch ansatz, with (1,2)
and (2,1) entries having the same magnitude, but with the (2,3) and (3,2) entries unrelated.

A model of the type just described has been constructed in Ref. \cite{bk}.  It is based on the
dihedral group $Q_6$ which contains pseudo--real as well as real doublets.  Most interestingly, if
the origin of CP violation is taken to be spontaneous, then the phase matrix $P$ appearing in
Eq. (\ref{phase}) will have the form $P = diag.\{e^{-i \phi},\, e^{i \phi},\,1\}$.  (This happens
since $M_u$ and $M_d$ each will have a single phase, appearing in the (2,3) and (3,2) entries,
apart from irrelevant overall phases, if all the Yukawa couplings are assumed to be real by virtue
of CP invariance.)  Such a model
will have one true prediction, since now there are nine parameters describing ten observables.  It was found in Ref. \cite{bk} that this prediction, which relates $\overline{\eta}$ with $\overline{\rho}$, is fully consistent with data.

We shall return to mass matrix ``textures" of the type described here when discussing fermion
masses in the context of Grand Unification in Sec. \ref{sec5}.  The nearest neighbor
interaction, not necessarily symmetrical, will find useful applications.

\section{Froggatt--Nielsen mechanism for mass hierarchy \label{sec4}}

The hierarchy in the masses and mixings of quarks and leptons can
be understood by assuming a flavor $U(1)$ symmetry under which the
fermions are distinguished.  In this approach developed by Froggatt
and Nielsen \cite{refFN}, there is a ``flavon" field $S$, which is a scalar,
usually a SM singlet field, which acquires a VEV and breaks the
$U(1)$ symmetry. This symmetry breaking is communicated to the
fermions at different orders in a small parameter $\epsilon = \left
\langle S \right \rangle/M_*$. Here $M_*$ is the scale of flavor
dynamics, and usually is associated with some heavy fermions which
are integrated out.  The nice feature of this approach is that the
mass and mixing hierarchies will be explained as powers of the
expansion parameter $\epsilon$ without assuming widely
different Yukawa couplings.  The effective theory below $M_*$
is rather simple, while the full theory will have many heavy
fermions, called Froggatt--Nielsen fields.

\subsection{A two family model}

Let me illustrate this idea with a two family example which is
realistic when applied to the second and third families of quarks.
Consider $M_u$ and $M_d$ for the ($c,~t$) and $(s,~b)$ sectors
given by
\begin{eqnarray}
\label{FG}
M_u = \left(\begin{matrix} \epsilon^4 & \epsilon^2 \\ \epsilon^2 &
1 \end{matrix}\right) v_u\,,~~~M_d= \left(\begin{matrix} \epsilon^3 & \epsilon^3\\
\epsilon & \epsilon \end{matrix}\right) v_d \,.
\end{eqnarray}
Here $\epsilon \sim 0.2$ is a flavor symmetry breaking parameter.
Every term in Eq. (\ref{FG}) has an order one coefficient which is not displayed. We
obtain from Eq. (\ref{FG}) the following relations for quark masses and
$|V_{cb}|$:
\begin{equation}
\frac{m_c}  {m_t} \sim \epsilon^4\,, ~ \frac{m_s} {m_b} \sim
\epsilon^2\,,~ |V_{cb}| \sim \epsilon^2\,.
\end{equation}
All of these relations work well, for $\epsilon \sim 0.2$.
Although precise predictions have not been made, one has a qualitative
understanding of the hierarchies.

How do we arrive at Eq. (\ref{FG})?  We do it in two stages.  First, let
us look at the effective Yukawa couplings, which can be obtained
from the Lagrangian:
\begin{eqnarray}
\label{FGYuk}
{\cal L}_{FN}^{\rm eff} &=& \left[Q_3 u_3^c H_u + Q_2 u_3^c H_u S^2 + Q_3
u_2^c H_u S^2 + Q_2 u_2^c H_u S^4 \right] \nonumber \\
&+& \left[Q_3 d_3^c H_d S + Q_3 d_2^c H_d S + Q_2 d_2^c H_d S^3 +
Q_2 d_3^c H_d S^3 \right]  + h.c.
\end{eqnarray}
Here I assumed supersymmetry, so that there are two Higgs doublets
$H_{u,d}$.  It is not necessary to assume SUSY, one can simply
identify $H_u$ as $H$ of SM, and replace $H_d$ by $\tilde{H}$.  In
Eq. (\ref{FGYuk}) all couplings are taken to be of order one.  The symmetry of
Eq. (\ref{FGYuk}) is a $U(1)$ with the following charge assignment.
\begin{equation}
\{Q_3, u_3^c\}: 0;~~\{Q_2, u_2^c\}: 2;~\{d_2^c, d_3^c\}: 1;~\{H_u,
H_d\}:0;~ S:-1\,.
\end{equation}

\begin{figure}[!htb]
\begin{center}
\begin{tabular}{cc}
\includegraphics[width=0.7\linewidth]{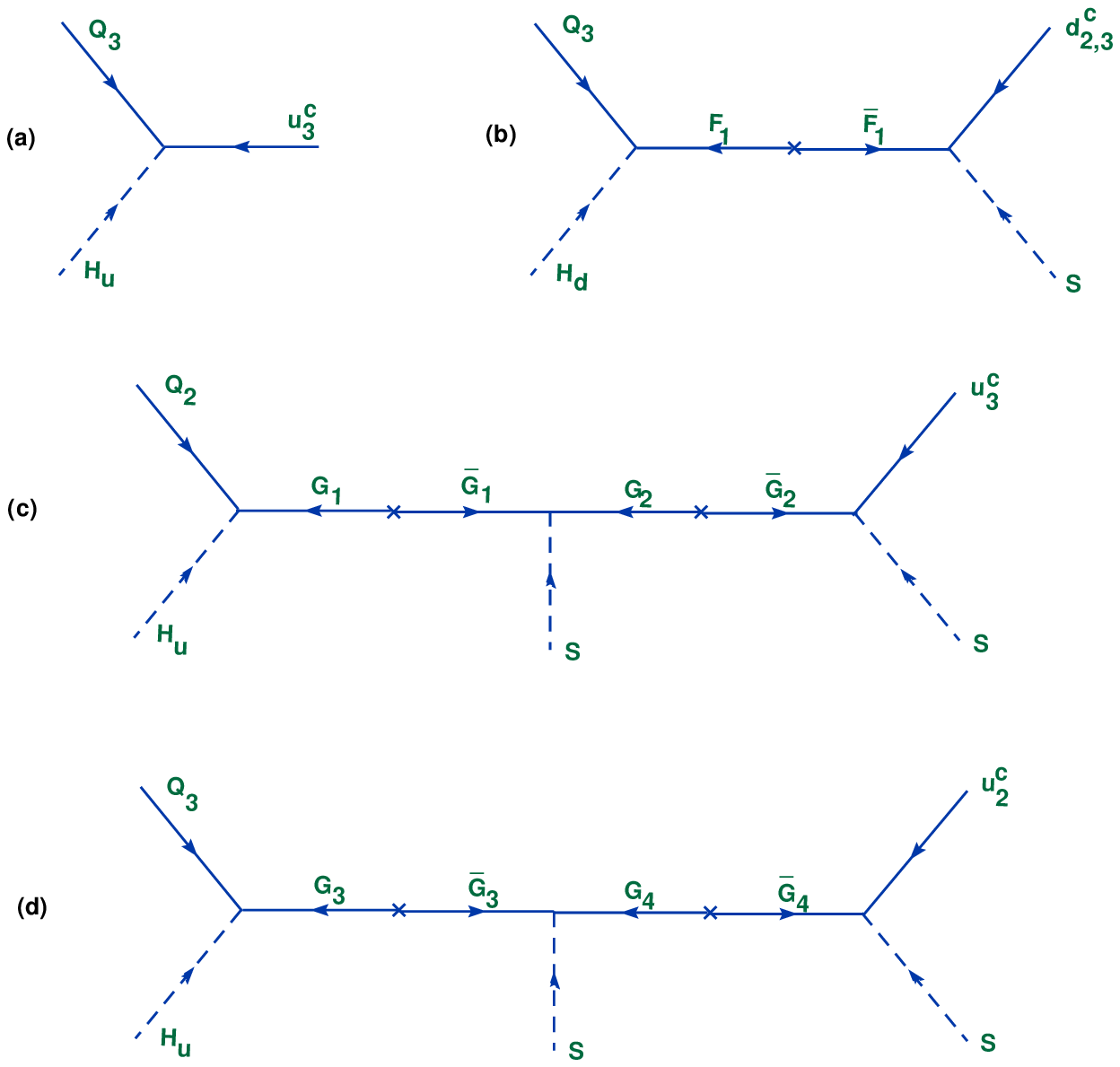} & ~ \\[0.1in]
\includegraphics[width=0.65\linewidth]{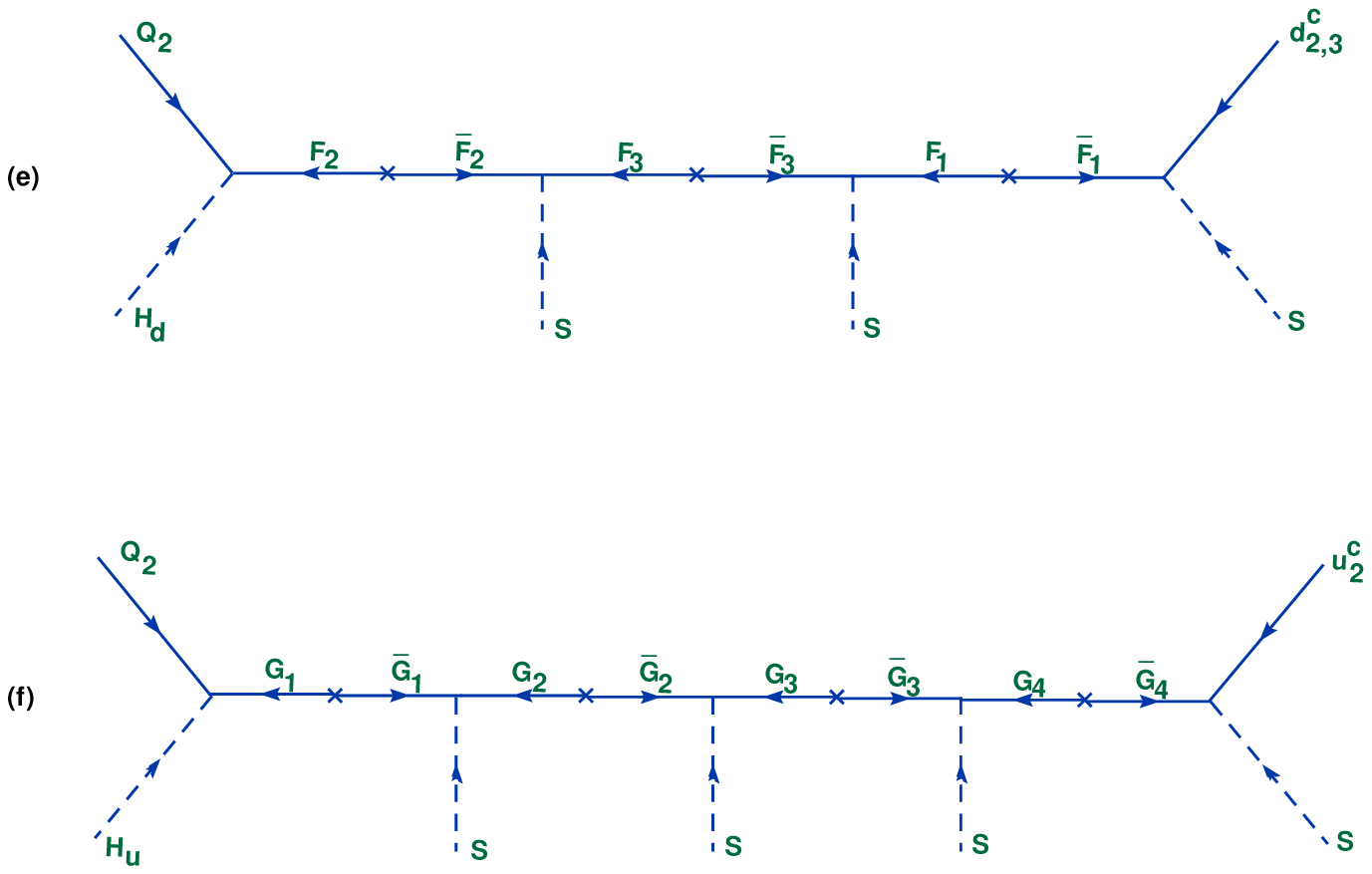} \\
\end{tabular}
\vspace*{0.1in} \caption{Froggatt--Nielsen fields generating effective Yukawa couplings of Eq.
(\ref{FGYuk}).}
\label{Fig:FGYuk}
\end{center}
\end{figure}

Now we wish to obtain Eq. (\ref{FGYuk}) by integrating out certain
Froggatt--Nielsen fields.  This is depicted in Fig. \ref{Fig:FGYuk}
 via a set of
``spaghetti" diagrams.  As you can see, there are a variety of
fields denoted by $G_i, \overline{G}_i$ ($i = 1-4$) for the
up--quark mass generation.  $G_i$ have the same gauge quantum
numbers as the $u^c$ quark of SM, while $\overline{G}_i$ have the
conjugate quantum numbers.  $F_i$ have the quantum numbers of
$d^c$ quark, while $\overline{F}_i$ the conjugate quantum numbers.

You can readily read off the flavor $U(1)$ charges of the various
$F_i$ and $G_i$ fields from the spaghetti diagrams.  For example,
the charge of $G_1$ is $-2$, while that of $\overline{G}_1$ is $+2$.
The charges of $G_2$ is $-1$ and that of $\overline{G}_2$ is $+1$.

All flavor dynamics in this class of models could occur near the Planck scale.  As long
as the hierarchy between $\left\langle S \right\rangle$ and the
masses of the Froggatt--Nielsen fields is not too strong, realistic
fermion masses will be generated.  Consider for example, Fig. \ref{Fig:FGYuk}
(b) which induces the $b$--quark mass.  The effective interaction from this
diagram goes as ${\cal L}^{\rm eff}_b = Y_1 Y_2\, (Q_3 d_3^c H_d)\, (S/M_{F_1})$,
where $Y_{1,2}$ are order one Yukawa couplings. If $\left\langle S \right
\rangle/M_{F_1} \sim 0.2$ or so, realistic $b$--quark mass is obtained
(with $\tan\beta \sim 10$).
This allows for both $\left\langle S \right \rangle$ and $M_{F_1}$ to
be near the Planck scale.  From Fig. \ref{Fig:FGYuk} (f), one can read off the
effective Lagrangian inducing the $c$--quark mass: ${\cal L}^{\rm eff}_c = \Pi_{i=1}^5 Y_i'\,
(Q_2 u_3^c H_u)\, (S^4/M_G^4)$. Here $Y_i'$ are order one Yukawa couplings, and we assumed
that all of $G_i$ ($i=1-4)$ appearing in Fig. \ref{Fig:FGYuk} (f) have a common mass $M_G$.
With all couplings being order one, $m_c/m_t \sim 1/400$ can be reproduced, with $\epsilon \sim 0.2$.
It should be emphasized that, although there are various Yukawa couplings, all of them can take order
one values.

\clearpage

\subsection{A realistic three family Froggatt--Nielsen model}

Actually the flavor $U(1)$ that we used in the previous section is anomalous.  String
theory, when compactified to four dimension, generically gives an anomalous $U(1)_A$ with anomaly
cancelation occuring by the Green--Schwartz mechanism \cite{GS}.  In this
case, we can get rid of the complicated Froggatt--Nielsen fields,
and simply write down higher dimensional operators suppressed by
the string scale.  A bonus in this approach is that the small
expansion parameter $\epsilon$ can be computed in specific models,
where it tends to come out close to 0.2, of order the Cabibbo angle.

An explicit and complete anomalous $U(1)$ model that fits well all
quark and lepton masses and mixings is constructed below. Consider
the quark and lepton mass matrices of the following form \cite{be}:
\begin{eqnarray}\label{massM1}
&&M_u\sim \langle
H_u\rangle \begin{pmatrix}\epsilon^{\,8}&\epsilon^{\,6}&\epsilon^{\,4}\\
\epsilon^{\,6}&\epsilon^4&\epsilon^2\\\epsilon^{\,4}&\epsilon^2&1\end{pmatrix}\,,\hspace{1.cm}
M_d\sim \langle H_d\rangle\epsilon^p
\begin{pmatrix}\epsilon^{\,5}&\epsilon^{\,4}&\epsilon^{\,4}\\
\epsilon^3 &\epsilon^2&\epsilon^2\\\epsilon&1&1 \end{pmatrix},\nonumber\\
\nonumber\\
\vspace{.5cm} &&M_e\sim \langle
H_d\rangle\epsilon^p\begin{pmatrix}\epsilon^{\,5}&\epsilon^3&\epsilon\\
\epsilon^{\,4}&\epsilon^2&1\\\epsilon^{\,4}&\epsilon^2&1 \end{pmatrix}\,,\hspace{1.cm}
M_{\nu_D}\sim \langle H_u\rangle\epsilon^s
\begin{pmatrix}\epsilon^2&\epsilon &\epsilon \\ \epsilon
&1&1\\\epsilon&1&1 \end{pmatrix},\nonumber\\
\nonumber\\
 \vspace{.5cm} &&M_{\nu^c}\sim M_R
\begin{pmatrix}\epsilon^2&\epsilon&\epsilon\\
\epsilon&1&1\\\epsilon&1&1 \end{pmatrix}\,\hspace{.5cm}\Rightarrow
\hspace{.5cm} M^{light}_\nu\sim \frac{{\langle
H_u\rangle}^2}{M_R}\epsilon^{2s}
\begin{pmatrix}\epsilon^2&\epsilon&\epsilon\\
\epsilon&1&1\\\epsilon&1&1\end{pmatrix}\,.
\end{eqnarray}
Here we work with the MSSM gauge group with supersymmetry realized
at the TeV scale. Each entry has an order one pre-factor in the matrices of Eq. (\ref{massM1}), which is
not explicitly shown. These matrices can be obtained by the $U(1)$ charge assignment of
Table \ref{Table:u1charge}.   In Eq. (\ref{massM1}), the integer $p$ is allowed to
take values 0, 1 or 2, corresponding to $\tan\beta$ taking large, medium or small values.
The integer $s$ only enters into neutrino masses.  Green--Schwarz
anomaly cancelation condition requires $s=p$ in the simplest scheme.  With $s=p$,
the charge assignment of Table {\ref{Table:u1charge} will be compatible with
$SU(5)$ unification.  That is to say that the $\{Q_i,\, u^c_i,\,e^c_i\}$ fields
of a given generation all have the same $U(1)_A$ charge, and similarly, the
$\{d^c_i,\,L_i\}$ fields of a given family have the same charge.  As we discuss
in Sec. \ref{sec5}, the former set of SM particles are grouped into a ${\bf 10}$ of
$SU(5)$, while the latter set forms a ${\bf \overline{5}}$.

In the last line of Eq. (\ref{massM1}), $M_{\nu^c}$ stands for the heavy $\nu^c$ Majorana
mass matrix.  When the seesaw formula is applied one obtains the light neutrino mass matrix
$M_\nu^{light}$, shown also in the last line of Eq. (\ref{massM1}).

\begin{table}[t]
\begin{center}
\begin{tabular}{|c|c|c|}\hline
\rule[5mm]{0mm}{0pt}Field& $U(1)_A$ Charge& Charge
notation\\\hline
\rule[5mm]{0mm}{0pt}$Q_1$, $Q_2$, $Q_3$& \,\,$ 4, \,2, \,0$ &$q^Q_i$\\
\rule[5mm]{0mm}{0pt}$L_1$, $L_2$, $L_3$&\,\,$ 1+s,\, s,
\,s$&$q^L_i$\\
\rule[5mm]{0mm}{0pt}$u^c_1$, $u^c_2$, $u^c_3$&$ 4,\, 2,\, 0$&$q^u_i$\\
\rule[5mm]{0mm}{0pt}$d^c_1$, $d^c_2$, $d^c_3$&$1+p,\, p,\,
p$& $q^d_i$\\
\rule[5mm]{0mm}{0pt}$e^c_1$,$e^c_2$,$e^c_3$&\,\,$ 4+p-s,\, 2+p-s,\, p-s$\,\,&$q^e_i$\\
\rule[5mm]{0mm}{0pt}$\nu^c_1$, $\nu^c_2$,
$\nu^c_3$ &$ 1,\, 0,\, 0$&$q^\nu_i$\\
\rule[5mm]{0mm}{0pt}$H_u$, $H_d$, $S$&$ 0,\, 0,\, -1$&$(h, \bar{h}, q_s)$\\
\hline
\end{tabular}
\caption{The flavor $U(1)_A$ charge assignment for
the MSSM fields and the flavon field $S$.}
\label{Table:u1charge}
\end{center}
\end{table}

All the qualitative features of quark and lepton masses and mixings
are reproduced by these matrices.  These include small quark
mixings and large neutrino mixings.  The mass ratios in the up--quark
sector scale as $m_u:m_c:m_t \sim \epsilon^8: \epsilon^4:1$, while
those in the down quarks scale as $m_d:m_s:m_b \sim \epsilon^5:\epsilon^2:1$
with an identical scaling for the charged lepton mass ratios.  (See the diagonal
entries of $M_{u,d,e}$ in Eq. (\ref{massM1}).) These are
all consistent with experimental data.  The quark mixing angles scale roughly
as the down quark mass ratios, which is also reasonable.  In the charged lepton
sector, the mixing angles are larger, compared to the quark sector.  This arises
because of the lopsided structure of $M_d$ and $M_e$ with $M_d \sim M_e^T$.
This is a feature of $SU(5)$ grand unification, where left--handed lepton doublets
are paired with the conjugate of the right--handed down quarks.  As a result,
the left--handed leptonic mixing angles will be related to the right--handed down
quark mixing angles, which are allowed to be large since they are unobservable
in the SM \cite{lopsided}.  Note also that the hierarchy between light neutrino
masses is weaker, $(m_1:m_2:m_3) \sim (\epsilon^2: 1 : 1),$ compared with the charged fermion mass hierarchy.
This feature is also consistent with neutrino oscillation data.

A variety of models based on anomalous $U(1)$ flavor symmetry have been proposed in
the literature.  A cross section of these models can be found in Ref. \cite{IbanezU1,Kobayashi,beg}.

\subsubsection{More about anomalous $U(1)$ flavor symmetry}

To see the consistency of the three family model described above, and to see how it may be subject to experimental
scrutiny, let us explore the structure of anomalous $U(1)$ flavor symmetry and its applications a little further.
This will also enable us to compute the small parameter $\epsilon$ in the model of Table \ref{Table:u1charge}.

In heterotic string theory the $U(1)_A$ anomalies are canceled by the Green--Schwarz
mechanism \cite{GS} which requires
\begin{eqnarray}\label{AnomalyGS}
\frac{A_1}{k_1}=\frac{A_2}{k_2}=\frac{A_3}{k_3}=\frac{A_{F}}{3k_F}=\frac{A_{gravity}}{24}\,.
\end{eqnarray}
Here $A_1$, $A_2$, $A_3$, $A_F$ and $A_{gravity}$ are the
$U(1)_Y^2\times U(1)_A$, $SU(2)_L^2\times U(1)_A$,
$SU(3)_C^2\times U(1)_A$, $U(1)^3_A$ and $(Gravity)^2\times
U(1)_A$ anomaly coefficients. (The subscript $F$ is used to indicate the anomalous $U(1)$ flavor
symmetry group.)   All other anomalies (such as
$U(1)_A^2\times U(1)_Y$) must vanish. $k_i\,(i=1,2,3)$, $k_F$ are
the Kac-Moody levels.  The non--Abelian levels $k_2$ and $k_3$
must be integers. The factor $1/3$ in front of the cubic anomaly
$A_F$ has a combinatorial origin owing to the three identical
$U(1)_A$ gauge boson legs.

Even without a covering grand unified group, string theory predicts unification of
all gauge couplings, including that of the $U(1)_A$ and $g_F$, at the fundamental scale $M_{st}$
\cite{Ginsparg,cvetic}:
\begin{eqnarray}
\label{UnifGS}
k_ig_i^2=k_Fg_F^2=2g_{st}^2.
\end{eqnarray}
Here $g_i$ are the $U(1)_Y$, $SU(2)_L$ and $SU(3)_C$ gauge couplings
for $i=1,\,2,\,3$.

With $k_2=k_3=1$ we find from Table \ref{Table:u1charge}, $A_2=(19+3s)/2$
and $A_3=(19+3p)/2$. Eq. (\ref{AnomalyGS}) then requires $p=s$,
i.e., a common exponent for the charged lepton and the neutrino
Dirac Yukawa coupling matrices. With $p=s$, the condition
$A_1/k_1=A_2/k_2$ fixes $k_1$ to be $5/3$, which is consistent with $SU(5)$
unification. Note also that the charges given in Table
\ref{Table:u1charge} become compatible with $SU(5)$ unification. Since
${\rm Tr}(Y) = 0$ for the fermion multiplets of $SU(5)$, and since
the Higgs doublets carry zero $U(1)_A$ charge, the anomaly
coefficient $[U(1)_A]^2 \times U(1)_Y$ vanishes, as required. The
last equality in Eq. (\ref{AnomalyGS}) requires
\begin{eqnarray}\label{Agravity}
A_{gravity}=\mbox{Tr}\left(q\right)=12(19+3p).
\end{eqnarray}
This cannot be satisfied with the MSSM fields alone, since
$\mbox{Tr}(q)_{MSSM}=5(13+3p)$, which does not match Eq.
(\ref{Agravity}). We cancel this anomaly by introducing MSSM
singlet fields $X_k$ obeying
$\mbox{Tr}\left(q\right)_X=A_{gravity}-\mbox{Tr}\left(q\right)_{MSSM}=163+21p$.
If all the $X_k$ fields have the same charge equal to $+1$, they
will acquire masses of order $M_{st}\epsilon^2$ through the coupling
$X_kX_kS^2/M_{st}$ and will decouple from low energy theory.
We will assume that these fields $X_k$  have charge $+1$.

With the charges of all fields fixed, we are now in a position to
determine the $U(1)_A$ charge normalization so that
$g_F^2=g_2^2=g_3^2$ at the string scale, (We take $k_2=k_3=1$.)
This normalization factor, which we denote as $|q_s|$, is given by
$|q_s|=1/\sqrt{k_F}$. All the charges given in Table \ref{Table:u1charge}
are to be multiplied by $|q_s|$. From the Green--Schwarz anomaly
cancelation condition $A_F/(3k_F)=A_2/k_2$, we have
\begin{eqnarray}\label{Ax}
\frac{\mbox{Tr}\left(q^3\right)}{3k_F}=\frac{19+3p}{2k_2},
\end{eqnarray}
from which we find the normalization of the $U(1)_A$ charge
$|q_s|=1/\sqrt{k_F}$ to be
\begin{eqnarray}\label{anomalyGS}
|q_s|=\left(0.179,\,0.186,\,0.181\right)\,\, \mbox{for $p=(0, 1,
2$)}\, .
\end{eqnarray}

The Fayet--Iliopoulos term for the anomalous $U(1)_A$, generated
through the gravitational anomaly, is given by \cite{DSW}
\begin{eqnarray}
\xi=\frac{g_{st}^2M_{st}^2}{192\pi^2}|q_s|A_{gravity}\,,
\end{eqnarray}
where $g_{st}$ is the unified gauge coupling at the string scale
(see Eq. (\ref{UnifGS})). By minimizing the potential from the
$U(1)_A$ $D$--term
\begin{eqnarray}\label{Dterm}
V=\frac{|q_s|^2g_F^2}{8}\left(\frac{\xi}{|q_s|}-|S|^2+\sum_aq_a^f|\tilde{f}_a|^2+\sum_k
q_k^X|X_k|^2\right)^2,
\end{eqnarray}
in such a way that supersymmetry remains unbroken ($\tilde{f}_a$ are the MSSM sfermions
and $X_k$ are the singlet fields, which do not acquire VEVs), one finds for
the VEV of $S$
\begin{eqnarray}\label{Eps}\epsilon=\langle
S\rangle/M_{st}=\sqrt{g_{st}^2A_{gravity}/192\pi^2}.\end{eqnarray}
For the fermion mass texture in Eq. (\ref{massM1}), corresponding
to the $U(1)_A$ charges given in Table \ref{Table:u1charge}, we find
\begin{eqnarray}\label{epsilon}
\epsilon=\left(0.177,\,0.191,\,0.204\right)\,\, \mbox{for $p=(0,
1, 2$)}\, .
\end{eqnarray}
This shows that the small expansion parameter can indeed be calculated
in string--inspired models. It should be noted that this is a bottom--up
approach to model building, it would of course be desirable to start from
string theory and arrive at the spectrum and charges listed in Table \ref{Table:u1charge}.

The masses of the $U(1)_A$ gauge boson and the corresponding
gaugino are obtained from $M_F=|q_s|g_F\langle S\rangle/\sqrt{2}$
and found to be
\begin{eqnarray}\label{MAM1}
M_F=\left(\frac{M_{st}}{54.5},\,
\frac{M_{st}}{52.5},\,\frac{M_{st}}{53.9}\right)\, \mbox{for
$p=(0, 1, 2$)}\, .
\end{eqnarray}
In the momentum range below $M_{st}$ and above $M_F$, these gauge
particles will be active and will induce flavor dependent
corrections to the sfermion soft masses and the $A$--terms.
Implications of these effects have been studied in Ref. \cite{be,beg},
where it has been shown that the process $\mu \rightarrow e \gamma$ in
this class of models is very close to the current experimental limits.
Ongoing MEG experiment should be able to probe the entire allowed
parameter space of these models, provided that the SUSY particles have
masses not exceeding about 1 TeV.

\subsection{The SM Higgs boson as the flavon}

Can the SM Higgs field itself be the flavon field?  Clearly, then
new flavor dynamics must happen near the TeV scale.  This is
apparently possible with significant consequences for Higgs boson
physics, as I shall now outline \cite{bn,gl}.

Consider an expansion in $H^\dagger H/M^2$, which is a SM singlet
that can play the role of $S$.  Here $H$ is the SM Higgs doublet and
$M$ is the scale of new physics.  Immediately you may wonder how
this is possible, since $H^\dagger H$ cannot carry any $U(1)$
quantum number.  But think of SUSY at the TeV scale.  SUSY has
two Higgs doublets, $H_u$ and $H_d$, in which case the combination
$H_uH_d$ can carry $U(1)$ charge.  When reduced to SM this expansion in
terms of $H^\dagger H$ can be consistent.

Consider the following mass
matrices for quarks in terms of the expansion parameter
\begin{equation}
\epsilon = \frac{v} {M}\,.
\end{equation}
\begin{eqnarray}
\label{hier}
M_u = \left(\begin{matrix}h_{11}^u \epsilon^6 & h_{12}^u \epsilon^4 &
h_{13}^u \epsilon^4 \\ h_{21}^u\epsilon^4 & h_{22}^u \epsilon^2 &
h_{23}^u \epsilon^2 \\ h_{31}^u \epsilon^4 & h_{32}^u \epsilon^2
& h_{33}^u \end{matrix}\right)v\,, ~~~~~ M_d = \left(\begin{matrix}h_{11}^d \epsilon^6
& h_{12}^d \epsilon^6 & h_{13}^d \epsilon^6 \\ h_{21}^d\epsilon^6
& h_{22}^d \epsilon^4 & h_{23}^d \epsilon^4 \\ h_{31}^d
\epsilon^6 & h_{32}^d \epsilon^4 & h_{33}^d \epsilon^2 \end{matrix}\right)v\,.
\end{eqnarray}
The charged lepton mass matrix is taken to have a form similar to $M_d$, with
the couplings $h_{ij}^d$ replaced by $h_{ij}^\ell$.
These matrices give good fit to masses and mixings, as in the case
of anomalous $U(1)$ model with  $\epsilon \sim 1/7$ and all the
couplings $h^{u,d}_{ij}$ being of order one.
The masses of the quarks and leptons can be read off from Eq. (\ref{hier}) in the
approximation $\epsilon \ll 1$:
\begin{eqnarray}
\{m_t, m_c, m_u\}  &\simeq& \{|h^u_{33}|,~ |h^u_{22}|\epsilon^2,~ |h^u_{11} -
h^u_{12}h_{21}^u/h_{22}^u|\epsilon^6\}v ,\nonumber \\
\{m_b, m_s, m_d\} &\simeq& \{|h_{33}^d| \epsilon^2, ~|h_{22}^d| \epsilon^4,~
|h_{11}^d|\epsilon^6\}v, \nonumber \\
\{m_\tau, m_\mu, m_e\} &\simeq& \{|h_{33}^\ell| \epsilon^2,~ |h_{22}^\ell| \epsilon^4,~
|h_{11}^\ell|\epsilon^6\}v.
\end{eqnarray}
The quark mixing angles are found to be:
\begin{eqnarray}
|V_{us}| &\simeq& \left|\frac{h_{12}^d} {h_{22}^d}-\frac{h_{12}^u} {h_{22}^u}\right| \epsilon^2,
\nonumber \\
|V_{cb}| &\simeq& \left|\frac{h_{23}^d} {h_{33}^d}-\frac{h_{23}^u} {h_{33}^u}\right|
\epsilon^2 ,\nonumber \\
|V_{ub}| &\simeq& \left|\frac{h_{13}^d} {h_{33}^d}-\frac{h_{12}^u h_{23}^d} {h_{22}^uh_{33}^d}
- \frac{h_{13}^u}{h_{33}^u}\right|\epsilon^4.
\end{eqnarray}

With $\epsilon =1/6.5$ and with all couplings $h_{ij}^{u,d}$ being of order one, excellent
fits to the quark masses and CKM mixing angles can be found.  As an example, take the
couplings to be
\begin{eqnarray}
\{|h_{33}^u|, |h_{22}^u|, |h_{11}^u-h_{12}^u h_{21}^u/h_{22}^u|\} &\simeq& \{0.96, 0.14,
0.95\}, \nonumber \\
\{|h_{33}^d|, |h_{22}^d|, |h_{11}^d|\} &\simeq& \{0.68, 0.77, 1.65\}, \nonumber \\
\{|h_{33}^\ell|, |h_{22}^\ell|, |h_{11}^\ell|\} &\simeq& \{0.42, 1.06, 0.21\}.
\end{eqnarray}
The corresponding quark masses at $\mu = m_t(m_t)$ are:
\begin{eqnarray}
\{m_t, m_c, m_u\} &\simeq& \{166,~ 0.60,~ 2.2 \times 10^{-3}\}~{\rm GeV}, \nonumber \\
\{m_b, m_s, m_d\} &\simeq& \{2.78,~ 7.5 \times 10^{-2},~ 3.8 \times 10^{-3}\}~{\rm GeV},
\nonumber \\
\{m_\tau, m_\mu, m_e\} &\simeq& \{1.75,~ 0.104,~ 5.01 \times 10^{-4}\}~{\rm GeV}.
\end{eqnarray}
All these are in agreement with values quoted in Table \ref{Table:masses}.  Furthermore,
the CKM mixing angles are also reproduced correctly with this choice of couplings.

In this scheme, the Yukawa coupling matrices of the physical quark fields are no longer
proportional to the corresponding mass matrices.  We obtain for the Yukawa
couplings,
\begin{eqnarray}
Y_u = \left(\begin{matrix}7h_{11}^u \epsilon^6 & 5h_{12}^u \epsilon^4 &
5h_{13}^u \epsilon^4 \\ 5h_{21}^u\epsilon^4 & 3h_{22}^u
\epsilon^2 & 3h_{23}^u \epsilon^2 \\ 5h_{31}^u \epsilon^4 &
3h_{32}^u \epsilon^2 & h_{33}^u \end{matrix}\right)\,, ~~~~~ Y_d =
\left(\begin{matrix}7h_{11}^d \epsilon^6 & 7h_{12}^d \epsilon^6 &
7h_{13}^d \epsilon^6 \\ 7h_{21}^d\epsilon^6 & 5h_{22}^d
\epsilon^4 & 5h_{23}^d \epsilon^4 \\ 7h_{31}^d \epsilon^6 &
5h_{32}^d \epsilon^4 & 3h_{33}^d \epsilon^2 \end{matrix}\right)\,.
\end{eqnarray}
Take for example, the (3,3) entry in $M_d$. It arises from the operator
$h^d_{33} Q_3 d_3^c \tilde{H} (H^\dagger H)/M^2$.  The contribution to the
mass matrix from this operator is $h^d_{33} v \epsilon^2$, while the
contribution to the Yukawa coupling is $(h/\sqrt{2})h^d_{33} (3 \epsilon^2)$.
The flavor factors (3 in this example) are not the same for various entries,
and would result in flavor violation in Higgs interactions.

There is a tree--level contribution mediated by the Higgs boson for $K^0-\bar{K^0}$
mass difference in this scheme.
The new contribution, $\Delta m_K^{\rm Higgs}$, is given by
\begin{eqnarray}
\Delta m_K^{\rm Higgs} &\simeq& \frac{4} {3} \frac{f_K^2 m_K B_K} {m_{h^0}^2}\epsilon^{12}\left[ \right.
\{ \frac{1}{6} \frac{m_K^2} {(m_d+m_s)^2} + \frac{1} {6}\} {\rm Re}\left[\left(\frac{h_{12}^d+
h_{21}^{d*}}{\sqrt{2}}\right)^2\right] \nonumber \\
&-& \{ \frac{11} {6} \frac{m_K^2} {(m_d+m_s)^2} + \frac{1} {6}
\} {\rm Re}\left[\left(\frac{h_{21}^d-h_{12}^{d*}}{\sqrt{2}}\right)^2\right]\left.\right]\,.
\end{eqnarray}
Here $B_K$ is the bag parameter. Using $B_K = 0.75, f_K \simeq 160$ MeV, $\epsilon \simeq 1/6.5$ $m_s(1 ~{\rm GeV}) = 175$ MeV,
$m_d(1~{\rm GeV}) = 8.9$ MeV, and with  $h_{12}^d = 1,  h_{21}^d = 0.5$,
we obtain $\Delta m_K^{\rm Higgs} \simeq 3.1 \times 10^{-17}$ GeV, for $m_{h^0} = 100$ GeV.
This is two orders of magnitude below the experimental value.   We see broad consistency with
data, primarily because of the appearance of high powers of $\epsilon$ in processes involving
the light generations.  For heavy flavors, this suppression is not that strong.  For example,
the $\overline{t} c h^0$ vertex has a coefficient
\begin{equation}
{\cal L}_{\rm FCNC}^{t-c} = \frac{ 2 \epsilon^2 h^0}{\sqrt{2}} (h_{23}^u\, c\, t^c + h_{32}^u\,
t\, c^c) + h.c.
\end{equation}
This can lead to a branching ratio for $t \rightarrow c h^0$ at the level of $(0.1-1)\%$, depending
on the actual value of the order one coupling $h_{ij}^u$.  This decay may be observable at the LHC.

The most striking signature of this scenario is that the decay branching ratios
of the Higgs boson will be modified considerably compared to the SM.  Decays into
light fermions are enhanced, while decay into $W$ pair is not.  For a specific
set of flavor quantum numbers, the decay branching ratios are shown in Fig. \ref{Fig:guidice}, adopted
from Ref. \cite{gl}.
The solid lines correspond to branching ratios in the present model, while the dashed
lines are the corresponding ones in the SM.
Note that the branching ratio for $h \rightarrow b \overline{b}$ is enhanced.  While
the $h \rightarrow W W^*$ decay rate becomes comparable to $h \rightarrow b \overline{b}$
in the SM for a Higgs boson mass of 135 GeV, this crossover occurs at $m_h = 175$ GeV
in the present case.  Branching ratio for $h \rightarrow \mu^+ \mu^-$ has increased,
while the branching ratio for $h \rightarrow \gamma \gamma$ has diminished.  These
predictions are readily testable at the LHC once the Higgs boson is detected.

\begin{figure}[ht]
\begin{center}
\includegraphics[width=0.5\linewidth]{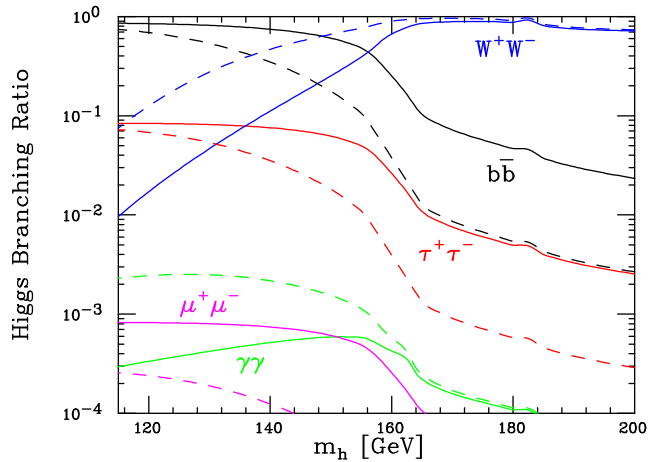}
\vspace*{0.1in} \caption{Higgs branching ratios with the SM Higgs
as a flavon field \cite{gl}.  The solid lines correspond to branching ratios
with Higgs as a flavon, while the dashed lines are the corresponding
SM branching fractions.}
\label{Fig:guidice}
\end{center}
\end{figure}

\section{Grand Unification and the flavor puzzle \label{sec5}}

In this section we will develop ideas of Grand Unification which can
provide significant insight into the flavor puzzle.  When assisted by
flavor symmetries, grand unified theories (GUTs) have great potential for
addressing many of the puzzles.

Grand Unification is an ambitious program that attempts to unify
the strong, weak and electromagnetic interactions \cite{ps,gg,gqw}.  It is strongly
suggested by the unification of gauge couplings that happens in
the minimal supersymmetric standard model.  This is shown
in Fig. \ref{Fig:unif}, where the three gauge couplings of the standard model
are extrapolated to high energies assuming weak scale supersymmetry.
It is clear that data supports the merging of all three couplings to
a common value. Besides its aesthetic appeal, in practical terms, grand unified theories
reduce the number of parameters.  For example, the three
gauge couplings of the SM are unified into one at a very high
energy scale $\Lambda_{\rm GUT} \simeq 2 \times 10^{16}$ GeV.  The
apparent differences in the strengths of the various forces is
attributed to the spontaneous breakdown of the GUT symmetry to the
MSSM and the resulting renormalization flow of the gauge
couplings. SUSY GUTs are perhaps the best motivated extensions of
the SM.  They explain the quantization of electric charge, as well
as the quantum numbers of quarks and leptons.  They provide ideal
settings for understanding the flavor puzzle, which will be the
focus of this discussion.

\begin{figure}[ht]
\begin{center}
\includegraphics[width=0.6\linewidth]{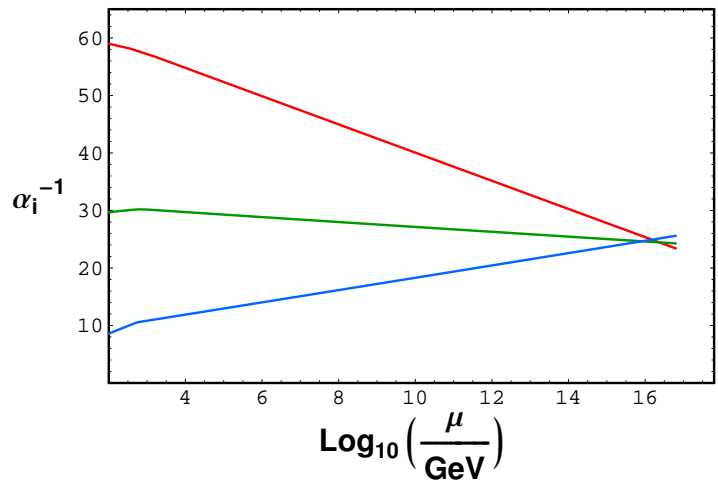}
\caption{Evolution of the inverse gauge couplings ($\alpha_1^{-1},~
\alpha_2^{-1},~\alpha_3^{-1}$) (from top to bottom) in the MSSM
as a function of momentum.}
\label{Fig:unif}
\end{center}
\end{figure}

The simplest GUT model is based on $SU(5)$ \cite{gg}.  I will assume low
energy supersymmetry, motivated by the gauge coupling unification
and a solution to the hierarchy problem.  For an understanding of
quark--lepton masses and mixings SUSY is not crucial, but within
the context of SUSY there will be many interesting flavor
violating processes. In $SU(5)$, the fifteen components of one
family of quarks and leptons are organized into two multiplets: A
${\bf 10}$--plet and a ${\bf \overline{5}}$--plet.  The ${\bf
\overline{5}}$ is of course the anti--fundamental representation
of $SU(5)$, while the ${\bf 10}$ is the anti-symmetric second rank
tensor.  These are represented by the following matrices:
\begin{eqnarray}
\label{su5form}
\psi({\bf 10}): \frac{1}{\sqrt{2}} \left(
\begin{matrix}
 0 & u^{c}_3 & -u^{c}_2 & u_1 & d_1 \\
  -u^{c}_3 & 0 & u^{c}_1 & u_2 & d_2 \\
  u^{c}_2 & -u^{c}_1 & 0 & u_3 & d_3 \\
  -u_1 & -u_2 & -u_3 & 0 & e^c \\
  -d_1 & -d_2 & -d_3 & -e^c & 0
\end{matrix}
\right)\,,~~~\chi(\bar{\bf 5}):
\left(\begin{matrix}d^{c}_1 \\ d^{c}_2\\ d^{c}_3\\ e\\
-\nu_e
\end{matrix} \right)\,.
\end{eqnarray}
Each family of quarks and leptons is organized in a similar form.
It is very nontrivial that this assignment of fermions
under $SU(5)$ is anomaly free.  The anomaly from the ${\bf \overline{5}}$--plet
is canceled by the anomaly from the ${\bf 10}$--plet.
Note that quarks and leptons are unified into common multiplets.  Furthermore,
particles and antiparticles are also unified. These features imply that baryon
number, which is a global symmetry of the SM, is violated, and that proton
will decay.   Because the unification scale is rather large, $\Lambda_{\rm GUT}
\approx 2 \times 10^{16}$ GeV,
the decay rate of the proton is very slow, with a lifetime of order $10^{35}$ years.
This is consistent with, but not very far from current experimental limits.
Note that there is no $\nu^c$ field in the
simplest version of $SU(5)$, but it can be added as a gauge
singlet, as in the SM.

The symmetry breaking sector consists of two types of Higgs
fields. One is an adjoint ${\bf 24}_H$--plet $\Sigma$, which acquires vacuum
expectation value and breaks $SU(5)$ down to the SM gauge
symmetry.  The VEV of this traceless hermitian matrix is chosen as
\begin{equation}
\label{su5vev}
\left\langle \Sigma \right \rangle = V. {\rm diag} \left\{1,\,1,\,1,\,-\frac{3}{2},\,-\frac{3}{2}\right\}\,.
\end{equation}
Under $SU(5)$ gauge transformation $\Sigma \rightarrow U \,\Sigma\, U^\dagger$.  It is then clear
that the VEV structure of Eq. (\ref{su5vev}) will leave invariant an $SU(3) \times SU(2) \times
U(1)$ subgroup, identified as the SM gauge symmetry.  12 of the 24 gauge bosons of $SU(5)$ will
acquire mass of order $V \sim \Lambda_{\rm GUT} \approx 2 \times 10^{16}$ GeV, leaving the remaining
12 SM gauge bosons massless.

$\Sigma$ cannot couple to the fermions.  A pair of $\{{\bf 5}_H+ {\bf \overline{5}}_H\}$ Higgs
fields, denoted as ($H + \overline{H}$),
 are used for generating fermion masses and for electroweak symmetry breaking.
$H$ contain the $H_u$ field of MSSM, while $\overline{H}$ contains the $H_d$ field.
These $(H+\overline{H}$) fields also contain color--triplet components, which must acquire GUT--scale
masses, since they mediate proton decay.  In minimal SUSY $SU(5)$ this splitting of color triplets
and weak doublets is done by a special arrangement,
by precisely tuning the mass term $M_H H \overline{H}$ and the coupling $\lambda H \overline{H} \Sigma$
of the superpotential, so that
the $SU(2)_L$ doublet components remain light, while their color--triplet partners acquire large masses.
This is possible, since the VEV of $\Sigma$ breaks the $SU(5)$ symmetry.

The Yukawa couplings of fermions and the $(H\,,\overline{H})$ fields are obtained from the superpotential
\begin{equation}
\label{su5yuk}
W_{\rm Yuk} = \frac{(Y_u)_{ij}}{4}\psi_i^{\alpha \beta} \psi_j^{\gamma \delta} H^\rho \epsilon_{\alpha \beta \gamma
\delta \rho} +
\sqrt{2}\,(Y_d)_{ij} \psi_i^{\alpha \beta} \chi_{j \alpha} \overline{H}_\beta\,.
\end{equation}
Here $(i,j)$ are family indices, and $(\alpha,\,\beta ...$) are $SU(5)$ indices
with $\epsilon$ being the completely antisymmetric Levi--Cevita tensor.  The $\overline{H}$
field has components similar to $\chi$ of Eq. (\ref{su5form}), so that its fifth component
is neutral and acquires a VEV: $\left\langle \overline{H}_5 \right\rangle = v_d$.  Similarly,
the fifth component of $H$ acquires a VEV: $\left\langle H_5 \right\rangle = v_u$.  When these
VEVs are inserted in Eq. (\ref{su5yuk}), the following mass terms for quarks and leptons are generated:
\begin{equation}
{\cal L}_{\rm mass} = \frac{1}{2}(Y_u)_{ij}\,v_u\, (u_i u^c_j + u_j u^c_i) + (Y_d)_{ij}\, v_d\, (d_i d^c_j + e_i^c e_j) + h.c.
\end{equation}
This leads to the following fermion mass matrices:
\begin{equation}
M_u = Y_u \,v_u\,,~~~M_d = Y_d \,v_d\,, ~~~M_\ell = Y_d^T \, v_d \,.
\end{equation}
Note that $M_u$ is a symmetric matrix in family space.  Furthermore,
there are only two Yukawa coupling matrices describing charged fermion masses,
unlike the three matrices we have in the SM. The reason for this reduction of parameters is
the higher symmetry and the unification of quarks with leptons.  Specifically, we have the relation
\begin{equation}
M_d = M_\ell^T\,.
\end{equation}
This  identity leads to the asymptotic (valid at the GUT scale) relations
for the mass eigenvalues
\begin{equation}
\label{asym}
m_b^0 = m_\tau^0,~ m_s^0 = m_\mu^0,~ m_d^0 = m_e^0\,,
\end{equation}
where the superscript $^0$ is used to indicate that the relation
holds at the GUT scale.

In order to test the validity of the prediction of minimal SUSY
$SU(5)$, we have to extrapolate the masses from GUT scale to low
energy scale where the masses are measured.  This is done by the
renormalization group equations.  The evolution of the $b$--quark
and $\tau$--lepton Yukawa couplings ($\lambda_b$ and $\lambda_\tau$),
which are proportional to $b$--quark and $\tau$--lepton masses,
is shown in Fig. \ref{Fig:btau} for two differen
values of $\tan\beta = (1.7,~50)$.  $\tan\beta = v_u/v_d$ is the ratio of
the two Higgs VEVs in MSSM.  Here we have extrapolated the Yukawa couplings
derived from the observed masses from low scale to the GUT scale.  It is remarkable
that unification of masses occurs in this simple context.  The
main effect on the evolution comes from QCD enhancement of $b$
quark mass as it evolves from high energy to low energy scale,
which is absent for the $\tau$ lepton.

\begin{figure}[ht]
\begin{center}
\begin{tabular}{cc}
\includegraphics[width=0.4\linewidth]{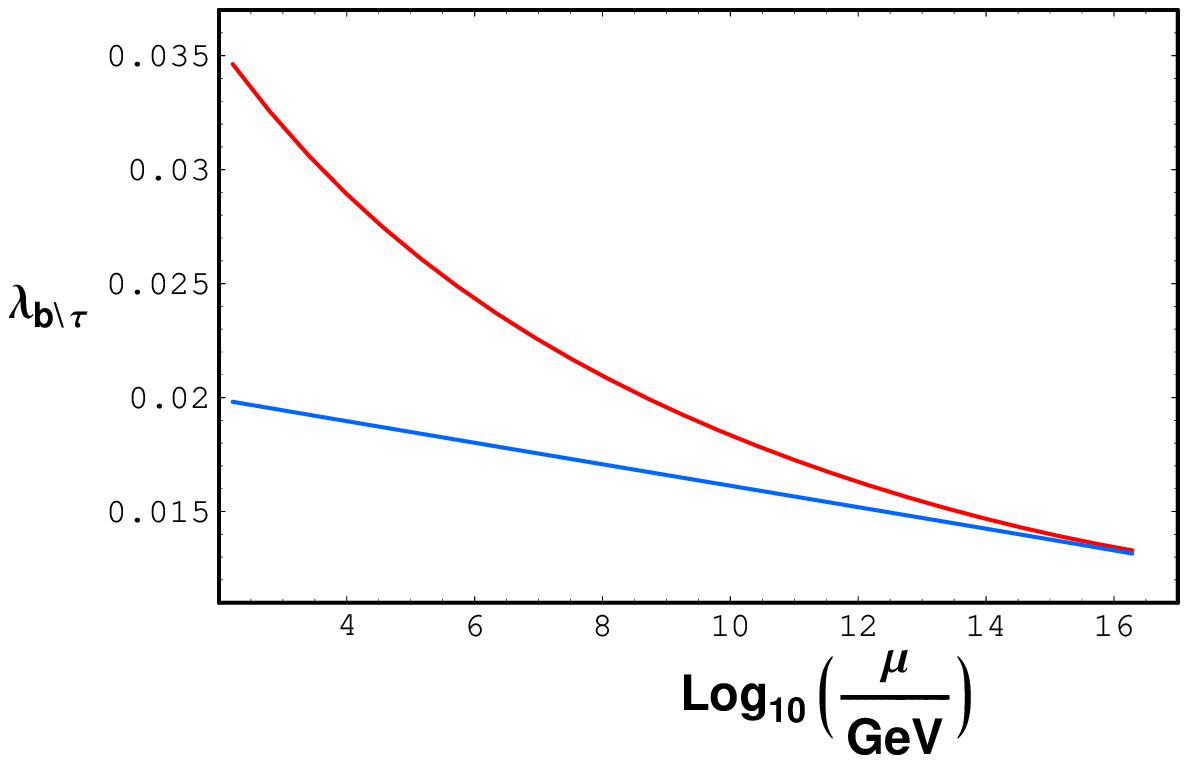} &
\includegraphics[width=0.4\linewidth]{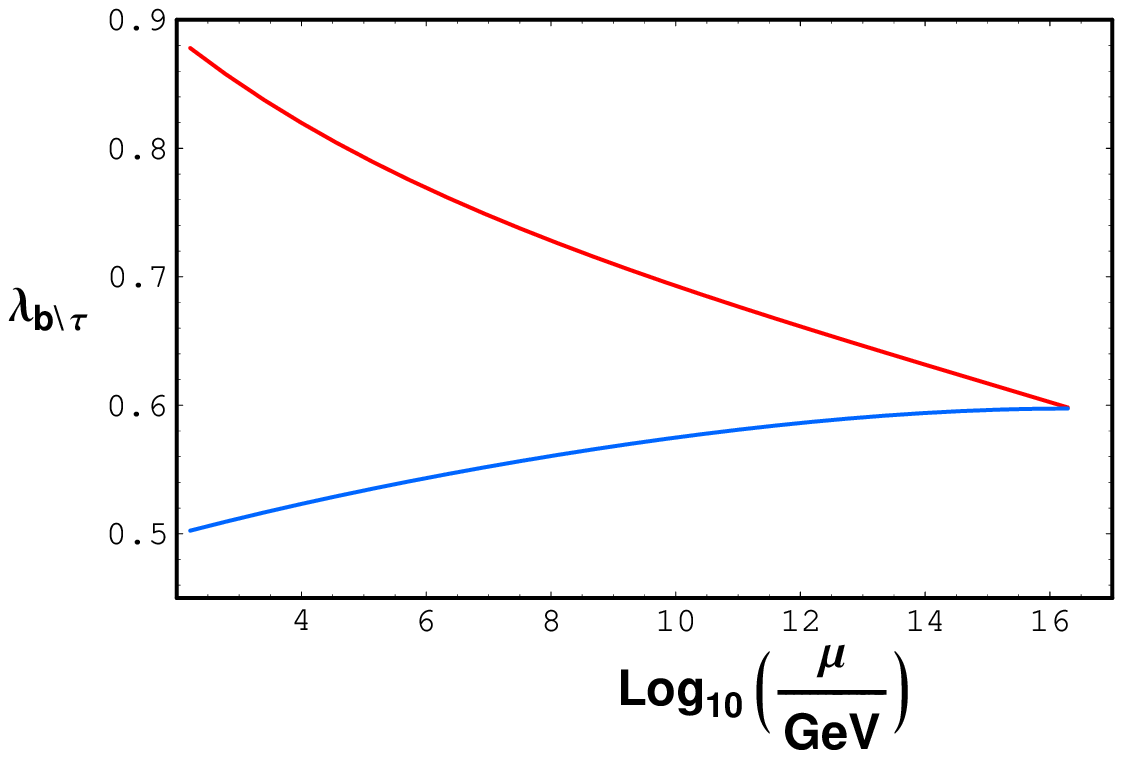} \\
\end{tabular}
\caption{Evolution of the $b$--quark and $\tau$--lepton Yukawa couplings in
the MSSM for $\tan\beta = 1.7$ (left panel) and 50 (right panel).
$m_b(m_b) = 4.65$ GeV has been used here. }
\label{Fig:btau}
\end{center}
\end{figure}

Why these two specific values of $\tan\beta$?  As it turns out,
$b-\tau$ mass unification occurs only for specific values of $\tan\beta$,
either for large values, or very small values.  In Fig. \ref{Fig:btaukolda} we plot
the allowed values of $\tan\beta$ as a function of the strong coupling
$\alpha_s$ \cite{bk1}.  From this figure, it is clear that intermediate values of
$\tan\beta$ would lead to deviation from $m_b^0 = m_\tau^0$ by as much
as 25\%.  For low and large values of
$\tan\beta$, there is always good solution for $m_b(m_b)$, while
for intermediate values there is no acceptable solution.  It
should be mentioned that there are significant finite corrections
to the $b$--quark mass from loops involving the gluino, which is
not included in the RGE analysis.  These graphs, while loop
suppressed, are enhanced by a factor of $\tan\beta$, and thus can
be as large 30-40\% for $m_b(m_b)$ \cite{hrs}.  So even intermediate values
of $\tan\beta$ are not totally excluded.

\begin{figure}[!htb]
\begin{center}
\includegraphics[width=0.6\linewidth]{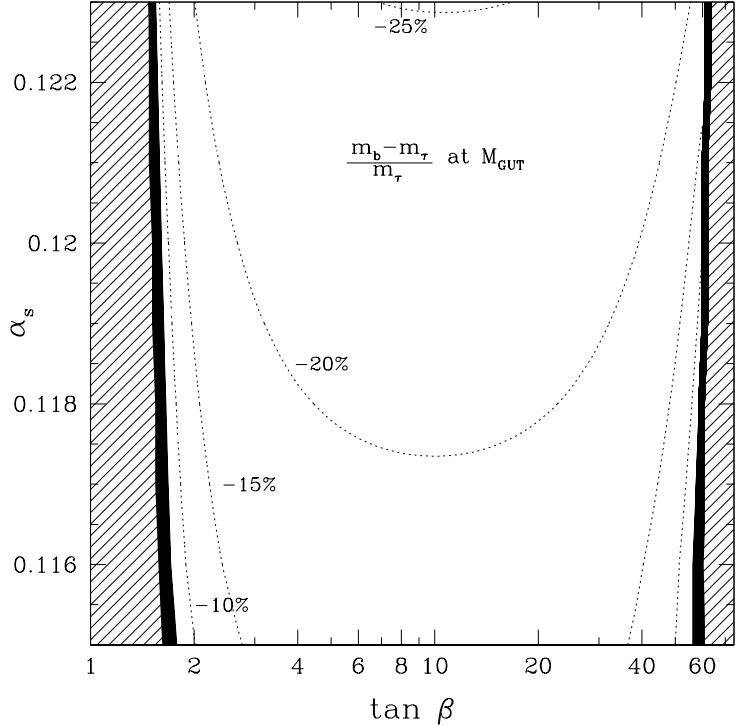}
\caption{Deviation in the asymptotic relation $m_b^0 = m_\tau^0$ as a function
of $\tan\beta$ and $\alpha_s$ \cite{bk1}. }
\label{Fig:btaukolda}
\end{center}
\end{figure}

The last two relations of Eq. (\ref{asym}) turn out to be not acceptable
when compared to low energy values of the masses.  One can see
this without going through the RGE evolution.  Eq. (\ref{asym}) implies
$m_s^0/m_d^0 = m_\mu^0/m_e^0$.  These mass ratios are RGE
independent, so one can compare them directly with observations.
We have seen that $m_s/m_d \simeq 20$, while $m_\mu/m_e \simeq
200$.  So this relation is off by an order of magnitude.

There is an elegant way of fixing the light fermion masses in
$SU(5)$.  Consider modifying Eq. (\ref{asym}) to the following relations:
\begin{equation}
\label{GJ}
m_b^0 = m_\tau^0,~ m_s^0 = \frac{1} {3}\, m_\mu^0,~ m_d^0 = 3 \,m_e^0\,.
\end{equation}
These relations were proposed by Georgi and Jarlskog and are known
as the GJ relations \cite{gj}.  The factors of 3 that appears in Eq. (\ref{GJ})
have a simple group theoretic understanding in terms of $B-L$,
under which lepton charges are ($-3$) times that of quark charges. The
RGE independent quantity from Eq. (\ref{GJ}) gives us
\begin{equation}
\frac{m_s}  {m_d} = \frac{1}  {9} \,\frac{m_\mu}  {m_e}\,,
\end{equation}
which is in good agreement with observations.  There is one
other prediction, which can be taken to be the value of
$m_d(1~{\rm GeV}) \simeq 8$ MeV, which is also is good agreement
with data, although recent lattice calculations prefer somewhat
smaller values of $m_d$.

\subsection{A predictive GUT framework for fermion masses}

How would one go about deriving the Georgi--Jarlskog mass
relations?  We invoke a flavor $U(1)$ symmetry as before. Consider
the following mass matrices for up quarks, down quarks and charged
leptons \cite{gj,hrr,dhr,bmgj}.
\begin{eqnarray}
\label{mat}
M_u~=\left(\begin{array}{ccc}
0 & a & 0\\
a & 0 & b\\
0 & b & c \end{array}\right)\,,~ M_d~=\left(\begin{array}{ccc}
0 & d e^{i\phi} & 0\\
d e^{-i\phi} & f & 0\\
0 & 0 & g\end{array} \right)\,,~ M_\ell~=\left(\begin{array}{ccc}
0 & d & 0\\
d & -3f & 0\\
0 & 0 & g \end{array}\right)\,.
\end{eqnarray}

\noindent The factor ($-3$) in charged lepton versus down quark mass
matrix is attributed to the
$B-L$ quantum number, and the zeros are enforced by a flavor
symmetry. In $SU(5)$ GUT, the (1,2) and the (2,1) entries of
$M_d$ (and $M_\ell$) are unrelated, but in $SO(10)$ GUT discussed in
the next subsection, they can be related, as in Eq. (\ref{mat}).
All parameters are complex to begin with, but after
field redefinitions, only a single complex phase survives. There
are 7 parameters in all to fit the 13 observables (9 masses, 3
mixing angles and one CP phase), thereby resulting in six
predictions. Three of these predictions are the $b,s$ and
$d$--quark masses. We write them at the low energy scale by
incorporating factors denoted as $\eta$ which are the RGE
evolution factors to go from the weak scale to the GUT scale.
For light quark masses, there is a further evolution to go down
from the weak scale to their respective mass (or hadron) scale.
The predictions of the model for the quark masses are given by:
\begin{eqnarray}
m_b = \eta^{-1}_{b/\tau} m_\tau; ~ \frac{{m_d/m_s}}
{{(1-m_d/m_s)^2}} = 9 \frac{{m_e/m_\mu}}
{{(1-m_e/m_\mu)^2}};~(m_s-m_d) = \frac{1}  {3}
\eta_{s/\mu}^{-1}(m_\mu-m_e)\,.
\end{eqnarray}
The other three predictions are for the quark mixing angles and
the CP phase $J$.  $J$ is the rephasing invariant CP violation
parameter (Jarlskog invariant) which can be defined as
\begin{equation}
J = {\rm Im}(V_{us} V_{cb} V_{ub}^* V_{cs}^*)
\end{equation}
and has a value of $J \simeq 2.8 \times 10^{-5}$.  We have for the
remaining three predictions \cite{bmgj}
\begin{eqnarray}
\label{vcbpred}
|V_{cb}| = \eta_{KM}^{-1}\eta_{u/t}^{1/2}\sqrt{\frac{{m_c}}
{{m_t}}}~; {}~~~~~~~~\frac{{|V_{ub}|}}  {{|V_{cb}|}} =
\sqrt{\frac{{m_u}} {{m_c}}} \,;
\end{eqnarray}
\begin{equation}
J = \eta_{KM}^{-2} \eta_{u/t} \sqrt{\frac{{m_d}} {{m_s}}}
\sqrt{\frac{{m_c}} {{m_t}}}\sqrt{\frac{{m_u}} {{m_t}}} \left[1-\frac{1}
 {4} \left(\sqrt{\frac{{m_u}} {{m_c}}} \sqrt{\frac{{m_s}}
{{m_d}}}+\sqrt{\frac{{m_c}} {{m_u}}} \sqrt{\frac{{m_d}} {{m_s}}}-
\sqrt{\frac{{m_c}} {{m_u}}}\sqrt{\frac{{m_s}}
{{m_d}}}|V_{us}|^2\right)^2 \right]^{\frac {1} {2}}\,. \nonumber
\end{equation}
Here $\eta_{ct} = [(m_c^0/m_t^0)/(m_c(m_t)/m_t(m_t))]$,
$\eta_{KM} = |V_{cb}^0|/|V_{cb}|$, etc.
One can write down semi--analytic results for the RGE factors, if
the bottom--quark Yukawa coupling $h_b$ is much smaller than the
top Yukawa coupling $h_t$, (corresponding to tan$\beta
\stackrel{_<}{_\sim} 10$ or so). These RGE factors can be expressed
then as
\begin{eqnarray}
\label{RGE}
\eta_{KM} &=& \eta_{d/b}=\left(1-\frac{Y_t} {Y_f}\right)^{\frac{1}
{12}};~ \eta_{u/t} = \left(1-\frac{Y_t} {Y_f}\right)^{\frac{1}
 {4}};~ \eta_{s/\mu} = \left(\frac{\alpha_1}
{\alpha_G}\right)^{-10/99}
\left(\frac{\alpha_3} {\alpha_G}\right)^{-8/9} \nonumber \\
\eta_{b/\tau} &=& \left(\frac{\alpha_1}
{\alpha_G}\right)^{-10/99} \left(\frac{\alpha_3}
{\alpha_G}\right)^{-8/9}\left(1-\frac{Y_t}
{Y_f}\right)^{-1/12}\,.
\end{eqnarray}
Here $\alpha_G$ is the unified gauge coupling strength, $Y_t =
h_t^2$ at the weak scale and $Y_f$ is the fixed point value of
$Y_t$.  That is, $Y_f$ is the largest value $Y_t$ can take consistent
with perturbation theory being valid upto the GUT scale.
Numerically, $Y_f \simeq 1.2$.  $Y_t$ is of course obtained from
$Y_t = [m_t(m_t)/v_u]^2$, which for $M_t = 172.5$ GeV is $Y_t \simeq
0.876$.  Note that the CKM mixing parameters and the mass ratios
in the same charge sector evolve only due to Yukawa couplings.
The mass ratio $m_s/m_\mu$ does change with momentum proportional
to the gauge interaction strength.

While five of the six predictions of this model agree well with
experiments, the relation for $|V_{cb}|$ of Eq. (\ref{vcbpred})
would imply that either the top quark mass is much higher than its observed
value, or that the value of $|V_{cb}|$ is much larger than allowed.
Indeed, if we use an acceptable value of $M_t = 172.5$ GeV, wiht
$Y_f = 1.2$, Eq. (\ref{vcbpred}) would lead to $|V_{cb}| \simeq 0.053$,
which is more than 10 standard deviations away from its central value.
If $|V_{cb}|$ is to be decreased down to any acceptable value, top quark
mass will have to be very close to its perturbative upper limit, around
200 GeV, which is also excluded by experiments.

We conclude that, although very predictive and simple, the ansatz of
Eq. (\ref{mat}) is excluded by data.  It is interesting that while the
original Fritszch ansatz of Eq. (\ref{frit}) was excluded since top quark
mass was predicted to be too low, the present ansatz, which was very popular
until a few years ago, is excluded for its prediction of top mass that is
too large.

\subsection{Fermion masses in a predictive $SO(10)$ model}

Now let us turn to an even more interesting class of GUTs, those
based on the gauge symmetry $SO(10)$ \cite{so10}.  All members of a family are
unified into a ${\bf 16}$ dimensional spinor representation of
$SO(10)$.  This requires the existence of right--handed neutrino
$\nu^c$, leading naturally to the seesaw mechanism and small
neutrino masses.  $SU(5)$ has the option of having neutrino mass,
but in that context there is no compelling argument for its existence.
$SO(10)$ models are the canonical grand unified models, owing to the
observed neutrino masses, and the fact that all members of a family
are unified into a single {\bf 16}--dimensional spinor multiplet in $SO(10)$.

The spinor of $SO(10)$ breaks down under $SU(5)$ (which is one of its subgroups) as
\begin{equation}
{\bf 16} = {\bf 10} + {\bf \overline{5}} + {\bf 1}\,,
\end{equation}
where the ${\bf 1}$ is the $\nu^c$ field.  The {\bf 10} and the ${\bf \overline{5}}$
fields are identical to the case of $SU(5)$.  We shall again assume low energy supersymmetry.
Gauge symmetry breaking is accomplished in the SUSY limit by
introducing Higgs fields in the adjoint ${\bf 45_H}$, spinor
\{${\bf 16_H} + {\bf \overline{16}_h}$\} and vector ${\bf 10_H}$
representations.  Because there is more symmetry in $SO(10)$, more scalars
are needed to achieve symmetry breaking down to the SM.  The spinor Higgs fields break
$SO(10)$ down to $SU(5)$ changing the rank of the gauge group, while the adjoint
${\bf 45_H}$--plet breaks this symmetry down to the SM.  The vector ${\bf 10_H}$--plet
is used for fermion mass generation and for electroweak symmetry breaking.
The MSSM Higgs doublets $H_{u,d}$ are contained
partially in the ${\bf 10_H}$ but can be partially also in the ${\bf
16_H}$.

Let me work out a specific flavor model based on $SO(10)$ supplemented by a
$U(1)$ symmetry \cite{bpw}.  While this model will not be as predictive as the ansatz that
generated the GJ relations in the previous subsection, there are still a number
of predictions, and these predictions are consistent with data.  Several variations of the theme can be
found in the literature \cite{albrightbarr}, but here I confine the discussions to the mass matrices
of Ref. \cite{bpw} and its slight generalization studied in Ref. \cite{bpr}.

The mass matrices for up and down quarks, and Dirac neutrino and charged leptons
take the form:
\begin{eqnarray}
\label{eq:mat}
\begin{array}{cc}
M_u=\left[
\begin{array}{ccc}
0&\epsilon'&0\\-\epsilon'&\zeta_{22}^u&\sigma+\epsilon\\0&\sigma-\epsilon&1
\end{array}\right]{\cal M}_u^0;&
M_d=\left[
\begin{array}{ccc}
0&\eta'+\epsilon'&0\\
\eta'-\epsilon'&\zeta_{22}^d&\eta+\epsilon\\0& \eta-\epsilon&1
\end{array}\right]{\cal M}_d^0\\
&\\
M_\nu^D=\left[
\begin{array}{ccc}
0&-3\epsilon'&0\\3\epsilon'&\zeta_{22}^u&\sigma-3\epsilon\\
0&\sigma+3\epsilon&1\end{array}\right]{\cal M}_u^0;& M_\ell=\left[
\begin{array}{ccc}
0&\eta'-3\epsilon'&0\\
\eta'+3\epsilon'&\zeta_{22}^d&\eta-3\epsilon\\0& \eta+3\epsilon&1
\end{array}\right]{\cal M}_d^0 \,.\\
\end{array}
\end{eqnarray}
Here $M_\nu^D$ is the Dirac neutrino mass matrix.

Notice the various correlations
in these matrices. The overall scale associated
with $M_u$ and $M_{\nu}^D$ are identical, while those for $M_d$ and $M_\ell$ are
the same. The ``1" entry in all matrices have a common origin, arising
from the operator ${\bf 16}_3 {\bf 16}_3 \,{\bf 10_H}$.  The $\epsilon$ entry appears with coefficient
$1$ in the up and down quark matrices, and with coefficient $-3$ in the leptonic
mass matrices.  This factor $(-3)$ is the ratio of the $B-L$ charge of leptons versus quarks.
Specifically, the $\epsilon$ entry arises from an operator ${\bf 16}_2 {\bf 16}_3\, ({\bf 10_H} \times
{\bf 45_H})/M$.  Here the adjoint ${\bf 45_H}$, which is a second rank antisymmetric
tensor of $SO(10)$,  acquires a VEV in a $B-L$ conserving direction:
\begin{equation}
\label{b-l}
\left \langle {\bf 45_H} \right \rangle = i\tau_2 \times {\rm diag}.(a,\, a,\, a,\, 0,\, 0)~.
\end{equation}
In the product ${\bf 10_H} \times {\bf 45_H}$, two fragments, an effective ${\bf 10_H}$ and
an effective ${\bf 120_H}$, couple to the fermions.  However, when the VEV of ${\bf 45_H}$ from
Eq. (\ref{b-l}) is inserted, only the effective ${\bf 120_H}$ is non-vanishing, leading to the relative
factor of $(-3)$ between leptons versus quarks.
Note that the $\epsilon$ entry arises suppressed by $1/M$ so that $\epsilon \ll 1$,
an idea familiar from the Froggatt--Nielsen mechanism.  In an analogous fashion, the $\epsilon'$
entry arises from the operator ${\bf 16}_1 {\bf 16}_2 \,({\bf 10_H} \times {\bf 45_H})\, S/M^2$, where $S$
is an $SO(10)$ singlet flavon filed carrying a flavor $U(1)$ charge.  This entry is then
more suppressed compared to the $\epsilon$ entry.  The $\sigma$ entry originates from the
operator ${\bf 16}_2 {\bf 16}_3\,{\bf 10_H}\, S/M$, and enters into all matrices with equal
coefficient, just as the ``1" entry.  An operator ${\bf 16}_2 {\bf 16}_3 {\bf 16_H} {\bf 16_H}/M$
contributes equally to the down quark and charged lepton mass matrices, but not to $M_u$ and
$M_\nu^D$, since ${\bf 16_H}$ contains only an $H_d$--type field, and not an $H_u$--type field.
The $\eta$ entry in $M_d$ and $M_\ell$ is the sum of the last two operators.  The entry
$\eta'$ originates from ${\bf 16}_1 {\bf 16}_2 {\bf 16_H} {\bf 16_H}\, S^2/M^3$ operator.

These are precisely the operators one would obtain when the three families of fermions and
the Higgs fields are assigned the following $U(1)$ charges:
\begin{equation}
\begin{array}{cccccccc}
\mathbf{16}_3  & \mathbf{16}_2 & \mathbf{16}_1 & \mathbf{10}_H &
\mathbf{16}_H & \overline{\mathbf{16}}_{H} &
\mathbf{45}_H & \mathbf{S} \\
a & a+1 & a+2 & -2a & -a-1/2 & -a & 0 & -1
\end{array}.
\label{eqn:charges}
\end{equation}

In Ref. \cite{bpw}, where for simplicity, CP violation was ignored,
the diagonal (2,2) entries were not introduced.  In subsequent work
these (2,2) entries, especially $\zeta_{22}^d$, were used to accommodate
CP violation.  Here we present the predictions of the model as given
in Ref. \cite{bpw}.  An acceptable fit to all mass and mixing parameters
is obtained by the following choice of parameters at the GUT scale:
\begin{eqnarray}
\sigma &=& -0.1096, \,\,\, \eta = -0.1507, \,\,\,
\epsilon = 0.0954, \,\,\, \nonumber \\
\epsilon' &=& 1.76 \times 10^{-4},\,\,\, \eta' = 4.14 \times 10^{-3}\,.
\end{eqnarray}
With these input, one obtains the following predictions:
\begin{eqnarray}
m_b(m_b) &=& 4.9~{\rm GeV},\,\,\, m_s(1~{\rm GeV}) = 116 \,{\rm MeV},\,\,\, m_d(1~{\rm GeV}) = 8~{\rm MeV} \nonumber
\\
\theta_C &\simeq& \left|\sqrt{\frac{m_d}{m_s}} - e^{i \phi} \sqrt{\frac{m_u}{m_c}}\right|\,, \,\,\, \frac{|V_{ub}|}{|V_{cb}|}
\simeq \sqrt{\frac{m_u}{m_c}} \simeq 0.07\,.
\end{eqnarray}
These predictions are in general agreement with data.  When the (2,2) entries
are included in the mass matrices, realistic CP violation phenomenology also follows \cite{bpr}.

Light neutrino masses are generated in this scheme via the seesaw mechanism.  Note that
the Dirac neutrino mass matrix elements are completely fixed, because of $SO(10)$ symmetry,
from the charged fermion sectors.  The mechanism that generates heavy Majorana neutrino
masses for the $\nu^c$ fields should be specified.  The model already contains operators
that do this, as given by
\begin{equation}
\label{Majorana}
W_{\rm Maj} = {\bf 16}_i {\bf 16}_j \,({\bf \overline{16}_H} {\bf \overline{16}_H})/M \,.
\end{equation}
The natural scale of the cut--off $M$ is $M = M_{\rm Planck} = 2 \times 10^{18}$ GeV.
Then with order one couplings in Eq. (\ref{Majorana}), one would obtain, for the (third family)
right--handed Majorana mass, $M_{\nu_3^c} \sim \Lambda_{\rm GUT}^2/M_{\rm Planck} \sim 10^{14}$ GeV.
This in turn leads to the light neutrino mass $m_\nu \sim m_t^2/M_{\nu_3^c} \simeq 0.05$ eV, nicely
consistent with the value desired for atmospheric neutrino oscillation data.

In Ref. \cite{bpw}, it was shown, with a specific choice of the flavor structure of $M_{\nu^c}$, that
large neutrino oscillation angles arise naturally, while preserving the smallness of quark mixing angles.
Specifically, while $|V_{cb}| \simeq 0.041$, $\sin^22 \theta_{23} \simeq (0.9 -0.99)$ was obtained,
as a function of the light neutrino mass ratio $m_2/m_3$.

\subsubsection{Flavor violation in SUSY GUTs}

How do we go about testing ideas of grand unification in the flavor sector?
Since the GUT scale is below the Planck scale, even though the flavor symmetry is broken
near the GUT scale, soft SUSY breaking  parameters can remember flavor
violating interaction due to their running  between the Planck scale and the GUT scale.
Such running is expected in supergravity models, where the messengers of SUSY breaking
have masses at the Planck scale. The most significant flavor violation in the model of
Ref. \cite{bpw}  arises due to
the splitting of the third family sfermions from those of the
first two families.  This is seen by the solution to the RGE
equations for these masses \cite{bpr1}.
\begin{eqnarray}
\label{FCNC}
\label{eq:deltambr} \Delta\hat{m}_{\tilde{b}_{L}}^2 =
\Delta\hat{m}_{\tilde{b}_{R}}^2 =
\Delta\hat{m}_{\tilde{\tau}_{L}}^2 =
\Delta\hat{m}_{\tilde{\tau}_{R}}^2 \equiv\Delta\approx
-\bigl(\frac{30m_o^2}{16\pi^2}\bigr) h_t^2\ {\rm log}(M^{*}/M_{GUT})\,.
\end{eqnarray}
Here $M^*$ is the fundamental scale where SUSY breaking messengers
reside, with $M^* > M_{\rm GUT}$.  $h_t$ is the top quark Yukawa
coupling.  Note that leptons also feel the effect of top Yukawa,
because leptons and quarks are unified.  In Eq. (\ref{FCNC}) $m_0$ is the universal
SUSY breaking scalar mass parameter.  One sees that, because of
the GUT threshold, universality is not preserved in this type of
models.  In going from gauge basis to the mass eigenbasis for the
fermions, Eq. (\ref{FCNC}) would imply that there will be flavor changing
scalar interactions.  Because SUSY particles have masses of order
TeV, these flavor violation can manifest in the MSSM sector via SUSY
loops.

The most constraining FCNC process in the present model turns out
to be $\mu \rightarrow e \gamma$.  The diagrams inducing such
processes in SUSY GUT models are shown in Fig. \ref{Fig:muegampenguin}.  In the present
case it turns out that the decay $\tau \rightarrow \mu \gamma$ is
not very significant, while the new contributions to $b \rightarrow s \gamma$
is not negligible. Predictions for the branching ratio for the
decay $\mu \rightarrow e \gamma$
are depicted in Fig. \ref{Fig:muegam} as a function of slepton mass \cite{bpr1}.  Part of the
parameter space is already ruled out, so there is a good chance
that this process will be discovered at the MEG experiment at PSI.

\begin{figure}[ht]
\begin{center}
\includegraphics[width=0.6\linewidth]{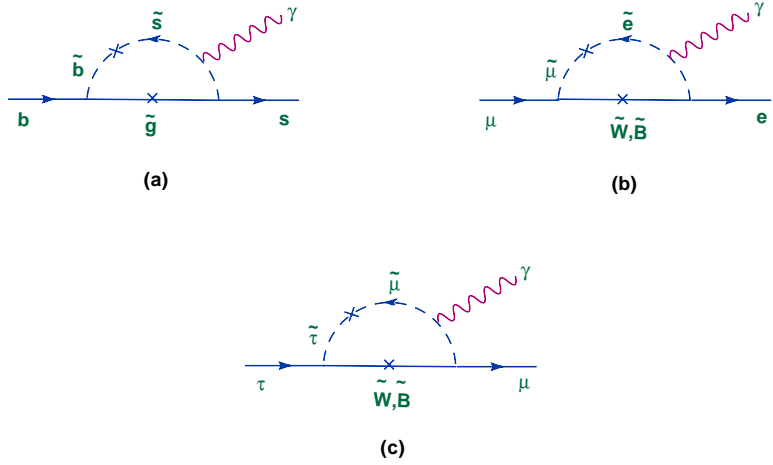}
\vspace*{-0.2in}\caption{Rare decays induced by penguin diagrams via the exchange
of SUSY particles.  The flavor mixing occurs during the RGE flow between
$M_{\rm GUT}$ and $M_*$.}
\label{Fig:muegampenguin}
\end{center}
\end{figure}

\begin{figure}[!htb]
\begin{center}
\includegraphics[width=0.7\linewidth]{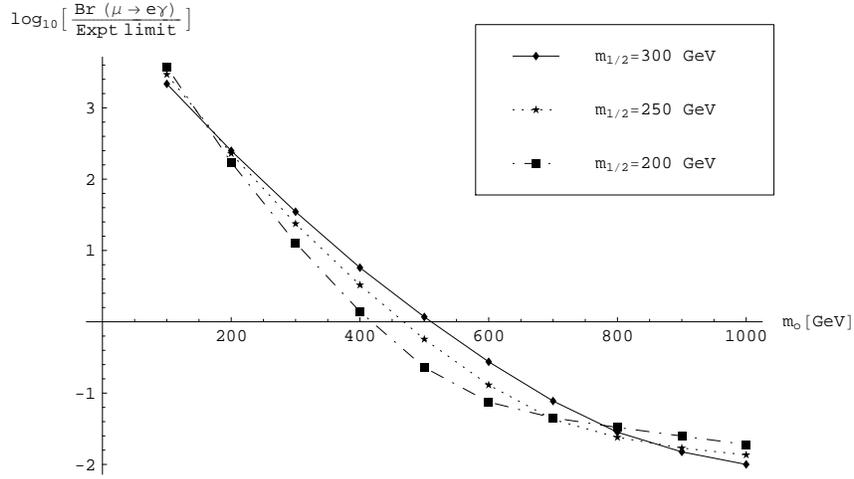}
\caption{Prediction for the branching ratio for $\mu \rightarrow e
\gamma$ in the SUSY $SO(10)$ model as a function of slepton mass.
The horizontal line indicates current experimental limit \cite{bpr1}.}
\label{Fig:muegam}
\end{center}
\end{figure}

There are other sources of flavor violation in SUSY GUTs.  A widely discussed
process is the $\ell_i \rightarrow \ell_j \gamma$ decay arising from neutrino
mass physics \cite{borzu}.  The heavy right--handed neutrino mass is expected to be in the
range $(10^{10}-10^{14})$ GeV in SUSY GUTs.  Even when the supergravity boundary
conditions on the soft SUSY breaking parameters are valid at the GUT scale (and not
the Planck scale), there
is a momentum regime $\mu$,  $M_{\nu^c} \leq \mu \leq \Lambda_{\rm GUT}$, where the
$\nu^c$ fields are active.  In this momentum regime the neutrino Dirac Yukawa couplings will
affect the RGE evolution of the soft slepton mass parameters and generate lepton flavor
violation. The FCNC effect in the slepton soft squared mass is given by
\begin{eqnarray}
(\Delta m^2_{\tilde{L}})_{ij} \simeq -\frac{{\rm log}(\Lambda_{\rm GUT}/M_{\nu^c})}{8 \pi^2}
\left\{3 \,m_0^2 (Y_\nu^\dagger Y_\nu)_{ij} + (A_\nu^\dagger A_\nu)_{ij}\right\}\,.
\end{eqnarray}
Here $Y_\nu$ is the neutrino Dirac Yukawa coupling, while $A_\nu$ is the corresponding
soft trilinear $A$--term.

In the MSSM, or in the SUSY $SU(5)$ model, the Yukawa coupling $Y_\nu$ cannot be determined
from neutrino oscillation data. This is because the seesaw formula for light neutrinos goes
as $m_\nu \sim Y_\nu^2 v^2/M_{\nu^c}$, and knowing $m_\nu$ does not determine $Y_\nu$
uniquely.  However, if some of the entries of $Y_\nu$ are of order $(10^{-2}-1)$, then
the decay rate $\mu \rightarrow e \gamma$ will be within reach of ongoing experiments.

In SUSY $SO(10)$ there is a more crisp prediction for $\mu \rightarrow e \gamma$ arising
from the neutrino sector.  This happens because $SO(10)$ symmetry relates $Y_\nu$ with
the up--quark Yukawa couplings.  Specifically, for the third family, we have $(Y_\nu)_{33}
= Y_t$, the top quark Yukawa coupling.  Since $Y_t$ is of order one, the FCNC effects from
the neutrino sector in SUSY $SO(10)$ are {\it predicted} to be significant. That is, they
cannot be tuned to disappear, unlike in the SUSY $SU(5)$ model.

\section{ Radiative fermion mass generation \label{sec6}}

The hierarchical structure of the quark and lepton masses and the
quark mixing angles can be elegantly understood by the mechanism of radiative
mass generation.  This is an alternative to the
Froggatt--Nielsen mechanism.  Here the idea is that only the
heaviest fermions (eg. the third family quarks) acquire tree level masses.  The next heaviest
fermions (second family quarks) acquire masses as one loop radiative corrections, which
are suppressed by a a typical loop factor $\sim 1/(16 \pi^2) \sim
10^{-2}$ relative to the heaviest fermions.  The lightest fermions ($u$ and $d$ quarks)
acquire masses as two loop radiative corrections, which are then a
factor $\sim [1/(16 \pi^2)]^2 \sim 10^{-4}$ suppressed relative to
the heaviest fermions.  Thus, even without putting in small Yukawa
couplings one understands the hierarchy in the mass spectrum of the fermions.

There is another appeal to this idea.  If the electron mass is
radiatively generated from the muon mass, then there must be no
counter--term needed in the Lagrangian to absorb infinity
associated with the electron mass.  In other words, electron mass
is ``calculable", in terms of other parameters of the model.  This
idea was originally suggested by 'tHooft in his classic paper on
the renormalizability of non--Abelian gauge theories \cite{thooft}.  This also
implies that there must be some symmetry reason for the light
fermions not to have tree level masses, otherwise the idea cannot
be implemented consistently.  Early attempts along this line were
presented in Ref. \cite{early}.  More realistic models came along
somewhat later \cite{bala,bmrad,volkas,dob,barrnew}.

There is a resurgence of interest in this idea as the LHC turns
on, since new particles with specific properties which may be seen
at the LHC are predicted.  There exist rather nice models of this
type by Mohapatra and collaborators \cite{bala} from the late 80's.  Recently
Dobrescu and Fox have written a nice paper on the subject \cite{dob}, which I
recommend to you.  As in past examples, I will try to convey the
main idea, with the understanding that implementation can vary
considerably.  I will discuss an implementation which I worked out with
Mohapatra based on the permutation symmetry \cite{bmrad}.

Let us focus on the quark sector of the SM first.  We wish to have
a scenario where only the top quark and the bottom quark have tree
level masses.  In the same limit, there should be no CKM mixing
induced.  This can be realized if one has the following
``democratic" mass matrices for up and down quarks.
\begin{eqnarray}
\label{dem}
M_{u,d} = \frac{m_{t,b}}  {3}\left(\begin{matrix}1 & 1 & 1 \\ 1 & 1 & 1
\\ 1 & 1 & 1 \end{matrix}\right)~.
\end{eqnarray}
Of course, these matrices have rank 1, implying that only the top
and the bottom acquire masses from here.  A common unitary matrix will diagonalize
$M_u$ and $M_d$,  so there is no CKM mixing induced at this
stage.

How do we obtain democratic mass matrices of Eq. (\ref{dem})?  It turns
out that the symmetry of these matrices is $S_{3L} \times S_{3R}$,
where $S_3$ is the group of permutation of three letters. The
Lagrangian that would generate Eq. (\ref{dem}) for $M_u$ is of the form
\begin{equation}
\label{s3lag}
 {\cal L}_{\rm Yukawa} = h_u(\overline{Q}_{1L}
+  \overline{Q}_{2L} + \overline{Q}_{3L})\tilde{H} (u_{1R} +
u_{2R}+u_{3R})\,
\end{equation}
which is manifestly symmetric under separate permutations of the
left--handed and the right--handed quark fields.  So it is tempting to
start with this symmetry group $S_{3L} \times S_{3R}$, but it is
not necessary to have the $S_{3R}$ group, since right-handed
rotations are un-physical in the SM.  So consider the following
Lagrangian which only has the $S_{3L}$ symmetry.
\begin{eqnarray}
\label{s3yuk}
{\cal L}_{\rm Yukawa} &=& (\overline{Q}_{1L} +  \overline{Q}_{2L}
+ \overline{Q}_{3L})\tilde{H} (h_1^u u_{1R} +h_2^u u_{2R}+h_3^u
u_{3R}) \nonumber \\
&+& (\overline{Q}_{1L} +  \overline{Q}_{2L} + \overline{Q}_{3L})H
(h_1^d d_{1R} +h_2^d d_{2R}+h_3^d d_{3R})\,.
\end{eqnarray}
By right--handed rotations on $u_R$ and $d_R$ fields, we can bring
Eq. (\ref{s3yuk}) into the form of Eq. (\ref{s3lag}). Two combinations of the $Q_{iL}$
and $(u_{iR}, d_{iR})$ fields orthogonal to Eq. (\ref{s3yuk}) will be
massless.

These massless $Q_{iL}$  modes actually form the ${\bf 2}$
dimensional representations of $S_3$.  It is convenient to
directly go to the irreducible representations of $S_3$.  They are
a true singlet ${\bf 1}$, an odd singlet ${\bf 1'}$ and a doublet
${\bf 2} = (x_1, x_2)$.  The product of two ${\bf 1'}$ gives a
${\bf 1}$, while the product of two ${\bf 2}$ gives ${\bf 1} +
{\bf 1'} + {\bf 2}$.  The Clebsch--Gordon coefficients for this
product (in a certain basis) are \cite{pakvasa}:
\begin{eqnarray}
\left(\begin{matrix} x_1 \cr x_2
\end{matrix}\right) \times \left(\begin{matrix} y_1 \cr
y_2\end{matrix}\right) = {\bf 1}: (x_1 y_1 + x_2 y_2);~~~ {\bf 1'}: (x_1 y_2
- x_2 y_1);~~~ {\bf 2}: \left(\begin{matrix}x_1 y_2 + x_2 y_1 \cr x_1
y_1 - x_2 y_2\end{matrix}\right)~.
\end{eqnarray}

Now, consider the following assignment of quarks and scalars under
$S_3$:
\begin{eqnarray}
\left(\begin{matrix}Q_{1L} \\ Q_{2L}\end{matrix}\right)&:&{\bf 2};~~~Q_{3L}:~ {\bf
1};~~~u_{iR}: {\bf 1}~, \nonumber \\
H&:& {\bf 1},~ \left(\begin{matrix}\omega_1 \\ \omega_2 \end{matrix}\right): {\bf
2},~~\omega_3: {\bf 1}~.
\end{eqnarray}
Here the gauge structure is simply that of SM with $H$ being the
SM Higgs doublet.  In order to radiatively generate light fermion
masses, new ingredients are needed.  The simplest possibility is
to introduce scalar fields which have Yukawa couplings connecting
the heavy (3rd generation) and the light fermions.  We have
assumed existence of $\omega_i(3,1,-1/3)$ fields, which can have
such Yukawa couplings, without inducing direct mass terms for the
light fermions.  Note that these $\omega_i$ fields are colored and
charged, so they do not acquire vacuum expectation values.

The most general Yukawa couplings allowed in this SM $\times S_3$
model is given by
\begin{eqnarray}
\label{s3lep}
{\cal L}_{\rm Yukawa} &=& h_t \overline{Q}_{3L} t_R \tilde{H} +
h_b \overline{Q}_{3L} b_R H + h_1 (Q_{1L}^T C Q_{3L} \omega_1 +
Q_{2L}^T C Q_{3L} \omega_2) \nonumber \\
&+& h_2(Q_{1L}^T C Q_{1L} + Q_{2L}^T C Q_{2L})\omega_3  + h_3
Q_{3L}^T C Q_{3L} \omega_3 \nonumber \\
 &+& h_4\{Q_{1L}^T C Q_{2L}
+Q_{2L}^T C Q_{1L})\omega_1 +(Q_{1L}^T C Q_{1L} -Q_{2L}^T C
Q_{2L})\omega_2\} + h.c.
\end{eqnarray}
Here we have redefined the combination of $u_R$ that couples to
$Q_{3L}$ as simply $t_R$ (and similarly for $b_R$).

Clearly, from Eq. (\ref{s3lep}), only the top and bottom quarks acquire
tree--level masses.  There is no tree--level CKM mixing angle. So
by symmetry reason, we have achieved the first stage of the
program.  Now, if $S_3$ is unbroken, none of the light fermions
will acquire masses, even though they have Yukawa couplings via
the $\omega_i$ fields.  We can break $S_3$ spontaneously, or by
soft bilinear terms in the Higgs potential:
\begin{equation}
V = \sum_{i,j=1}^{3} \mu_{ij}^2 \omega_{i}^* \omega_j + h.c.
\end{equation}
With these soft breaking terms, light fermion masses will be
induced.  In Fig. \ref{Fig:radiative} we have the one-loop  and the two--loop mass
generation diagrams.

\begin{figure}[ht]
\begin{center}
\begin{tabular}{cc}
\includegraphics[width=0.4\linewidth]{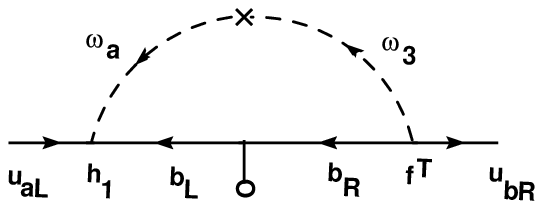} &
\includegraphics[width=0.4\linewidth]{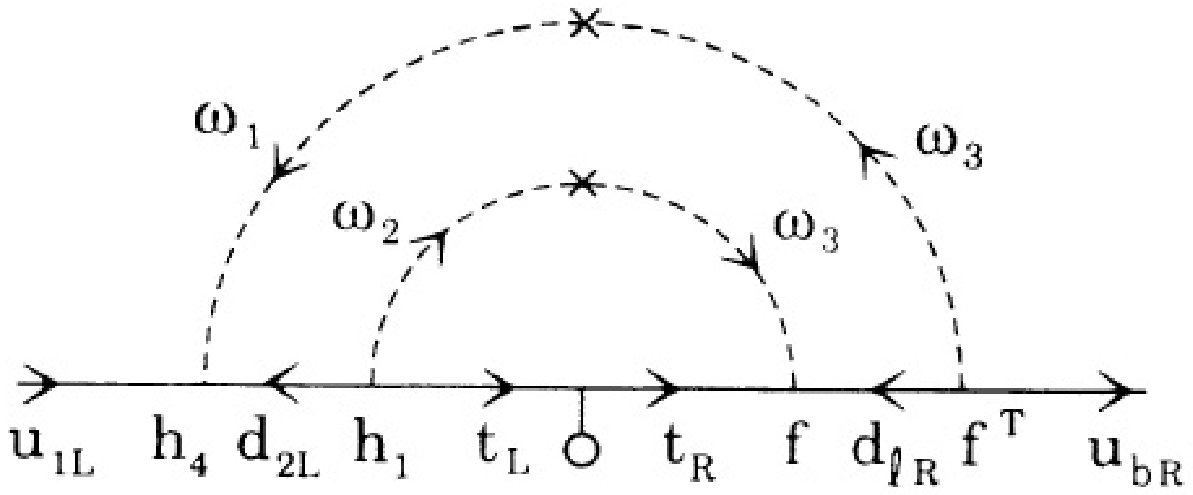} \\
\end{tabular}
\caption{One--loop diagram inducing charm quark mass (left) and
two--loop diagram inducing up quark mass (right). }
\label{Fig:radiative}
\end{center}
\end{figure}

The one--loop diagram of Fig. \ref{Fig:radiative} only generates charm quark mass,
and not the up quark mass.  This can be understood as follows.  At
tree--level, among the down quarks, only $b$ has a mass.  There is
a single linear combination of up quarks which couples to the $b$
quark via the $\omega_i$ fields.  It is this combination that
picks up mass at one--loop.  The orthogonal combination remains
massless at this order.  Now, the two--loop diagram connects up
quarks to both $b$ and $s$ quarks. The inner loop of the two--loop
diagram is the one--loop diagram that generates the $s$ quark
mass.  As a result, $u$ quark will acquire a mass proportional to
the $s$ quark mass at two--loop.

Including the one--loop diagram, the mass matrix for the $(c,t)$
sector has the form
\begin{eqnarray}
M_u^{\rm 1-loop} = \left(\begin{matrix}\epsilon & a \epsilon \\ 0 &
m_t^0 \end{matrix}\right)
\end{eqnarray}
where $a$ is of order one and the small parameter $\epsilon$ is
found to be
\begin{equation}
\epsilon \simeq \left(\frac{h_1 f}  {8
\pi^2}\right)m_b\left(\frac{\mu_{a3}^2}  {M_\omega^2}\right){\rm
log}\left(\frac{M_\omega^2}  {m_b^2}\right)\,.
\end{equation}
With the Yukawa couplings being order one, we can explain why the
charm is much lighter than the top.  The mixing angle $V_{cb}$ is
of order $m_s/m_b$, in agreement with observations. The two--loop
diagrams which induce the up and down quark masses also induce the
mixings of the first family.  There is a natural hierarchy of
mixing angles where $|V_{us}| \gg |V_{cb}| \gg |V_{ub}|$.

It is straightforward to extend the $S_3$ model to the leptonic
sector.  Consider the following assignment of leptons and
$\omega_\ell$ fields under $S_3$, where $\omega_\ell$ are $(3^*,1,
-1/3)$ scalar fields.  (These are not the conjugates of the
$\omega_i$ fields from the quark sector, or else, there will be
proton decay mediated by these scalars.  We assume separate baryon
number conservation, so the proton is stable.)
\begin{eqnarray}
\left(\begin{matrix}\psi_{1L} \\ \psi_{2L} \end{matrix}\right)&:& {\bf
2},~~\psi_{3L}: 1;~ e_{iR}: {\bf 1'} \nonumber \\
\omega_\ell&:& {\bf 1},~ \omega_{\ell}': {\bf 1'}
\end{eqnarray}

The general Yukawa coupling of leptons is given by
\begin{eqnarray}
{\cal L'}_{\rm Yukawa} &=& h_1' Q_{3L}^T C \psi_{3L} \omega_\ell +
h_2'(Q_{1L}^T C \psi_{1L} + Q_{2L}^T C \psi_{2L})\omega_\ell  \nonumber \\
&+& h_3'(Q_{1L}^T C \psi_{2L} - Q_{2L}^TC  \psi_{1L})\omega_\ell'
+ f_{ab}' u_{aR}^T C e_{bR} \omega_\ell' + h.c.
\end{eqnarray}
Note that all leptons are massless at the tree level.  The
one--loop diagram shown in Fig. \ref{Fig:radiativelep} will induce the $\tau$ lepton
mass, and is proportional to the top quark mass with a loop
suppression.  Only $\tau$ acquires a one--loop mass.  The muon
mass arises from the two--loop diagram of Fig. 14.  The electron
remains massless at this order, and acquires a mass only via a
three--loop diagram.

\begin{figure}[!htb]
\begin{center}
\begin{tabular}{cc}
\includegraphics[width=0.4\linewidth]{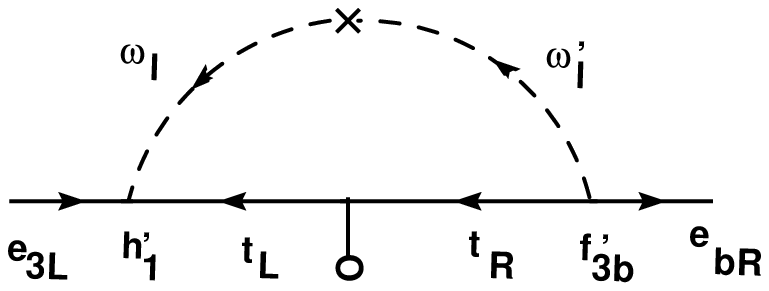} &
\includegraphics[width=0.4\linewidth]{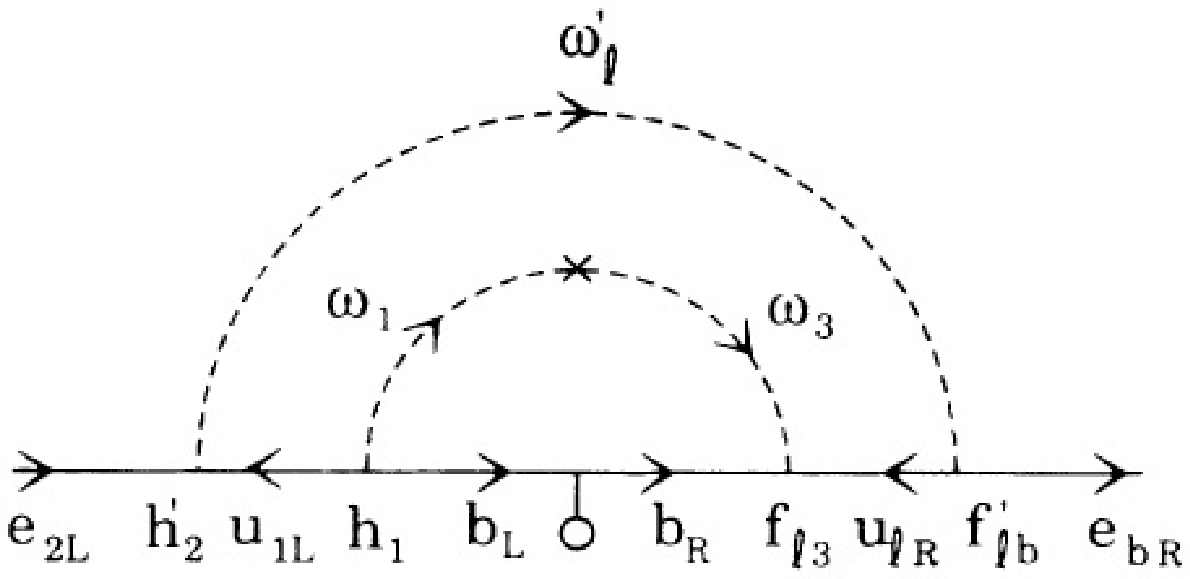} \\
\end{tabular}
\caption{One loop diagram inducing $\tau$ lepton mass (left) and
two--loop diagram inducing the muon mass (right).}
\label{Fig:radiativelep}
\end{center}
\end{figure}

Note that we cannot constrain the masses of the $\omega$ fields
from this process, since by taking the masses of the $\omega$ fields and
the soft breaking $\mu^2$ term to large values, the light fermion masses
will remain unchanged.

However, in the supersymmetric version of the radiative mass
generation mechanism, the new scalars should remain light, to
about 1 TeV, since the superpotential is un-renormalized.  That is
to say that in a SUSY context, in the exact SUSY limit, the
frmionic and bosonic loop diagrams add up to give zero.  Once SUSY breaking terms are turned
on, these diagrams will no longer cancel, and will generate finite
quark and lepton masses.  Thus, there is a
prediction in this scenario.  In addition to SUSY particles, LHC
should discover these $\omega_i$ particles and their
superpartners.

\section{The strong CP problem and its resolution \label{sec7}}

There is no indication of CP violation in strong interactions.  Yet,
the QCD Lagrangian admits a term
 \begin{equation}
{\cal L}_{QCD} = \frac{\theta\, g^2} {32 \pi^2} G_{\mu \nu}^a
\tilde{G}^{a \mu \nu}
\label{strong}
 \end{equation}
 which is $P$ and $T$ violating, and thus, owing to  $CPT$ invariance
$CP$ violating as well.  In Eq. (\ref{strong}), $\tilde{G}^{a \mu\nu} =
\frac{1}  {2} \epsilon^{\mu \nu \rho \sigma} G^a_{\rho \sigma}$ is the dual
field strength for the gluon.  The Lagrangian in Eq. (\ref{strong}) is
a total divergence, since $G_{\mu \nu}^a \tilde{G}^{a \mu \nu} = \partial_\mu K^\mu =
\partial_\mu[\epsilon^{\mu \nu \rho \sigma} A_\nu^a(F^a_{\rho \sigma} - \frac{2}{3}
\epsilon^{abc}A_\rho^b A_\sigma^c)]$.  In a $U(1)$ gauge theory, the resulting
surface term in the action would vanish for finite energy configurations.
Thus a term analogous to Eq. (\ref{strong})
does not lead to $P$ or $T$ violation in QED.  However, in QCD, the surface term
gives rise to non-zero contributions, owing to finite energy ``instanton"
configurations, causing $P$ and $T$ violation.

It is not the parameter $\theta$ in Eq. (\ref{strong}) that is physical.
Recall that the QCD Lagrangian also contains quark mass matrices $M_u$
and $M_d$, which are generated after electroweak symmetry breaking.  These
matrices are complex, and generate the KM phase for CP violation in weak interactions.
As discussed in Sec. \ref{sec2}, one makes bi--unitary transformations to bring
these matrices into diagonal form: $U_L^{u \dagger} M_u U_R^u = {\rm diag}(m_u,\,
m_c,\,m_t)$, and similarly for $M_d$.  If $U_L$ and $U_R$ belong to the global
$SU(N_f)_L \times SU(N_f)_R$ chiral symmetry ($N_f$ is the number
of quark flavors), which has no QCD anomaly, the diagonal quark masses cannot
be made real.  Specifically, Det($M_u) \rightarrow$ Det$(M_u)$ under such a
special bi--unitary transformation.  If the phases of the quark masses are denoted as
$\theta_{u,c,t}$ and $\theta_{d,s,b}$, the combination
\begin{equation}
\theta_{\rm QFD} = \theta_u+\theta_c+\theta_t+\theta_d+\theta_s+\theta_b = {\rm Arg}[{\rm Det}(M_q)]
\end{equation}
cannot be removed by anomaly--free rotations. A chiral rotation on the quark fields is necessary in order
to remove this phase.  This however will generate an anomaly term in the Lagrangian,
of the same form as in Eq. (\ref{strong}).  The physical parameter is then
\begin{equation}
\overline{\theta} = \theta + {\rm Arg}[{\rm Det}M_q]\,.
\end{equation}

With $\overline{\theta}$ physical, there will be CP violation in strong interactions.  However,
there are stringent constraints on the value of $\overline{\theta}$ from experimental limits on the
electric dipole moment (EDM) of the neutron:  $\overline{\theta} < 10^{-10}$.  This arises since in the
presence of $\overline{\theta}$ neutron EDM can be shown to have a non-zero value given by
\begin{equation}
d_n \simeq \left[10^{-16}\times \overline{\theta}\right]~ {\rm
e-cm}\,.
\end{equation}
From the experimental limit on neutron EDM, $d_n < 10^{-26}$ e-cm, one obtains the limit
$\overline{\theta} < 10^{-10}$. Why is it that a fundamental dimensionless parameter of the Lagrangian,
which should naturally be
of order one, so small is the strong CP problem.  If CP were a good symmetry
of the entire Lagrangian, small $\overline{\theta}$ would have been quite
natural.  However, weak interactions do break $CP$ invariance, which makes
the strong CP problem acute.

There are various proposed solutions to the problem. At some point
in time it was thought that the up quark mass may be zero.  If
true, that would solve the strong CP problem, since $\theta_u$ is
then un-physical, and therefore $\overline{\theta}$ can be
removed from the theory. But now we know, especially from lattice
gauge theory results, that $m_u=0$ is not an acceptable solution.

\subsection{Peccei--Quinn symmetry and the axion solution}

The most widely studied solution of the  strong CP problem is the
Peccei--Quinn (PQ) mechanism \cite{pq}, which yields a light
pseudo--Goldstone boson, the axion \cite{refww}. Here the parameter
$\overline{\theta}$ is promoted to a dynamical filed.  This field
acquires a non--perturbative potential induced by the QCD anomaly.
Minimization of the potential yields the desired solution $\overline{\theta}=0$,
solving the strong CP problem.

In the presence of the $\overline{\theta}$ term in the Lagrangian, non--perturbative
QCD effects will induce a vacuum energy given by
\begin{equation}
\label{energy}
E_{\rm vac} = \mu^4 \cos \overline{\theta}\,,
\end{equation}
where $\mu \sim \Lambda_{\rm QCD} \sim 100$ MeV.
This observation is crucially used in the PQ mechanism.  What if
$\overline{\theta}$ is a dynamical field?  Then this
non--perturbative potential will have to be minimized to locate the
ground state (unlike the case where $\overline{\theta}$ is a constant in
the Lagrangian).  Minimization of this potential will yield $\overline{\theta} = 0$,
as desired.

The essence of the PQ mechanism can be explained with a simple toy model \cite{barrtalk}.
Consider QCD with one quark flavor ($q$) and no weak interactions.  Suppose there is
a global $U(1)$ symmetry under which $q \rightarrow e^{-i \alpha \,\gamma_5/2} q$.
Such a symmetry has a QCD anomaly, and can only be imposed at the classical level.
A bare mass for $q$ is then forbidden.  Introduce now a complex color singlet scalar field $\phi$
which transforms under this $U(1)$ as $\phi \rightarrow e^{i \alpha} \phi$.  The following
Yukawa interaction is then allowed.
\begin{equation}
{\cal L}_{\rm Yuk} = Y \overline{q}_L \phi q_R + Y^*\overline{q}_R \phi^* q_L\,.
\end{equation}
The potential for $\phi$ also respects the $U(1)$ symmetry, and is given by
\begin{equation}
V(\phi) = -m_{\phi}^2 |\phi|^2 + \lambda |\phi|^4
\end{equation}
With a negative sign for $m_\phi^2$, the $\phi$ field will acquire a non-zero VEV, spontaneously breaking
the $U(1)$.  In this broken symmetric phase, we can parametrize $\phi$ as
\begin{equation}
\phi = \left[f_a + \tilde{\phi}(x^\mu) \right]e^{i a(x)/f_a}\,.
\end{equation}
Here $f_a$ is a real constant, while $\tilde{\phi}(x^\mu)$ and $a(x^\mu$) are dynamical (real)
fields.  The quark $q$ now acquires a mass, given by $M_q = Y f_a e^{i a(x)/f_a}$.  Making the quark
mass real by a field redefinition will induce a $\overline{\theta}$ given by
\begin{equation}
\overline{\theta}_{\rm eff} = \theta + {\rm Arg}[{\rm Det}\,Y] + \frac{1}{f_a} a(x^\mu) \,.
\end{equation}
The crucial point is that now $\overline{\theta}$ is a dynamical field, because of
the presence of the $a$ field, the axion.  Without non--perturbative  QCD effects, $a$
will be massless, since it is the Goldstone boson associated with the spontaneous breaking
of the global $U(1)$.  The vacuum energy analog of Eq. (\ref{energy}) is now
\begin{equation}
E_{\rm vac} = -\mu^4 \cos \overline{\theta}_{\rm eff}\,.
\end{equation}
Minimizing this potential with respect the dynamical $a$ field would yield $\overline{\theta}_{\rm eff} = 0$.

The field--dependent redefinition  on $q$, $q(x^\mu)  \rightarrow q(x^\mu)  e^{-i  (a(x^\mu)/f_a)(\gamma_5/2)}$
would remove the axion field from quark interactions except via derivatives, originating from the kinetic terms.
The axion also will have couplings to the gluon field strength.  These couplings are given by
\begin{equation}
\label{axionint}
{\cal L}_a = -\left(\frac{\partial_\mu a}{f_a}\right)\, \overline{q} \gamma_\mu \gamma_5 q  + \frac{g^2}{32 \pi^2}
\left(\frac{a}{f_a}\right) G \tilde{G} \,.
\end{equation}
It is the second term of Eq. (\ref{axionint}) that actually induces the potential for the axion.
Because of this potential, axion will have a mass of order $m_a \sim \Lambda_{\rm QCD}^2/f_a$.

The essentials of realistic axion model are already present in this toy model.  We need to turn on weak
interactions, and we need to add three families of quarks.  The straightforward implementation would involve
the SM extended to have two Higgs doublets, one coupling to the up--type quarks, and the other
coupling to the down--type quarks \cite{refww}.  A global $U(1)$ can then be defined classically, which has a
QCD anomaly.  The axion will now be part of the Higgs doublet, with the axion decay constant
$f_a \sim v \sim 10^2$ GeV.  The couplings of the axion to quarks, Eq. (\ref{axionint}), will
now be rather strong.  The decay $K^+ \rightarrow \pi^+ a$ will occur at an observable strength.
This process has been searched for, but has not been observed.  Negative results in searches for this and
other such processes have excluded the weak scale axion model.

Acceptable axion models of
the ``invisible" type \cite{dfsz,kim} involving high scale PQ symmetry
breaking are fully consistent.  In the model of Ref. \cite{dfsz}, in addition
to the two Higgs doublets, a complex singlet Higgs scalar $S$ is also introduced.  The
axion decay constant $f_a$ is now the VEV of $S$, which can be much above the weak scale.
The axion is primarily in $S$, with very weak couplings to the SM fermions.  There
are non--trivial constraints from astrophysics and cosmology on such a weakly interacting
light particle.  For example, axion can be produced inside supernovae.  Once produced,
they will escape freely, draining the supernova of its energy.  Consistency with supernova
observations requires that $f_a > 10^9$ GeV.  Cosmological abundance of the axion requires that
$f_a < 10^{12}$ GeV.

In the invisible axion model of Ref. \cite{kim}, there is only a single Higgs doublet of the
SM. A Higgs singlet $R$ and a heavy quark $Q$, which has vectorial properties under the SM, are
introduced.  The PQ $U(1)$ symmetry acts on $Q$ and the scalar $R$.  $Q$ acquires its mass
only via its Yukawa coupling with $R$.  (This example is essentially the same as the toy
model described above.)  The phase of $R$ is the axion in this case, with
phenomenology similar to, but somewhat different from, the axion model of Ref. \cite{dfsz}.

It should be noted that axion is a leading candidate for the cosmological dark matter.
For reviews of axion physics, astrophysics, cosmology, and detection techniques, see Ref. \cite{sikivi}.

\subsection{Solving strong CP problem with Parity symmetry}

There is another class of solution to the strong CP problem.  One
can assume Parity \cite{pcp,bdm} to set $\theta = 0$.  If the
fermion mass matrices have real determinant, then
$\overline{\theta}$ can be zero at the tree level.  Loop induced
$\overline{\theta}$ needs to be small, but this is not difficult
to realize.

Let me illustrate this idea with the left--right symmetric model
which has Parity invariance.  The Yukawa couplings are hermitian
in this setup.  To make the mass matrices also hermitian, we must
ensure that the VEVs of scalars are real.  This is easily done in
the SUSY version, which is what I will describe \cite{bdm}.
In SUSY models, one should also take into account the contributions
from the gluino to $\overline{\theta}$.

The model is the SUSY version of left--right symmetric model based
on the gauge symmetry $SU(3)_C \times SU(2)_L \times SU(2)_R \times
U(1)_{B-L}$ discussed in Sec. \ref{sec3}.  Two bi-doublet scalars $\Phi_i(1,2,2,0)$
($i=1,2$) are used to generate quark and lepton masses as well as
CKM mixings.  The relevant superpotential is given as
\begin{equation}
\label{W}
W = Y_u Q Q^c \Phi_u + Y_d Q Q^c \Phi_d\,.
\end{equation}
The Yukawa coupling matrices $Y_u$ and $Y_d$  will be hermitian, owing to
Parity invariance.  Parity also implies that the QCD Lagrangian parameter
$\theta =0$ and that the gluino mass is real. The soft SUSY breaking
$A$--terms, analogous to W in Eq. (\ref{W}) will also be hermitian. We shall
consider the case where these $A$ terms are proportional to the respective
Yukawa matrices.  Furthermore, we assume universal masses for the squarks,
as in minimal supergravity, or in gauge mediated SUSY breaking models.

The quark mass matrices
$M_{u,d}$ are hermitian at tree level since the VEVs of the bi-doublet
scalars turn out to be real. Therefore $\bar{\theta}=0$
at tree level.  We wish to demonstrate that loop induced contributions
to $\overline{\theta}$ are not excessive.  Note that this setup has two
hermitian matrices $Y_u$ and $Y_d$ which are complex, with all other
(flavor singlet) parameters being real.

Since parity is broken at a high scale (denoted as
$v_R$), a nonzero value of $\bar{\theta}$ will be induced at the
weak scale through renormalization group extrapolation below
$v_R$.  This is because the SM gauge symmetry does not permit the Yukawa
couplings to remain hermitian. The induced
$\bar{\theta}$ will have the general structure given by
\begin{eqnarray}
\delta \bar{\theta} = {\rm Im}{\rm Tr}[\Delta M_u M^{-1}_{u} +
\Delta M_d M^{-1}_{d}] -3\, {\rm Im}(\Delta M_{\tilde{g}}
M_{\tilde{g}}^{-1})
\end{eqnarray}
where  $M_{u,d,\tilde{g}}$ denote the tree level contribution to
the up--quark matrix, down--quark matrix  and the gluino mass
respectively, and $\Delta M_{u,d,\tilde{g}}$ are the loop
corrections.  To estimate the corrections from $\Delta M_u$ and
$\Delta M_d$, we note that the beta function for the evolution of
$Y_u$ below $v_R$ is given by $\beta_{Y_u} = Y_u/(16 \pi^2)(3
Y_u^\dagger Y_u + Y_d^\dagger Y_d + G_u)$ with the corresponding
one for $Y_d$ obtained by the interchange $Y_u \leftrightarrow
Y_d$ and $G_u \rightarrow G_d$.  Here $G_u$ is a
family--independent contribution arising from gauge bosons and the
Tr$(Y_u^\dagger Y_u)$ term. The $3Y_u^\dagger Y_u$ term and the
$G_u$ term cannot induce non--hermiticity in $Y_u$, given that
$Y_u$ is hermitian at $v_R$.  The interplay of $Y_d$ with $Y_u$
will however induce deviations from hermiticity.  Repeated
iteration of the solution with $Y_u \propto Y_uY_d^\dagger Y_d$
and $Y_d \propto Y_dY_u^\dagger Y_u$ in these equations will
generate the following structure:
\begin{equation}
\label{trace}
\delta \bar{\theta} \simeq \left(\frac{{\rm ln}(M_U/M_W)}  {16
\pi^2}\right)^4\left[ c_1 {\rm Im}{\rm Tr}\left(Y_u^2 Y_d^4 Y_u^4
Y_d^2 \right) +c_2 {\rm Im}{\rm Tr}\left(Y_d^2 Y_u^4 Y_d^4 Y_u^2
\right)\right]\,,
\end{equation}
where $M_U$ is the unification scale. Here $c_1$ and $c_2$  are
order one coefficients which are not equal.  To estimate the
induced $\bar{\theta}$, we choose a basis where $Y_u$ is diagonal,
$Y_u = D$ and $Y_d = V D' V^\dagger$ where $D_u v_u = {\rm
diag}(m_u,~m_c,~m_t)$, $D_dv_d = {\rm diag}(m_d,~m_s,~m_b)$ with
$V$ being the CKM matrix.  The Trace of the first term in Eq. (\ref{trace})
is then Im($D_i^2 D_k^4 D_j^{\prime^4} D_l^{\prime^2}
V_{ij}V_{kl}V_{il}^* V_{kj}^*)$.  The leading contribution in this
sum is $(m_t^4 m_c^2 m_b^4 m_s^2)/(v_{u}^{6}v_d^6){\rm
Im}(V_{cb}V_{ts}V_{cs}^* V_{tb}^*)$. The second Trace in Eq.(\ref{trace})
is identical, except that it has an opposite sign. Numerically
we find
\begin{equation}
\delta \bar{\theta} \sim  3 \times 10^{-27} ({\tan\beta})^6
(c_1-c_2)\,,
\end{equation}
which is well below the experimental limit of $10^{-10}$
from neutron EDM.

There are also finite corrections to the quark and gluino masses,
which are not contained in the RG equations.
Consider first the finite one loop corrections to the quark mass
matrices.  A typical diagram involving the exchange of squarks and
gluino is shown in Fig. \ref{Fig:strongcp}, where the crosses on the $\tilde{Q}$
and $\tilde{Q^c}$ lines represent (LL) and (RR) mass insertions
that will be induced in the process of RGE evolution. From this
figure we can estimate the form for  $\Delta M_u =
\frac{2\alpha_s}{3\pi} m^2_{\tilde{Q}} A_u m^2_{\tilde{u^c}}$
where $\tilde{Q}$ is the squark doublet and $\tilde{u^c}$ is the
right--handed singlet up squark. Without RGE effects, the trace of
this term will be real, and will not contribute to $\bar{\theta}$.
Looking at the RGE for $m^2_{\tilde{u^c}}$ upto two loop order, we
see that for the case of proportionality of $A_u$ and $Y_u$,
$m_{\tilde{u^c}}^2$ gets corrections having the form $m^2_0 Y^2_u$
or $m^2_0Y^4_u$ or $m^2_0 Y_uY^2_dY_u$. Therefore in $\Delta M_u
M^{-1}_u$, the $M^{-1}_u$ always cancels and we are left with a
product of matrices of the form
$Y^n_uY^m_dY^p_uY^q_d\cdot\cdot\cdot$. A similar comment applies
when we look at the RGE corrections for $m^2_{\tilde{Q}}$ or
$A_u$. If the product is hermitian, then its trace is real. So to
get a nonvanishing contribution to theta, we have to find the
lowest order product of $Y^2_u$ and $Y^2_d$ that is non--hermitian
and we get
\begin{equation}
\label{quarktheta}
\delta \bar{\theta} = \frac{2\alpha_s}{3\pi} \left(\frac{{\rm
ln}(M_U/M_W)}  {16 \pi^2}\right)^4 \left(k_1{\rm Im} {\rm
Tr}[Y^2_uY^4_dY^4_uY^2_d]+ k_2{\rm Im}{\rm
Tr}[Y^2_dY^4_uY^4_dY^2_u]\right)
\end{equation}
where $k_{1,2}$ are calculable constants.  The numerical estimate
of this contribution parallels that of the previous discussions,
$\delta \bar{\theta} \sim (k_1-k_2) \times 10^{-28}
(\tan\beta)^6$.  The contributions from the up--quark and down
quark matrices tend to cancel, but since the $\tilde{d^c}$ and the
$\tilde{u^c}$ squarks are not degenerate, $k_1 \neq k_2$ and the
cancellation is incomplete.

In Fig. \ref{Fig:strongcp} we have also displayed the one--loop contribution to the
gluino mass arising from the quark mass matrix.  Here again one
encounters the imaginary trace of two hermitian matrices $Y_u$ and
$Y_d$, in the case of universality and proportionality of SUSY
breaking parameters.  Our estimate for $\delta \bar{\theta}$ is
similar to that of the quark mass matrix of Eq. (\ref{quarktheta}).

This exercise shows that the strong CP problem can be consistently
resolved with the imposition of parity symmetry.

\begin{figure}[!htb]
\begin{center}
\begin{tabular}{cc}
\includegraphics[width=0.4\linewidth]{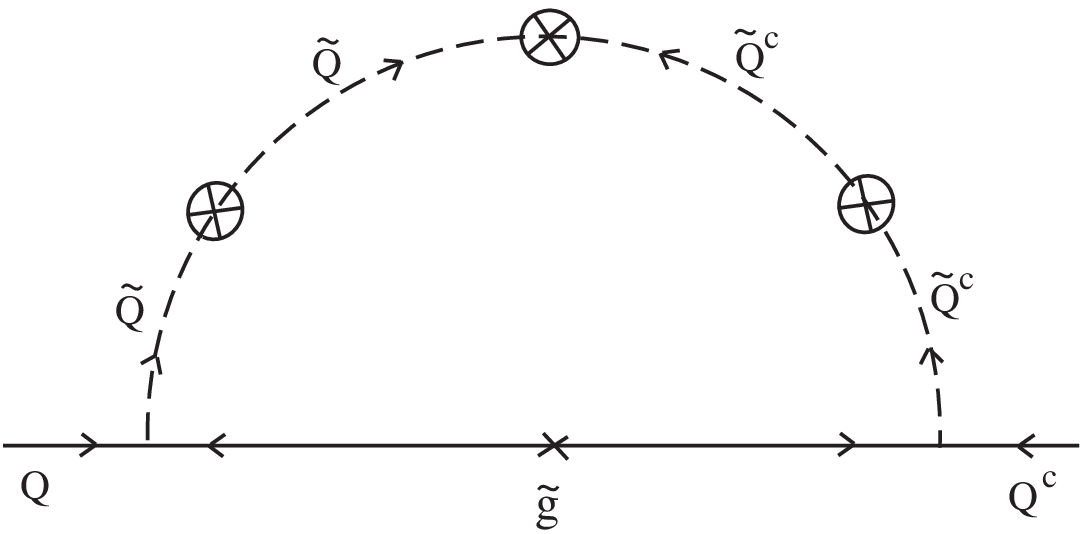} &
\includegraphics[width=0.4\linewidth]{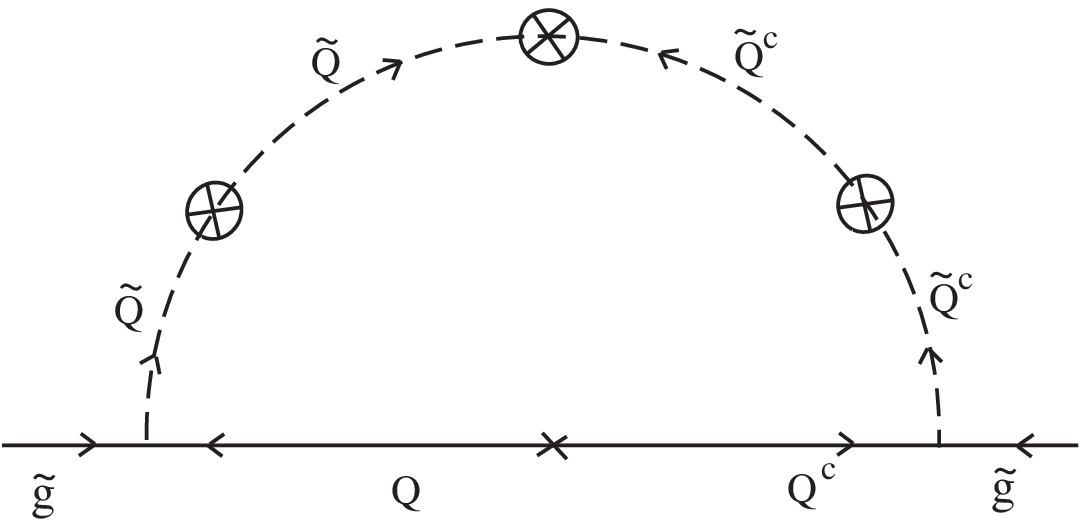} \\
\end{tabular}
\caption{One loop diagram inducing complex correction to the quark
mass (left) and to the gluino mass (right).}
\label{Fig:strongcp}
\end{center}
\end{figure}

\subsection{Solving the strong CP problem by CP symmetry}

The idea of Ref. \cite{nelson} is to use CP as a spontaneously broken symmetry.
The QCD $\theta$ is then zero.  In order to generate KM CP violation in
weak interactions, the mass matrices of the up and down quarks will have
to be complex.  This can be realized consistently, while keeping the
determinant of the quark mass matrix real by breaking CP spontaneously.
Then at tree--level $\overline{\theta}$ will
be zero.

A model of this type can be readily constructed.  Consider the addition of three
vector--like $D+D^c$ quarks to the SM.  These are $SU(2)$ singlets with $Y=\mp2/3$,
so that they can mix with the down--type quarks ($d,\,d^c$) of the SM.  Suppose there is a discrete
symmetry $Z_2$ under which the $d^c$ quarks reflect sign.  Several SM singlet Higgs scalar fields $S_i$
with $i \geq 2$ are also needed.  Under $Z_2$ these $S_i$ fields are odd.

The Yukawa Lagrangian of this theory is given by
\begin{equation}
\label{barrcp}
{\cal L}_{\rm Yuk} = Y_u Q u^c H + Y_D Q D^c \tilde{H} + M_D D d^c + F_i D D^c S_i + h.c.
\end{equation}
CP invariance implies that all the coupling matrices $(Y_u,\,Y_D,\,M_D,\,F_i)$ are real.
Complex phases appear only in the VEVs of the $S_i$ fields, which break CP spontaneously.
The down--type quark mass matrix arising from Eq. (\ref{barrcp}) is given by
\begin{eqnarray}
M_{d-D} =
\left(\begin{matrix}
0 & Y_D v \\ M_D & \sum_i F_i \left\langle S_i\right\rangle
\end{matrix} \right)\,.
\end{eqnarray}
When the heavy $D$ states are integrated out, the light $3 \times 3$ quark mass matrix
for the down quarks will have a complex form, yielding weak CP violation.  The determinant
of $M_{d-D}$ is real, owing to its structure (with all complex phases residing in the
lower right--hand block).  So $\overline{\theta} = 0$ in this model at tree level.

Loop corrections will induce non--zero $\overline{\theta}$ at the one loop level, which
has a magnitude of order $\overline{\theta} \sim F^2/(16 \pi^2)$.  For $F \sim 10^{-4}$,
this induced $\overline{\theta}$  will be within experimental limits.

\section{Rare $B$ meson decay and new physics \label{sec8}}

In this section we turn to specific processes where new physics
may show up at colliders.  It is quite likely that such processes
will show up first in the heaviest fermion ($t,\,b,\,\tau)$ systems.
Specifically, LHCb will be sensitive to such effects occurring in the
$B$ meson system.  We focus on this system here.

New physics may show up at the LHC in decays of the $B$ meson that
are rare or forbidden in the SM.  Low energy supersymmetry can provides such
possibilities.  Specifically, in the framework of SUSY with minimal flavor
violation \cite{mfv}, that is, flavor violation arising only via the MSSM Yukawa
couplings,  there are processes that are enhanced at large $\tan\beta$ which
can be in the observable range.

One such example is the rare
decay $B_{s,d} \rightarrow \mu^+\mu^-$ that has not been observed
so far.  In the SM, this process occurs via penguin and box
diagrams. The branching ratio has been calculated to be \cite{buras}
\begin{eqnarray}
Br(B_s \rightarrow \mu^+ \mu^-) &=& (3.35 \pm 0.32) \times 10^{-9}\,,
\nonumber \\
Br(B_d \rightarrow \mu^+ \mu^-) &=& (1.03 \pm 0.09) \times
10^{-10}\,.
\end{eqnarray}
This prediction is to be compared with the current experimental
limits from CDF and D0 \cite{PDG}
\begin{eqnarray}
Br(B_s \rightarrow \mu^+ \mu^-) &<& (5.8 \pm 0.32) \times 10^{-8}\,,
\nonumber \\
Br(B_d \rightarrow \mu^+ \mu^-) &<& (1.8 \pm 0.09) \times 10^{-8}\,.
\end{eqnarray}
There is a lot of room for new physics in these processes.  At the
LHC, sensitivity of the experiments will be better than the SM
prediction.

\subsection{$B_s \rightarrow \mu^+\mu^-$ in MSSM at large $\tan\beta$}

Minimal supersymmetry at large $\tan\beta$ can significantly enhance the
decay rate $B_s \rightarrow \mu^+ \mu^-$.  This occurs via exchange of Higgs
bosons of MSSM \cite{bk2}.  MSSM Yukawa couplings do preserve flavor at
the tree level, see Eq. (\ref{YukawaMSSM}).  That is, in the quark sector only
$H_u$ couples to the up--quarks, while only $H_d$ couples to the down--quarks.
There is no tree--level FCNC mediated by the Higgs boson.  However, this situation
changes once loop corrections to the Yukawa couplings are included.

To see this, let us begin by writing the
effective Lagrangian for the interactions of the two Higgs
doublets with the quarks in an arbitrary basis:
\begin{equation}
-{\cal L}_{ef\!f}=\overline{D}_R {\bf Y_D} Q_L H_d + \overline{D}_R {\bf Y_D}
\left[\epsilon_g+\epsilon_u {\bf Y_U^\dagger}{\bf Y_U}\right] Q_L H_u^*+h.c. \label{leff1}
\end{equation}
Here ${\bf Y_D}$ and ${\bf Y_U}$ are the $3\times3$ Yukawa matrices of
the microscopic theory, while the $\epsilon_{g,u}$ are the finite,
loop-generated non-holomorphic Yukawa coupling coefficients. The
leading contributions to $\epsilon_g$ and $\epsilon_u$ are generated by
the two diagrams in Fig. \ref{Fig:Bmumu}.

\begin{figure}[!htb]
\begin{center}
\includegraphics[width=0.8\linewidth]{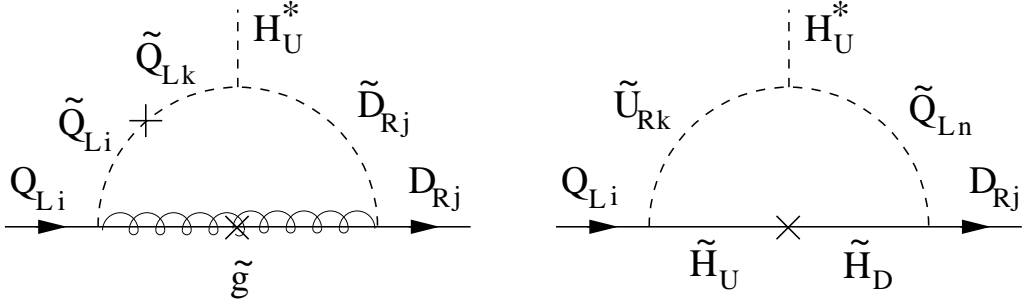}
\caption{One loop diagram inducing $\tau$ lepton mass (left) and
two--loop diagram inducing the muon mass (right).}
\label{Fig:Bmumu}
\end{center}
\end{figure}

Consider the first diagram in Fig. \ref{Fig:Bmumu}. If all $\tilde Q_i$ masses
are assumed degenerate at some scale $M_{\rm unif}$ then, at lowest
order, $i=k$ and the diagram contributes only to $\epsilon_g$:
\begin{equation}
\epsilon_g \simeq\frac{2\alpha_3}{3\pi}\mu^*M_3 f(M_3^2,m^2_{\tilde
  Q_L},m^2_{\tilde d_R})\,,
\end{equation}
where
\begin{equation}
f(x,y,z)=-\frac{xy\log(x/y)+yz\log(y/z)+zx\log(z/x)}{(x-y)(y-z)(z-x)}\,.
\end{equation}
Meanwhile, the second diagram of Fig. \ref{Fig:Bmumu} contributes to
$\epsilon_u$:
\begin{equation} \epsilon_u \simeq \frac{1}{16\pi^2}\mu^* A_U
f(\mu^2,m^2_{\tilde Q_L},m^2_{\tilde
  u_R})\,.
\label{epsu1}
\end{equation}
(We assume that the trilinear $A$-terms can be
written as some flavor-independent mass times ${\bf Y_U}$.) For typical
inputs, one usually finds $|\epsilon_g|$ is about 4 times larger than
$|\epsilon_u|$.

Owing to these loop corrections,
the CKM mixing angles receive finite corrections.  In particular,
\begin{equation}
V_{ub}\simeq V^0_{ub}\left[\frac{1+\epsilon_g\tan\beta}
{1+(\epsilon_g+\epsilon_u y_t^2)\tan\beta}\right]\,. \label{vub}
\end{equation}
The
same form also holds for the corrected $V_{cb}$, $V_{td}$ and
$V_{ts}$.

 For $\epsilon_u\neq0$, however, the rotation that
diagonalized the mass matrix does not diagonalize the Yukawa
couplings of the Higgs fields, leading to FCNC Higgs couplings
given by
\begin{equation}
{\cal L}_{FCNC}=\frac{\bar
y_bV^*_{tb}}{\sin\beta}\, \chi_{FC} \left[V_{td}\bar
b_Rd_L+V_{ts}\bar b_Rs_L\right] \left(\cos\beta H_u^{0*}-\sin\beta
H_d^0\right) +h.c. \label{final}
\end{equation}
with the quark fields in the
physical/mass eigenbasis, and defining
\begin{equation}
\chi_{FC}=\frac{-\epsilon_uy_t^2\tan\beta}{(1+\epsilon_g\tan\beta)
[1+(\epsilon_g+\epsilon_uy_t^2)\tan\beta]}
\end{equation}
to parameterize the amount
of flavor-changing induced.

We now consider the rare decay $B^0\to\mu^+\mu^-$. This occurs via
emission off the quark current of a single virtual Higgs boson
which then decays leptonically. The amplitude for the process
$B_{(d,s)}^0\to\mu^+\mu^-$ is given by:
\begin{equation}
{\cal
A}=\eta_{{}_{QCD}} \frac{\bar y_b y_\mu V_{t(d,s)}V^*_{tb}}{2
\sin\beta}\,\chi_{FC} \left\langle 0 |\bar b_R d_L|
B_{(d,s)}^0 \right\rangle
\left[\bar\mu\left(a_1+a_2\gamma^5\right)\mu\right]
\end{equation}
where
\begin{eqnarray}
a_1&=&\frac{\sin(\beta-\alpha)\cos\alpha}{m_H^2} -
\frac{\cos(\beta-\alpha)\sin\alpha}{m_h^2}\,,\nonumber\\
a_2&=&-\frac{\sin\beta}{m_A^2}\,.
\end{eqnarray}
The partial width is then
\begin{equation}
\Gamma(B^0_{(d,s)}\to\mu^+\mu^-)=\frac{\eta^2_{{}_{QCD}}}{128\pi}\,
m_B^3 f_B^2\, \bar y_b^2 y_\mu^2\, |V_{t(d,s)}^*
V_{tb}|^2\,\chi_{FC}^2 (a_1^2+a_2^2). \label{width}
\end{equation}

For SUSY scalar masses of order 500 GeV, we can estimate the
branching ratio to be near current experimental limit for
$\tan\beta $ larger than about 30.  The reason for this enhancement
has to do with the dependence of this rate on $\tan\beta$.  For
large values of $\tan\beta$, the rate scales as $(\tan\beta)^6$.
Two powers of $\tan\beta$ arise each from $\bar y_b^2$ and
$y_\mu^2$, while the remaining two powers arise from $\chi_{FC}^2$.

New physics contributions in $B$ meson system can arise in SUSY
GUTs \cite{chang,ciuchini}.  Generically, these models predict large $\tilde{b}_R -
\tilde{s}_R$ mixing, especially when large neutrino mixing angles
are induced.  As a result, there is a SUSY box diagram that
contributes to $B_s-\overline{B}_s$ mixing, shown in Fig. \ref{Fig:susyfcnc}.  This
contribution can be at the level of 30\% of SM box diagram. Now,
in the SM, CP violation arising from mixing in $B_s$ is very
small, but the new diagrams can significantly alter this scenario.
There is also new contribution to direct $B$ decays, which can
also be comparable to the SM contribution.  These ideas will
therefore be tested in the near future at the LHC.

\begin{figure}[h]
\begin{center}
\begin{tabular}{cc}
\includegraphics[width=0.4\linewidth]{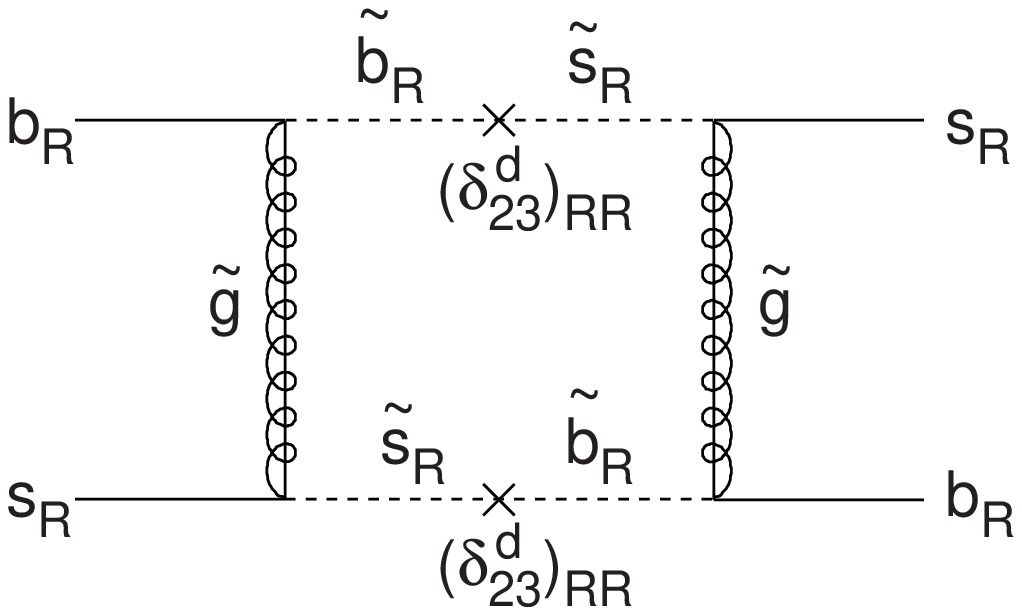} &
\includegraphics[width=0.4\linewidth]{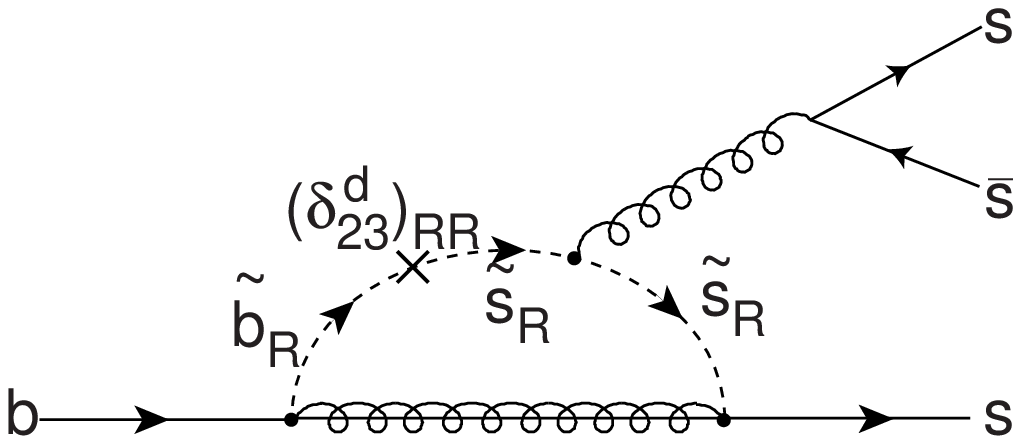} \\
\end{tabular}
\caption{New physics contributions to $B_s-\overline{B}_s$ mixing
and $b \rightarrow s \overline{s} s$ in SUSY GUTs. }
\label{Fig:susyfcnc}
\end{center}
\end{figure}

\section{Conclusion and Outlook}

Flavor physics is quite rich, in these lecture notes I have only scratched
the surface of a subset of the various issues.

It is a great triumph for experiment and theory, that we know so much about
the fundamental parameters of the flavor sector.  Even a few years ago, it
looked unlikely that so much would be learned with such high precision.  On
the experimental side, the two $B$ factories, BABAR and BELLE, have contributed
tremendously to the improved understanding.  We have seen substantial progress
on the theoretical understanding, especially from lattice gauge theory and Heavy Quark
Effective Theory in the last decade.  Both have played crucial roles in the precise
determination of the fundamental parameters of the quark flavor sector, viz.,  quark
masses, CKM mixing angles and CP violation.  While we have learned a great deal
about the fundamental parameters of the neutrino sector, in these lectures we
focused primarily on the quark sector.

Knowing the fundamental parameters precisely is only the start.  It is imperative that
we seek explanations to these observations.  Any such attempt will take us beyond
the realms of the standard model.  There is great hope that the LHC will actually
test some of the new ideas introduced to explain some the puzzles in the
flavor sector.  It should, however, be cautioned
that flavor dynamics could very well happen near the Planck scale, which would
mask its direct effects.  If there is low energy supersymmetry, there is a good
chance that flavor physics, even if it occurs at a very highs scale, transmits
information to the SUSY breaking sector, which may be observed.  The prime
candidate for these effects are rare decays of the type $\ell_i \rightarrow
\ell_j \gamma$. Observing such decays will show the existence of new flavor physic,
but it would be impossible, from these processes alone, to distinguish between
various possibilities.  We have seen that neutrino mass physics, GUT physics, and
flavor physics related to anomalous $U(1)$, all lead to the prediction that
$\mu \rightarrow e \gamma$ is in the observable range.

I have discussed at some length some, but not all, of the popular ideas that address the puzzles
from the flavor sector.  The mixing--mass sum rules in the quark sector appeared quite
promising, but with more precise data, many of the models in this class have already
been excluded.  It has become increasingly difficult to find precise patterns in the
masses and mixings that fit observations.  Perhaps the best setting to address these
issues is supersymmetric grand unification, supplemented by flavor symmetries.  SUSY
GUTs are well motivated on independent grounds, they have the power to shed light on
the flavor puzzle.  Some recent ideas along this line are discussed in Sec. \ref{sec5}.
I have also emphasized the close connection between the strong CP problem and the
flavor puzzle.  Axion solution to this problem is the most popular, but using P or
CP symmetries seem to work equally well.  These ideas may have collider signals, such
as the discovery of right--handed $W_R^\pm$ gauge bosons.

With some luck, the path chosen by Nature may be revealed at the LHC in the coming years.
Let us wait with hope.

\section*{Acknowledgments}

I wish to thank Tao Han for inviting me to lecture at TASI and
for his encouragement to write up these lecture notes.
I also wish to acknowledge many enjoyable discussions with
the participants at TASI 2008. It is a pleasure to thank K.T. Mahanthappa
and the University of Colorado physics department
for its warm hospitality. I have benefitted from discussions with
Zurab Tavartkiladze. This work is supported in part by DOE Grant Nos. DE-FG02-04ER41306
and DE-FG02-ER46140.


\newpage


\begin{thebibliography}{9}

\bibitem{C}
N.~Cabibbo,
  Phys.\ Rev.\ Lett.\  {\bf 10}, 531 (1963).

 \bibitem{KM}
 M.~Kobayashi and T.~Maskawa,
  Prog.\ Theor.\ Phys.\  {\bf 49}, 652 (1973).

\bibitem{refseesaw}
P.~Minkowski,
  Phys.\ Lett.\  B {\bf 67}, 421 (1977);\\
  M.~Gell-Mann, P.~Ramond and R.~Slansky, in {\it Supergravity} eds.
  P. van Nieuwenhuizen and D.Z. Freedman (North Holland, Amsterdam, 1979) p. 315;\\
  T.~Yanagida,
{\it In Proceedings of the Workshop on the Baryon Number of the Universe and Unified Theories, Tsukuba, Japan, 13-14 Feb 1979};\\
S.~L.~Glashow,
  NATO Adv.\ Study Inst.\ Ser.\ B Phys.\  {\bf 59}, 687 (1980);\\
R.~N.~Mohapatra and G.~Senjanovic,
  Phys.\ Rev.\ Lett.\  {\bf 44}, 912 (1980).

\bibitem{PMNS}
 B.~Pontecorvo,
  Sov.\ Phys.\ JETP {\bf 26}, 984 (1968)
  [Zh.\ Eksp.\ Teor.\ Fiz.\  {\bf 53}, 1717 (1967)];\\
  Z.~Maki, M.~Nakagawa and S.~Sakata,
  Prog.\ Theor.\ Phys.\  {\bf 28}, 870 (1962).


\bibitem{PDG}
C.~Amsler {\it et al.}  [Particle Data Group],
  Phys.\ Lett.\  B {\bf 667}, 1 (2008).

\bibitem{valle}
T.~Schwetz, M.~A.~Tortola and J.~W.~F.~Valle,
  New J.\ Phys.\  {\bf 10}, 113011 (2008).



\bibitem{tribi}
 P.~F.~Harrison, D.~H.~Perkins and W.~G.~Scott,
  Phys.\ Lett.\  B {\bf 530}, 167 (2002).

 \bibitem{A4}
  E.~Ma and G.~Rajasekaran,
  Phys.\ Rev.\  D {\bf 64}, 113012 (2001);\\
  K.~S.~Babu, E.~Ma and J.~W.~F.~Valle,
  Phys.\ Lett.\  B {\bf 552}, 207 (2003);\\
   G.~Altarelli and F.~Feruglio,
  Nucl.\ Phys.\  B {\bf 720}, 64 (2005);\\
   K.~S.~Babu and X.~G.~He,
  arXiv:hep-ph/0507217;\\
   W.~Grimus and L.~Lavoura,
  JHEP {\bf 0601}, 018 (2006);\\
  C.~Luhn, S.~Nasri and P.~Ramond,
  Phys.\ Lett.\  B {\bf 652}, 27 (2007);\\
  I.~de Medeiros Varzielas, S.~F.~King and G.~G.~Ross,
  Phys.\ Lett.\  B {\bf 644}, 153 (2007).




\bibitem{gasser} For a review see:
 J.~Gasser and H.~Leutwyler,
  Phys.\ Rept.\  {\bf 87}, 77 (1982).



\bibitem{manohar}
  A.~V.~Manohar and C.~T.~Sachrajda, ``Quark Masses,'' in {\it Review of particle physics},
  Phys.\ Lett.\  B {\bf 667}, 1 (2008).

 \bibitem{neubert}
For a review see:  M.~Neubert,
  Phys.\ Rept.\  {\bf 245}, 259 (1994).


\bibitem{refMILC}
  C.~Aubin {\it et al.}  [MILC Collaboration],
  Phys.\ Rev.\  D {\bf 70}, 114501 (2004).


\bibitem{refJLQCD}
  T.~Ishikawa {\it et al.}  [JLQCD Collaboration],
  Phys.\ Rev.\  D {\bf 78}, 011502 (2008).


\bibitem{refRBC}
  C.~Allton {\it et al.}  [RBC-UKQCD Collaboration],
  Phys.\ Rev.\  D {\bf 78}, 114509 (2008).

\bibitem{refHPQCD}
 Q.~Mason, H.~D.~Trottier, R.~Horgan, C.~T.~H.~Davies and G.~P.~Lepage
                  [HPQCD Collaboration],
  Phys.\ Rev.\  D {\bf 73}, 114501 (2006).



\bibitem{ramondrge}
See for example the compilation in H.~Arason, D.~J.~Castano, B.~Keszthelyi, S.~Mikaelian, E.~J.~Piard, P.~Ramond and B.~D.~Wright,
  Phys.\ Rev.\  D {\bf 46}, 3945 (1992).


\bibitem{xing}
 Z.~z.~Xing, H.~Zhang and S.~Zhou,
  Phys.\ Rev.\  D {\bf 77}, 113016 (2008).


\bibitem{wolf}
L.~Wolfenstein,
  Phys.\ Rev.\ Lett.\  {\bf 51}, 1945 (1983).

\bibitem{pdgmixing}
  A.~Ceccucci, Z.~Ligeti and Y.~Sakai, ``The CKM quark-mixing matrix,'' in
  {\it Review of particle physics},
 Phys.\ Lett.\  B {\bf 667}, 1 (2008).


\bibitem{flei} For a review see:  R.~Fleischer, ``Flavour Physics and CP Violation: Expecting the LHC,''
  arXiv:0802.2882 [hep-ph].

\bibitem{isgur}
 N.~Isgur and M.~B.~Wise,
  Phys.\ Lett.\  B {\bf 232}, 113 (1989); Phys.\ Lett.\  B {\bf 237}, 527 (1990).

\bibitem{Luke}
M.~E.~Luke,
  Phys.\ Lett.\  B {\bf 252}, 447 (1990).

\bibitem{inami}
T.~Inami and C.~S.~Lim,
  Prog.\ Theor.\ Phys.\  {\bf 65}, 297 (1981)
  [Erratum-ibid.\  {\bf 65}, 1772 (1981)].


\bibitem{refUTfit}
M.~Bona {\it et al.}  [UTfit Collaboration],
  Nuovo Cim.\  {\bf 123B}, 666 (2008).

\bibitem{refCKMfitter}
J.~Charles {\it et al.}  [CKMfitter Group],
  Eur.\ Phys.\ J.\  C {\bf 41}, 1 (2005).
 Updated fits from http://ckmfitter.in2p3.fr.



\bibitem{sumrule}
 S.~Weinberg,
  Trans.\ New York Acad.\ Sci.\  {\bf 38}, 185 (1977);\\
  F.~Wilczek and A.~Zee,
  Phys.\ Lett.\  B {\bf 70}, 418 (1977)
  [Erratum-ibid.\  {\bf 72B}, 504 (1978)];\\
 H.~Fritzsch,
  Phys.\ Lett.\  B {\bf 70}, 436 (1977).



\bibitem{LR}
 J.~C.~Pati and A.~Salam,
  Phys.\ Rev.\  D {\bf 8}, 1240 (1973);\\
   R.~N.~Mohapatra and J.~C.~Pati,
  Phys.\ Rev.\  D {\bf 11}, 566 (1975);\\
   G.~Senjanovic and R.~N.~Mohapatra,
  Phys.\ Rev.\  D {\bf 12}, 1502 (1975).



\bibitem{fritzsch}
H.~Fritzsch,
  Phys.\ Lett.\  B {\bf 73}, 317 (1978).


\bibitem{babushafi}
 See eg. K.~S.~Babu and Q.~Shafi,
  Phys.\ Rev.\  D {\bf 47}, 5004 (1993).


\bibitem{rasin}
L.~J.~Hall and A.~Rasin,
  Phys.\ Lett.\  B {\bf 315}, 164 (1993).


\bibitem{bk}
 K.~S.~Babu and J.~Kubo,
  Phys.\ Rev.\  D {\bf 71}, 056006 (2005).

 \bibitem{refFN}
 C.~D.~Froggatt and H.~B.~Nielsen,
  Nucl.\ Phys.\  B {\bf 147}, 277 (1979).


 \bibitem{GS}
 M.B. Green and J.H. Schwarz, Phys. Lett. {\bf B149}, 117
(1984); Nucl. Phys. {\bf B255}, 93 (1985);\\
M.B. Green, J.H. Schwarz and P. West, Nucl. Phys. {\bf B254}, 327 (1985).

\bibitem{be}
  K.~S.~Babu and T.~Enkhbat,
  Nucl.\ Phys.\  B {\bf 708}, 511 (2005).


\bibitem{lopsided}
K.~S.~Babu and S.~M.~Barr,
  Phys.\ Lett.\  B {\bf 381}, 202 (1996);\\
  C.~H.~Albright, K.~S.~Babu and S.~M.~Barr,
  Phys.\ Rev.\ Lett.\  {\bf 81}, 1167 (1998);\\
  J.~K.~Elwood, N.~Irges and P.~Ramond,
  Phys.\ Rev.\ Lett.\  {\bf 81}, 5064 (1998);\\
  J.~Sato and T.~Yanagida,
  Phys.\ Lett.\  B {\bf 430}, 127 (1998).


\bibitem{IbanezU1} L. E. Ibanez, G. G. Ross, Phys. Lett. {\bf B332}, 100 (1994);\\
P. Binetruy and P. Ramond, Phys. Lett. {\bf B350}, 49 (1995); \\
P. Binetruy, S. Lavignac and P. Ramond, Nucl. Phys. {\bf B477}, 353 (1996).

\bibitem{Kobayashi}T. Kobayashi, H. Nakano, H. Terao and K. Yoshioka, Prog. Theor. Phys. {\bf 110}, 247 (2003);\\
K.S. Babu, I. Gogoladze and K. Wang, Nucl. Phys. {\bf B660}, 322 (2003);\\
H.~K.~Dreiner, H.~Murayama and M.~Thormeier, Nucl.\ Phys.\  B {\bf 729}, 278 (2005).

\bibitem{beg} K.S. Babu, Ts. Enkhbat and I. Gogoladze, Nucl. Phys. {\bf B678}, 233 (2004).

\bibitem{Ginsparg} P. Ginsparg, Phys. Lett. {\bf B197}, 139 (1987);\\
V. S. Kaplunovsky, Nucl. Phys. {\bf B307}, 145 (1988),
Erratum-\textit{ibid.} {\bf B382}, 436 (1992).
\bibitem{cvetic}
M. Cvetic, L. L. Everett and J. Wang, Phys. Rev. {\bf D59}, 107901 (1999).

\bibitem{DSW}M. Dine, N. Seiberg and E. Witten, Nucl. Phys. {\bf B289}, 589
(1987);\\
J. Atick, L. Dixon and A. Sen, Nucl. Phys. {\bf B292}, 109 (1987).

\bibitem{bn}
  K.~S.~Babu and S.~Nandi,
  Phys.\ Rev.\  D {\bf 62}, 033002 (2000).

\bibitem{gl}
 G.~F.~Giudice and O.~Lebedev,
  Phys.\ Lett.\  B {\bf 665}, 79 (2008).


\bibitem{ps}
 J.~C.~Pati and A.~Salam,
  Phys.\ Rev.\  D {\bf 10}, 275 (1974)
  [Erratum-ibid.\  D {\bf 11}, 703 (1975)].

\bibitem{gg}
H.~Georgi and S.~L.~Glashow,
  Phys.\ Rev.\ Lett.\  {\bf 32}, 438 (1974).


\bibitem{gqw}
   H.~Georgi, H.~R.~Quinn and S.~Weinberg,
  Phys.\ Rev.\ Lett.\  {\bf 33}, 451 (1974).


\bibitem{bk1}
 K.~S.~Babu and C.~F.~Kolda,
  Phys.\ Lett.\  B {\bf 451}, 77 (1999).


\bibitem{hrs}
  L.~J.~Hall, R.~Rattazzi and U.~Sarid,
  Phys.\ Rev.\  D {\bf 50}, 7048 (1994).


\bibitem{gj}
  H.~Georgi and C.~Jarlskog,
  Phys.\ Lett.\  B {\bf 86}, 297 (1979).


\bibitem{hrr}
J.~A.~Harvey, D.~B.~Reiss and P.~Ramond,
  Nucl.\ Phys.\  B {\bf 199}, 223 (1982).


\bibitem{dhr}
   S.~Dimopoulos, L.~J.~Hall and S.~Raby,
  Phys.\ Rev.\ Lett.\  {\bf 68}, 1984 (1992); Phys.\ Rev.\  D {\bf 45}, 4192 (1992);\\
  G.~Anderson, S.~Raby, S.~Dimopoulos, L.~J.~Hall and G.~D.~Starkman,
  Phys.\ Rev.\  D {\bf 49}, 3660 (1994).


\bibitem{bmgj}
  K.~S.~Babu and R.~N.~Mohapatra,
  Phys.\ Rev.\ Lett.\  {\bf 74}, 2418 (1995).

%


\bibitem{so10}
H. Georgi, in Particles and Fields, Ed. by C. Carlson (AIP, NY, 1975);\\
 H.~Fritzsch and P.~Minkowski,
  Annals Phys.\  {\bf 93}, 193 (1975).

\bibitem{bpw}
  K.~S.~Babu, J.~C.~Pati and F.~Wilczek,
  Nucl.\ Phys.\  B {\bf 566}, 33 (2000).

 \bibitem{albrightbarr}
 C.~H.~Albright and S.~M.~Barr,
  Phys.\ Rev.\  D {\bf 58}, 013002 (1998);\\
  C.~H.~Albright, K.~S.~Babu and S.~M.~Barr,
  Phys.\ Rev.\ Lett.\  {\bf 81}, 1167 (1998);\\
  V.~Lucas and S.~Raby,
  Phys.\ Rev.\  D {\bf 55}, 6986 (1997);\\
   M.~C.~Chen and K.~T.~Mahanthappa,
  Int.\ J.\ Mod.\ Phys.\  A {\bf 18}, 5819 (2003).





\bibitem{bpr}
  K.~S.~Babu, J.~C.~Pati and P.~Rastogi,
  Phys.\ Rev.\  D {\bf 71}, 015005 (2005).


 \bibitem{bpr1}
   K.~S.~Babu, J.~C.~Pati and P.~Rastogi,
  Phys.\ Lett.\  B {\bf 621}, 160 (2005).

\bibitem{borzu}
F.~Borzumati and A.~Masiero,
  Phys.\ Rev.\ Lett.\  {\bf 57}, 961 (1986);\\
  For a more recent analysis see: J.~Hisano, T.~Moroi, K.~Tobe, M.~Yamaguchi and T.~Yanagida,
  Phys.\ Lett.\  B {\bf 357}, 579 (1995).



\bibitem{thooft}
  G.~'t Hooft,
  Nucl.\ Phys.\  B {\bf 35}, 167 (1971).


\bibitem{early}
  S.~Weinberg,
  Phys.\ Rev.\ Lett.\  {\bf 29}, 388 (1972);\\
 H.~Georgi and S.~L.~Glashow,
  Phys.\ Rev.\  D {\bf 7}, 2457 (1973);\\
S.~M.~Barr and A.~Zee,
  Phys.\ Rev.\  D {\bf 15}, 2652 (1977);\\
  L.~E.~Ibanez,
  Phys.\ Lett.\  B {\bf 117}, 403 (1982).

\bibitem{bala}
  B.~S.~Balakrishna, A.~L.~Kagan and R.~N.~Mohapatra,
  Phys.\ Lett.\  B {\bf 205}, 345 (1988);\\
 B.~S.~Balakrishna,
  Phys.\ Rev.\ Lett.\  {\bf 60}, 1602 (1988);\\
K.~S.~Babu and E.~Ma,
  Mod.\ Phys.\ Lett.\  A {\bf 4}, 1975 (1989);\\
H.~P.~Nilles, M.~Olechowski and S.~Pokorski,
  Phys.\ Lett.\  B {\bf 248}, 378 (1990);\\
 R.~Rattazzi,
  Z.\ Phys.\  C {\bf 52}, 575 (1991).


\bibitem{bmrad}
 K.~S.~Babu and R.~N.~Mohapatra,
  Phys.\ Rev.\ Lett.\  {\bf 64}, 2747 (1990).



 \bibitem{volkas}
  X.~G.~He, R.~R.~Volkas and D.~D.~Wu,
  Phys.\ Rev.\  D {\bf 41}, 1630 (1990).




\bibitem{dob}
  B.~A.~Dobrescu and P.~J.~Fox,
  JHEP {\bf 0808}, 100 (2008).



\bibitem{barrnew}
 S.~M.~Barr,
  Phys.\ Rev.\  D {\bf 76}, 105024 (2007);\\
  S.~M.~Barr and A.~Khan,
  Phys.\ Rev.\  D {\bf 79}, 115005 (2009).


\bibitem{pakvasa}
 S.~Pakvasa and H.~Sugawara,
  Phys.\ Lett.\  B {\bf 73}, 61 (1978).

\bibitem{pq}
 R.~D.~Peccei and H.~R.~Quinn,
  Phys.\ Rev.\ Lett.\  {\bf 38}, 1440 (1977);
  Phys.\ Rev.\  D {\bf 16}, 1791 (1977).

\bibitem{refww}
 S.~Weinberg,
  Phys.\ Rev.\ Lett.\  {\bf 40}, 223 (1978);\\
 F.~Wilczek,
  Phys.\ Rev.\ Lett.\  {\bf 40}, 279 (1978).

\bibitem{barrtalk} S. Barr, in {\it CP violation and the limits of the standard model},
TASI 94 Proceedings, ed. J.F. Donoghue, World Scientific Publication (1995).

\bibitem{dfsz}
  M.~Dine, W.~Fischler and M.~Srednicki,
  Phys.\ Lett.\  B {\bf 104}, 199 (1981);\\
 A.~R.~Zhitnitsky,
  Sov.\ J.\ Nucl.\ Phys.\  {\bf 31} (1980) 260
  [Yad.\ Fiz.\  {\bf 31} (1980) 497].

\bibitem{kim}
  J.~E.~Kim,
  Phys.\ Rev.\ Lett.\  {\bf 43}, 103 (1979);\\
  M.~A.~Shifman, A.~I.~Vainshtein and V.~I.~Zakharov,
  Nucl.\ Phys.\  B {\bf 166}, 493 (1980).

\bibitem{sikivi}
J.~E.~Kim,
  Phys.\ Rept.\  {\bf 150}, 1 (1987);\\
  P.~Sikivie,
  Phys.\ Rev.\ Lett.\  {\bf 51}, 1415 (1983)
  [Erratum-ibid.\  {\bf 52}, 695 (1984)].




\bibitem{pcp}
   R.~N.~Mohapatra and G.~Senjanovic,
  Phys.\ Lett.\  B {\bf 79}, 283 (1978);\\
  R.~N.~Mohapatra and A.~Rasin,
  Phys.\ Rev.\ Lett.\  {\bf 76}, 3490 (1996);\\
  R.~N.~Mohapatra, A.~Rasin and G.~Senjanovic,
  Phys.\ Rev.\ Lett.\  {\bf 79}, 4744 (1997).


\bibitem{bdm}
 K.~S.~Babu, B.~Dutta and R.~N.~Mohapatra,
  Phys.\ Rev.\  D {\bf 65}, 016005 (2002).




\bibitem{nelson}
 A.~E.~Nelson,
  Phys.\ Lett.\  B {\bf 136}, 387 (1984);\\
  S.~M.~Barr,
  Phys.\ Rev.\  D {\bf 30}, 1805 (1984).

\bibitem{mfv}
 G.~D'Ambrosio, G.~F.~Giudice, G.~Isidori and A.~Strumia,
  Nucl.\ Phys.\  B {\bf 645}, 155 (2002).


\bibitem{buras}
 A.~J.~Buras,
  Phys.\ Lett.\  B {\bf 566}, 115 (2003);\\
C.~Bobeth, M.~Bona, A.~J.~Buras, T.~Ewerth, M.~Pierini, L.~Silvestrini and A.~Weiler,
  Nucl.\ Phys.\  B {\bf 726}, 252 (2005).


\bibitem{bk2}
 K.~S.~Babu and C.~F.~Kolda,
  Phys.\ Rev.\ Lett.\  {\bf 84}, 228 (2000).


%

\bibitem{chang}
  D.~Chang, A.~Masiero and H.~Murayama,
  Phys.\ Rev.\  D {\bf 67}, 075013 (2003).



\bibitem{ciuchini}
  M.~Ciuchini, A.~Masiero, P.~Paradisi, L.~Silvestrini, S.~K.~Vempati and O.~Vives,
  Nucl.\ Phys.\  B {\bf 783}, 112 (2007).

\end{thebibliography}
\end{document}